\documentclass{aa}

\usepackage{graphicx}
\usepackage{txfonts}
\usepackage{subfig}
\usepackage{makecell}
\usepackage{gensymb} 
\begin{document}

\title{The luminosity history of fading local quasars over 10$^{4-5}$ years as observed by VLT/MUSE}

   \author{Finlez, C.
          \inst{1}
          \and
          Treister, E. \inst{2}
          \and 
          Bauer, F. \inst{2}
          \and
          Koss, M. \inst{3,4}
          \and
          Keel, W. \inst{5} 
          \and
          Maksym, W. P. \inst{6}
          \and
          Sartori, L. \inst{7}
          \and
          Venturi, G. \inst{8,9}
          \and
          Ricci, C. \inst{11}
          \and
          Nagar, N. \inst{10}
          \and
          Riesco, C. \inst{1}
          \and
          D\'iaz, Y. \inst{11}
          \and
          Parra, M. \inst{1}}
    
    \institute{Instituto de Astrof\'isica, Facultad de F\'isica, Pontificia Universidad Cat\'olica de Chile, Casilla 306, Santiago 22, Chile
    \and
    Instituto de Alta Investigaci{\'{o}}n, Universidad de Tarapac{\'{a}}, Casilla 7D, Arica, Chile
    \and
     Eureka Scientific, 2452 Delmer Street Suite 100, Oakland, CA 94602-3017, USA
    \and
     Space Science Institute, 4750 Walnut Street, Suite 205, Boulder, Colorado 80301
    \and
    Department of Physics and Astronomy, University of Alabama, Box 870324, Tuscaloosa, AL 35487, USA
    \and
    NASA Marshall Space Flight Center, Huntsville, AL 35812, USA
    \and
    ETH Z\"urich, Institute for Particle Physics and Astrophysics, Wolfgang-Pauli-Strasse 27, CH-8093 Z\"urich, Switzerland
    \and
    Scuola Normale Superiore, Piazza dei Cavalieri 7, I-56126 Pisa, Italy
    \and
    INAF-Osservatorio Astrofisico di Arcetri, Largo E. Fermi 5, 50125 Firenze, Italy
    \and
    Departamento de Astronom\'ia, Universidad de Concepci\'on, Concepci\'on, Chile 
    \and
    Instituto de Estudios Astrof\'isicos, Facultad de Ingenier\'ia y Ciencias, Universidad Diego Portales, Av. Ej\'ercito Libertador 441, Santiago, Chile }

\abstract
{
We present a comprehensive study of five nearby active galaxies featuring large (tens of kpc) extended emission-line regions (EELRs). The study is based on large-format integral field spectroscopic observations conducted with the Multi Unit Spectroscopic Explorer (MUSE) at the Very Large Telescope (VLT). The spatially resolved kinematics of the ionized gas and stellar components show signs of rotation, bi-conical outflows, and complex behavior likely associated with past interactions. Analysis of the physical conditions of the EELRs indicates that in these systems, the active galactic nucleus (AGN) is the primary ionization source. Using radiative transfer simulations, we compare the ionization state across the EELRs to estimate the required AGN bolometric luminosities at different radial distances.  Then, considering the projected light travel time, we reconstruct the inferred AGN luminosity curves. We find that all sources are consistent with a fading trend in intrinsic AGN luminosity by 0.2--3 dex over timescales of 40,000--80,000 years, with a time dependence consistent with previous studies of fading AGNs. These results support the hypothesis that most AGN undergo significant fluctuations in their accretion rates over multiple timescales ranging from 10,000 to 1,000,000 years, as proposed by existing theoretical models. These results provide new insights into the transient phases of AGN activity at previously unexplored scales and their potential long-term impact on their host galaxies through various feedback mechanisms.
}


\maketitle

\section{Introduction} 
\label{sec:intro}

A fundamental discovery in galaxy evolution is the tight correlation observed between the supermassive black holes (SMBHs) that reside at the center of most galaxies and the properties of the spheroidal component of the host galaxy \citep[][]{kormendy+2013}. This correlation implies a possible co-evolution between the SMBH and its host. The study of active galactic nuclei (AGN) can provide pivotal information about this co-evolution, as it reveals the SMBH actively accreting material. 

Stochastic variability is a defining feature of AGN \cite[e.g.][]{kozlowski+2016}. Variability has been observed on timescales ranging from minutes to days to decades \citep[e.g.,][]{ulrich+1997,padovani+2017} in all wavebands.
The variability time scales provide a powerful insight into the physical mechanisms beneath them. The minimum timescale observed at a specific band provides an estimate of the linear size of the coherent AGN component emitting at that waveband. 
Fast X-ray variability is thought to be produced in the innermost regions of the accretion flow. Longer time scales (days to months) are associated with UV and optical variability, which are highly correlated, indicating that they arise from similar physical regions \citep[from annuli of the accretion disk; e.g.][]{krolik+1991,arevalo+2008}.

An overall active accretion phase can last up to 10$^{7-9}$ yrs \citep[][]{marconi+2004}. However, simulations and observations \citep[]{novak+2011,schawinski+2015} indicate that this phase can be broken down in shorter ($\sim 10^{5}$ yr) cycles, causing the AGN to ‘’flicker’’ and implying that SMBHs grow episodically. 
These time scales are well beyond what can be directly observed in a human life. To fill in this timescale gap, we must look for probes of past activity, potentially imprinted on large physical scales.

Extended emission-line regions (EELRs) can be observed on some AGN, stretching tens of kpc from the nucleus. In some, the EELR exhibits evidence of photo-ionization by a powerful AGN, while the small-scale (10 pc) signatures of activity indicate a nuclear source that is several dex too faint to ionize the large-scale clouds. Considering the light-crossing time from the nucleus, this suggests an AGN fading in timescales of $\sim 10^{4-5}$ yr.
Altogether, $\sim$30 fading AGN candidates have been observed in recent years \cite[e.g.][]{lintott+2009, keel+2012a, schawinski+2015, gagne+2014, sartori+2016}.
AGN variability across a wide range of time scales has been uncovered through models \citep[][]{sartori+2018,sartori+2019} and simulations \citep[][]{novak+2011}.  \cite{sartori+2018} presents a framework to test whether variability can be explained based on the distribution of Eddington ratios among galaxy populations. Their analysis proposes that the observed AGN diversity is the result of generating light curves from the same set of statistical properties. Using a compilation of variability measurements from the literature, covering an extensive range of time lags and variability amplitudes, they create a structure function (SF) that provides a valuable overview of the AGN variability on a wide range of time scales. 

In this paper, we analyze five nearby active galaxies that present extended emission line regions. We create ionization models that are compared to our data to estimate the bolometric luminosity needed at to explain the photoionization state of the gas at different radii. Considering the light travel time from the central source to the ionized clouds, we can connect this to the AGN luminosity at different moments in the past. With this, we are able to infer the luminosity histories of the nuclear source. 

This paper is structured as follows: In Sect. \ref{sec:obs}, we provide details of the sample analyzed, the Multi Unit Spectroscopic Explorer (MUSE) observations, and the data reduction process. In Sect. \ref{sec:results}, we present our data analysis and results. In Sect. \ref{sec:discussion}, we discuss the reach and implications of our results, and conclusions are drawn in Sect. \ref{sec:conclusions}. Standard cosmological parameters (H $= 70$ km s$^{-1}$Mpc$^{-1}$, $\Omega_{\Lambda}=0.714$ and $\Omega_{m} = 0.28$) are adopted.

\section{Observations and data reduction} \label{sec:obs}
 
Our sample consists of five nearby active galaxies showing EELRs, taken from the \textit{voorwerpjes} sample  \citep[VPs; described in][]{keel+2012a}. The total VP sample was visually selected from an AGN catalog formed from the combination of optically selected AGN from SDSS, and the \cite{veron+1995} AGN catalog. This sample was complemented by serendipitously found objects from the Galaxy Zoo project \citep{lintott+2008}. The most robust candidates from this sample were followed up with H$\alpha$ and [OIII]$\lambda$5007 imaging, as well as optical long-slit spectra to confirm that the EELRs observed were mainly AGN photoionized. 

The five targets analyzed in the present paper were selected for further integral field spectroscopy observations based on: a) their redshift ($z <0.1$); b) the spatial extension of each galaxy and the associated EELRs, wihch should fit within the field of view (FOV) of the VLT/MUSE instrument, and c) the surface brightness of the clouds is high enough to achieve a sufficient signal-to-noise ratio in a reasonable exposure time. 
The galaxies of our sample were observed with the integral field spectrograph MUSE \citep[][]{bacon+2010} installed on the Very Large Telescope (VLT) using the wide field mode (WFM), which covers a 1’$\times$1’ field of view, has a wavelength range between $\sim$4600 and 9300 \AA, resolving power of R=1770-3590, and spatial sampling of 0\farcs2 per spaxel.
The observations were taken as part of program ID 0102.B-0107 (PI: L. Sartori), carried out between March 2019 and April 2020, for a cumulative on-target time of 11 hours and 27 minutes.
NGC 5972 and the Teacup galaxies were analyzed in depth by \cite{finlez+2022} and \cite{venturi+2023}, respectively. Therefore, we will focus here solely on processing the remaining sources. The observations for SDSS J151004.01+074037.1 (hereafter SDSS1510+07), UGC 7342, and SDSS J152412.58+083241.2 (hereafter SDSS 1524+08) were carried out in  3 observation blocks (OBs), with each OB consisting of 3 on-target observations of 950-second exposure each. The observations were dithered by 1\arcsec~(4 native pixels) in each direction.
Data were reduced using the ESO VLT/MUSE pipeline \citep[v2.8][]{weilbacher+2020}, under the ESO Reflex environment \citep{freudling+2013}. The reduction process consists, briefly, of the following steps: creation of a master bias and master flat and processing of the arc-lamp exposures. The calibrations are then applied to a standard star to generate a sensitivity function. Afterward, we applied the bias correction, flux, and wavelength calibrations to the raw science exposures. 
Given that the sources do not cover the entire FOV, a sky model is created from selected spaxels free of source emission. Then, this model is subtracted from the on-target exposures. 

To combine the dithered observations of each system, we identify point sources in the FOV to align the individual exposures, and then generate a stacked final data cube.

\begin{table*}
\caption{Sample observations}             
\label{table:obs_details}      
\centering          
\begin{tabular}{lcccccc}   
\hline\hline      
Source & RA, DEC & z & Scale (kpc/\arcsec) & FOV (kpc) & Res. (\arcsec) & Obs. time (s) \\
(1)   & (2)   & (3) & (4) & (5) & (6) &  (7) \\
\hline                    
SDSS J151004.01+074037.1 & 227.516728, 7.676991  & 0.04584 & 0.892 & 58$\times$57 & 0.85 & 8550 \\
UGC 7342                 & 184.580333, 29.253664 & 0.04770 & 0.926 & 60$\times$60 & 1.3  & 9500 \\
SDSS J152412.58+083241.2 & 231.052426, 8.544806  & 0.03712 & 0.732 & 47$\times$46 & 0.98 & 8820 \\
NGC 5972                 & 234.725692, 17.026192 & 0.02964 & 0.591 & 50$\times$50 & 0.78 & 5700 \\
The Teacup               & 217.624510, 13.653346 & 0.08506 & 1.563 & 50$\times$50 & 0.74 & 5700 \\
\hline                  
\end{tabular}
\tablefoot{Columns: (1) Target name, (2) Position in degrees, (3) Redshift, (4) Spatial scale in kpc/\arcsec, (5) Detector field-of-view in kpc, (6) Spatial resolution in arcseconds, and (7) Total observation time in seconds.}
\end{table*}

\section{Results}
\label{sec:results}

\begin{figure*}
    \centering
    \subfloat{
        \includegraphics[width=0.45\textwidth]{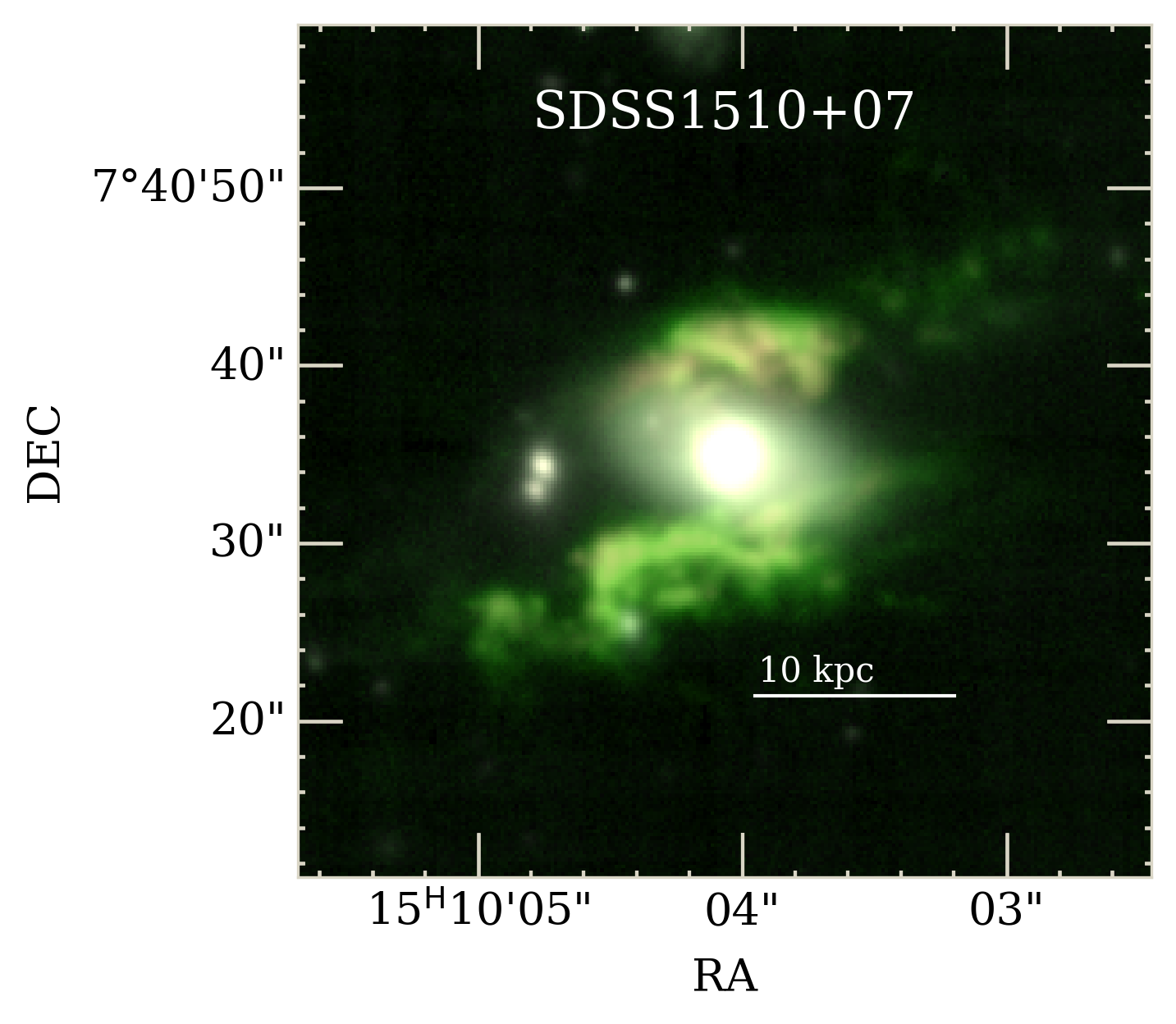}
        }
    \subfloat{
        \includegraphics[width=0.45\textwidth]{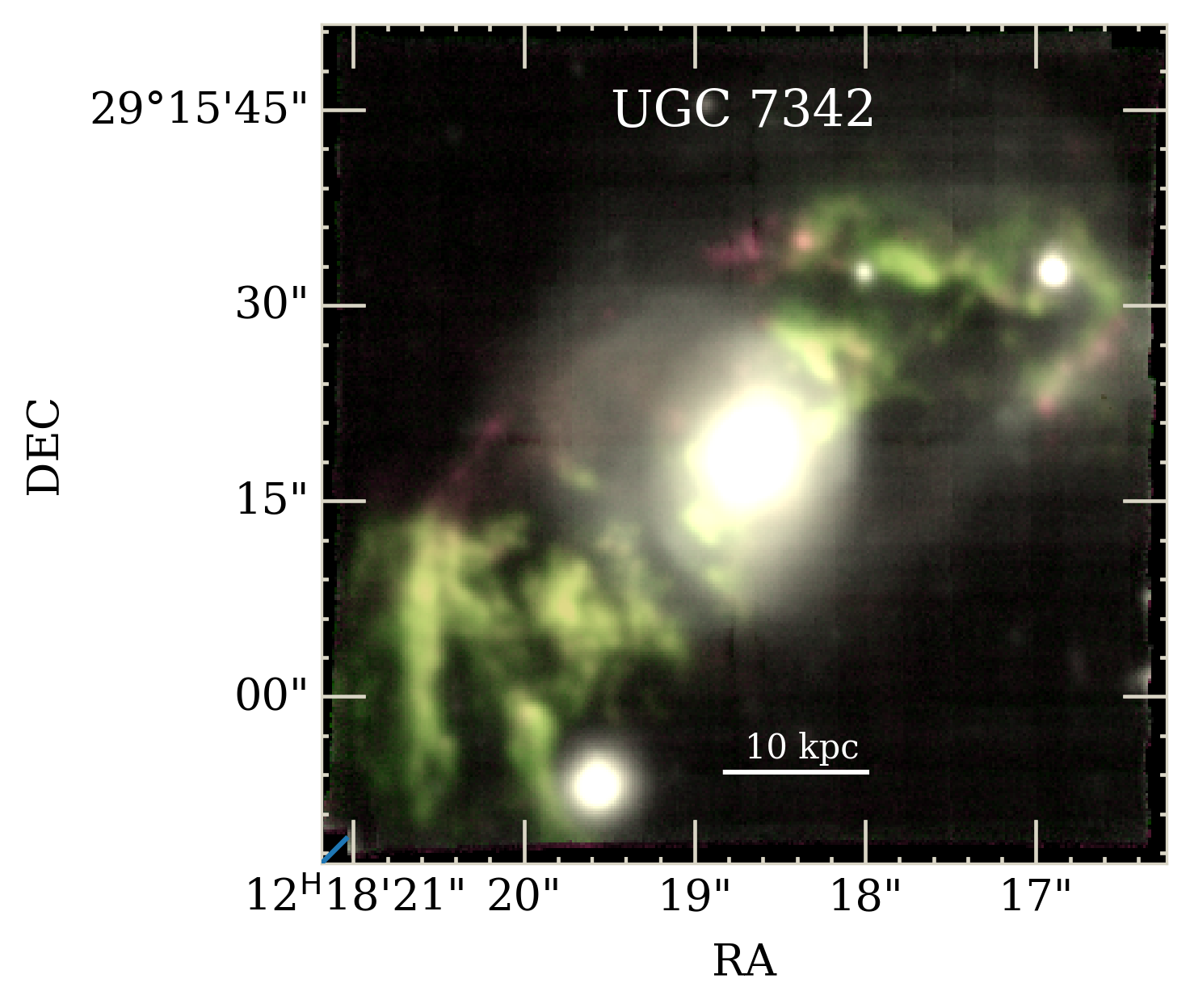}
    }
    
    \subfloat{
        \includegraphics[width=0.45\textwidth]{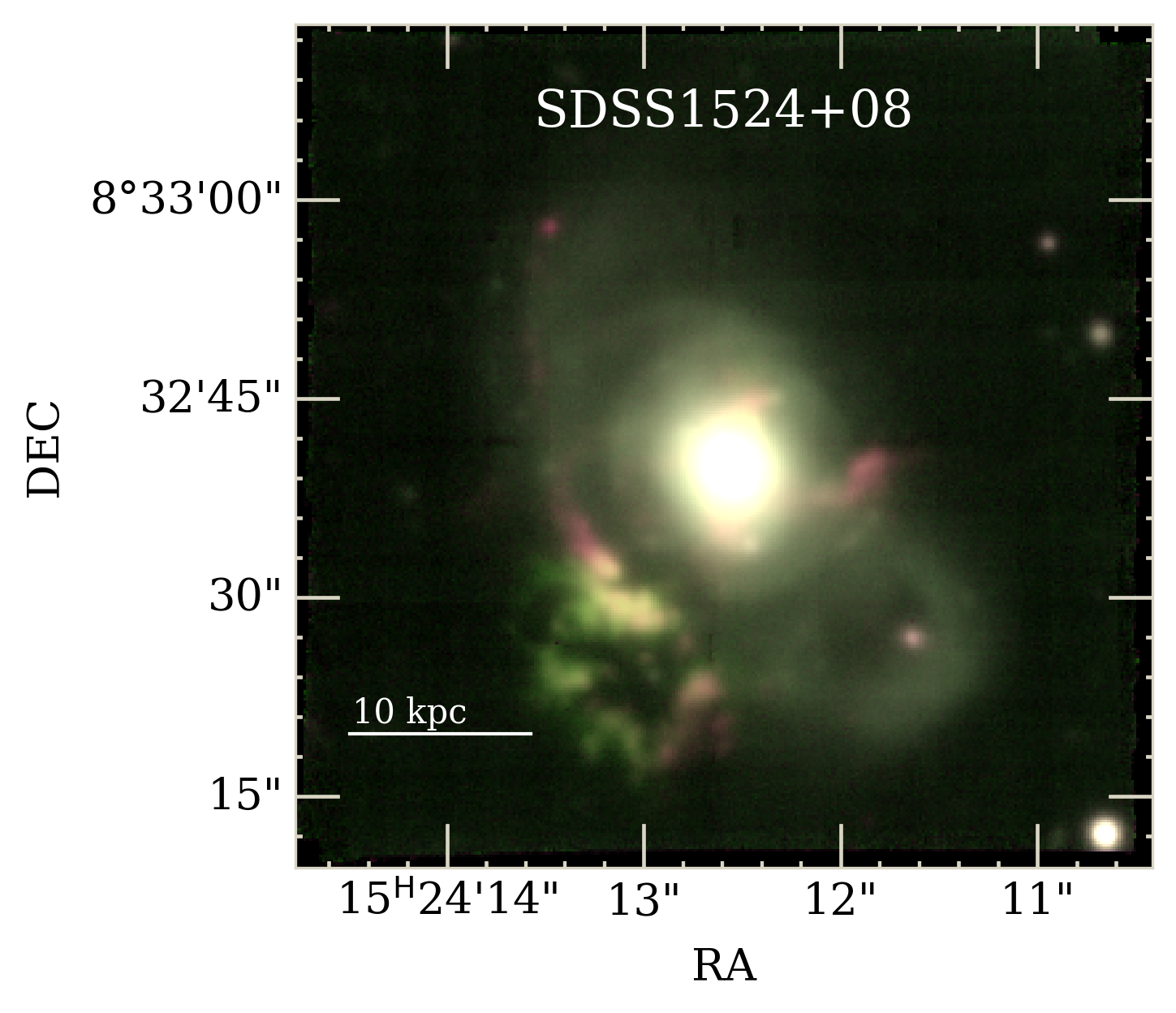}
    }
    \subfloat{
        \includegraphics[width=0.45\textwidth]{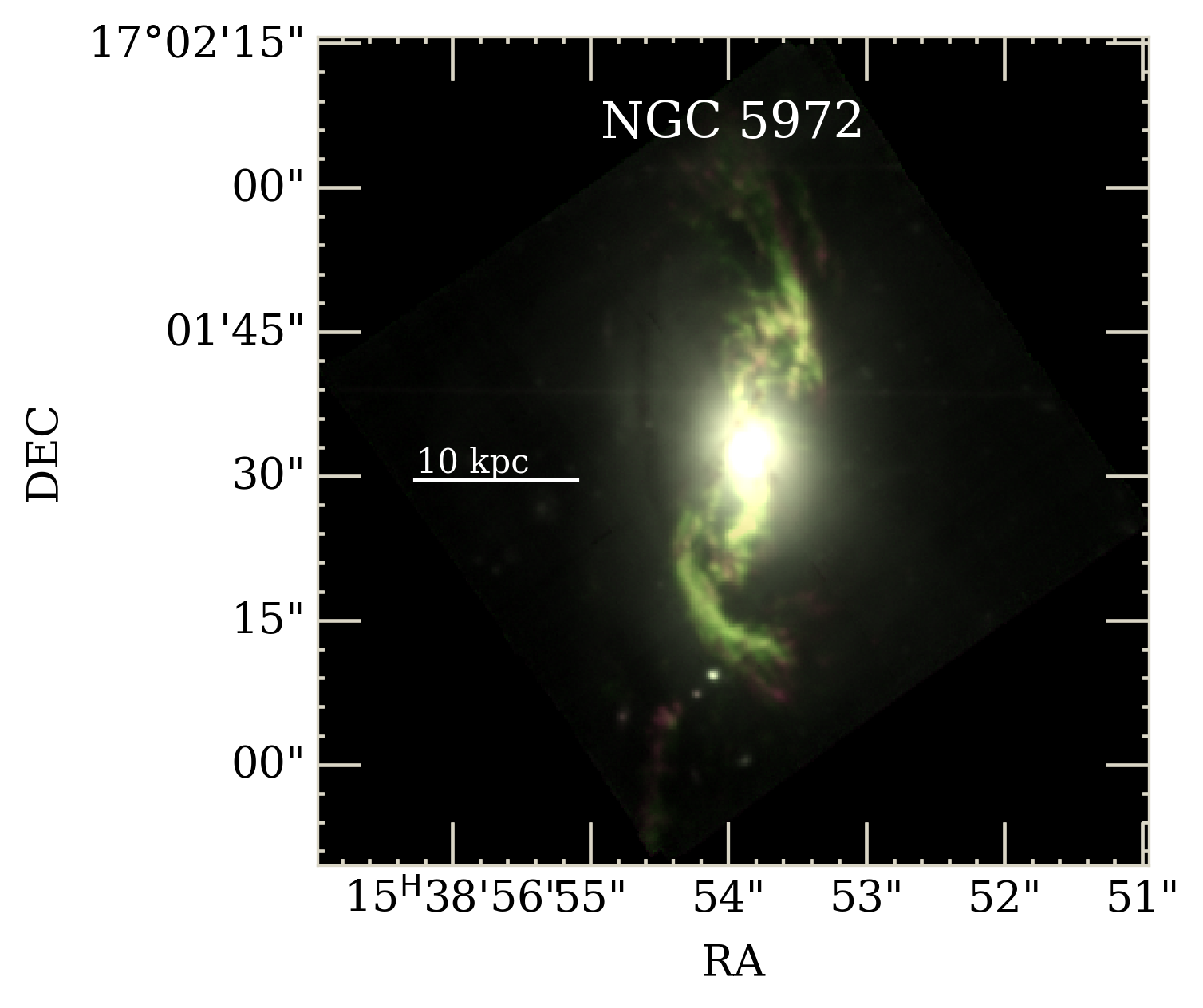}
    }\qquad   
    
    \subfloat{
        \includegraphics[width=0.45\textwidth]{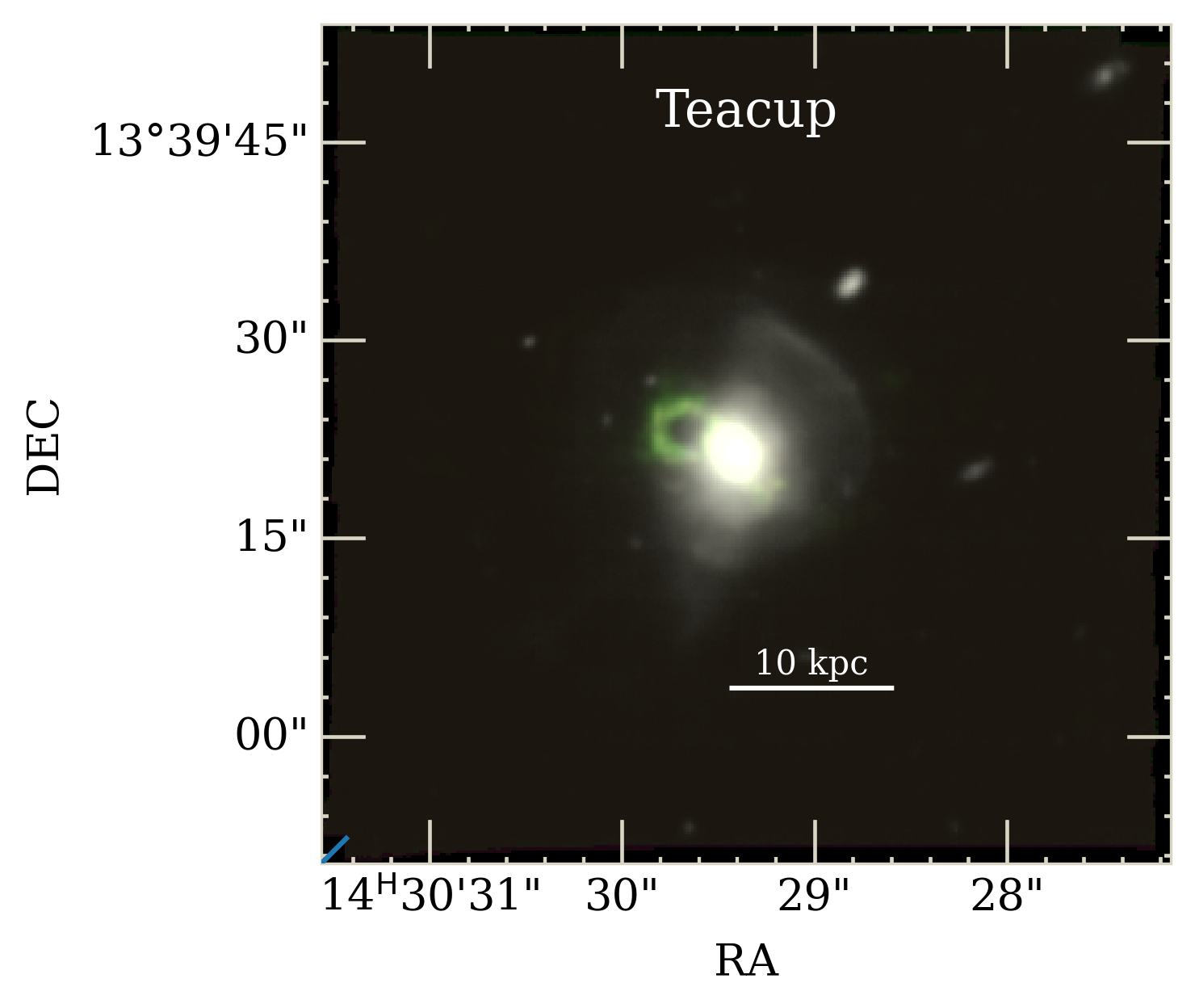}
    }\qquad
\caption{Composite color images for the targets analyzed in this work, with red, white, and green pseudo-colors. Images were created using three bands extracted from the data cube. Green was created with a spectral window that encloses the [OIII]5007 emission lines to represent the highly ionized gas; Red was created with a spectral window that encloses H$\alpha$ 6563\AA, while white represents the stellar continuum, with a spectral window of 7000-9000\AA. In all images, North is up and East to the left.}
 \label{fig:composites}
\end{figure*}

Color composite images of SDSS1510+07, UGC 7342, and SDSS1524+08 are shown in Fig. \ref{fig:composites}. These images were created by collapsing three pseudo-bands of the data cubes, encompassing [OIII] (in green pseudo-color, bandwidth of 40\AA), H$\alpha$ (in red, bandwidth of 20\AA), and the stellar continuum (in white) obtained by integrating the emission in the  7000-9000\AA\ range. 
In this figure, we can see the morphology of the EELRs. In the case of SDSS1510+07, the ionized gas extends N of the nucleus towards the W and S of the nucleus towards the E. The clouds appear to be somewhat symmetric, presenting a complex filamentary structure. 
UGC 7342 shows EELRs extending almost the entire FOV from SE to NW, also showing symmetry. As for SDSS524+08, the ionized cloud is observed mainly at the SE with no NW counterpart and, hence, highly asymmetric. At least partially, the EELR seems connected to a tidal arm illuminated by the nuclear ionizing source.  
NGC 5972 shows two large structures extending towards the N and S with a helix-type shape, as discussed extensively by \citet{finlez+2022}. The teacup shows a loop of ionized gas extending 15 kpc to the NE, nicknamed  "the handle". This source has been studied in detail by several authors, including \citet{gagne+2014,keel+2017,lansbury+2018,venturi+2023}.

\subsection{Stellar populations}
\label{sec:stars}

In this section, we describe the process we followed to fit and subtract the optical continuum, aimed to isolate the stellar component morphology and kinematics from the ionized gas component. Additionally, we outline the observed properties of the flux distribution, velocity, and velocity dispersion maps.  This process was applied to all the galaxies in the sample. 

To ensure a minimum signal-to-noise ratio (SNR) in the entire FOV for our analysis, we bin our data cube using a Python implementation of the Voronoi binning method \citep[vorbin,][]{cappellari+2003}, requiring a minimum SNR of 20.
To study the stellar kinematics and extract the stellar continuum from the data cube for the posterior emission line analysis, we use the \texttt{Python} version of the penalized spaxel fitting algorithm \citep[pPXF;][]{cappellari+2004}. 

A linear combination of template spectra and additive or multiplicative polynomials was used to fit the resulting spectrum for each bin. For the template spectra, we use the E-MILES \citep[][]{vazdekis+2016} library, covering the entire MUSE wavelength range. The line-of-sight velocity distribution (LOSVD) is parameterized by a Gauss-Hermite function. The outcome of this routine is a model cube for the stellar continuum, which is then subtracted from our original data cube to obtain a continuum-free spectrum for every spaxel.

\begin{figure}
    \centering
    \subfloat{
        \includegraphics[width=0.95\columnwidth]{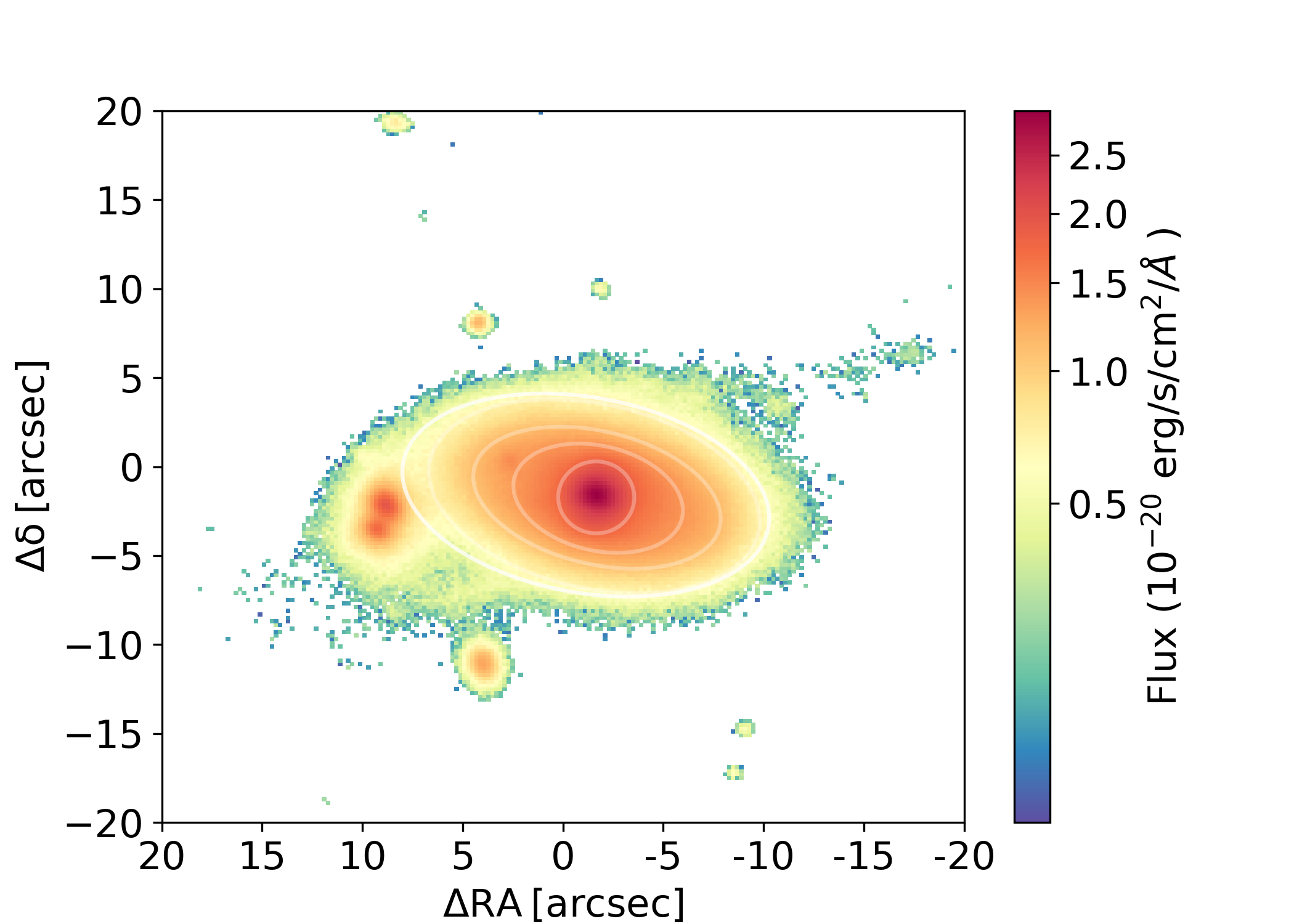}
    }\qquad   
    \subfloat{
        \includegraphics[width=0.95\columnwidth]{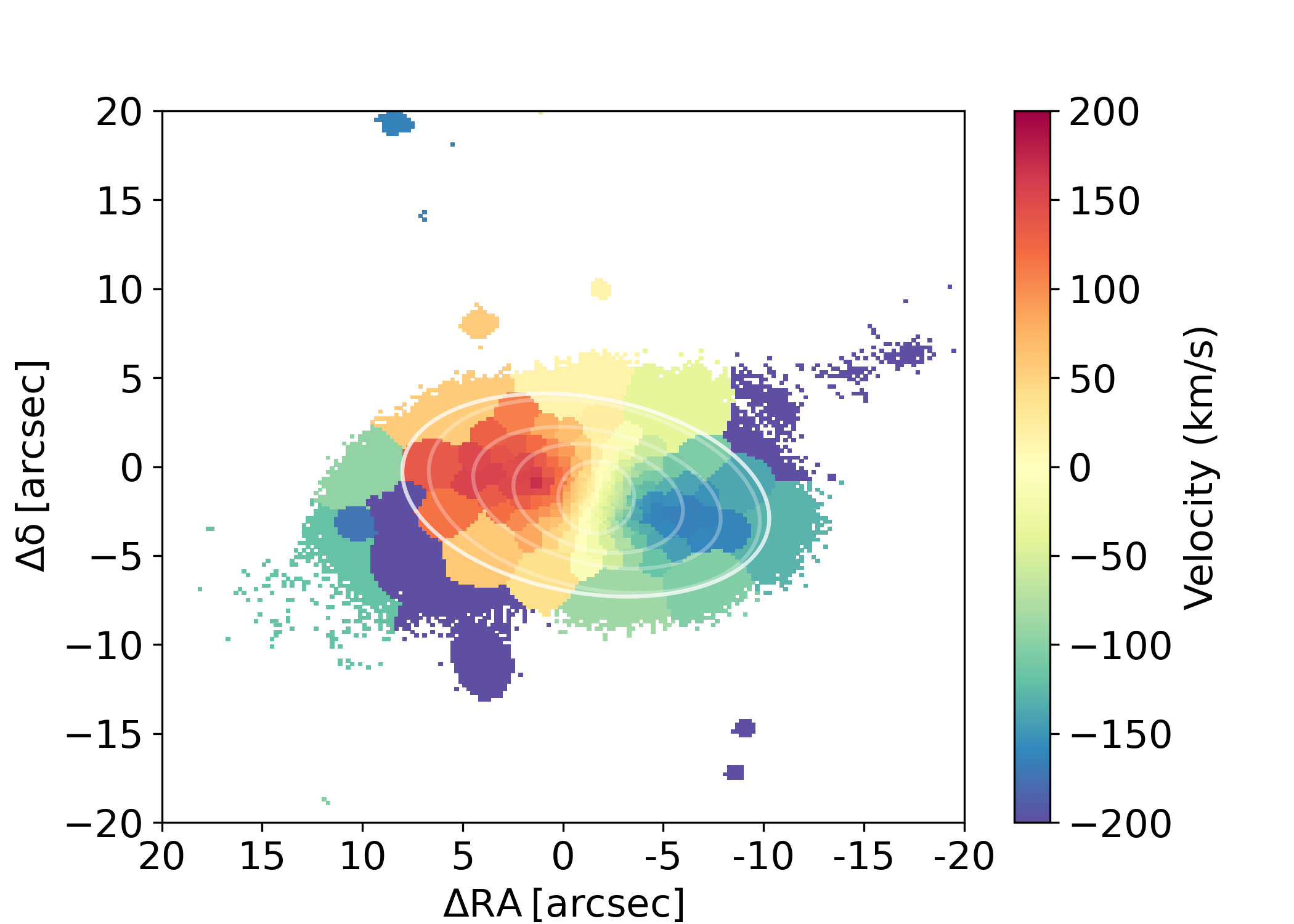}
    }\qquad     
    \subfloat{
        \includegraphics[width=0.95\columnwidth]{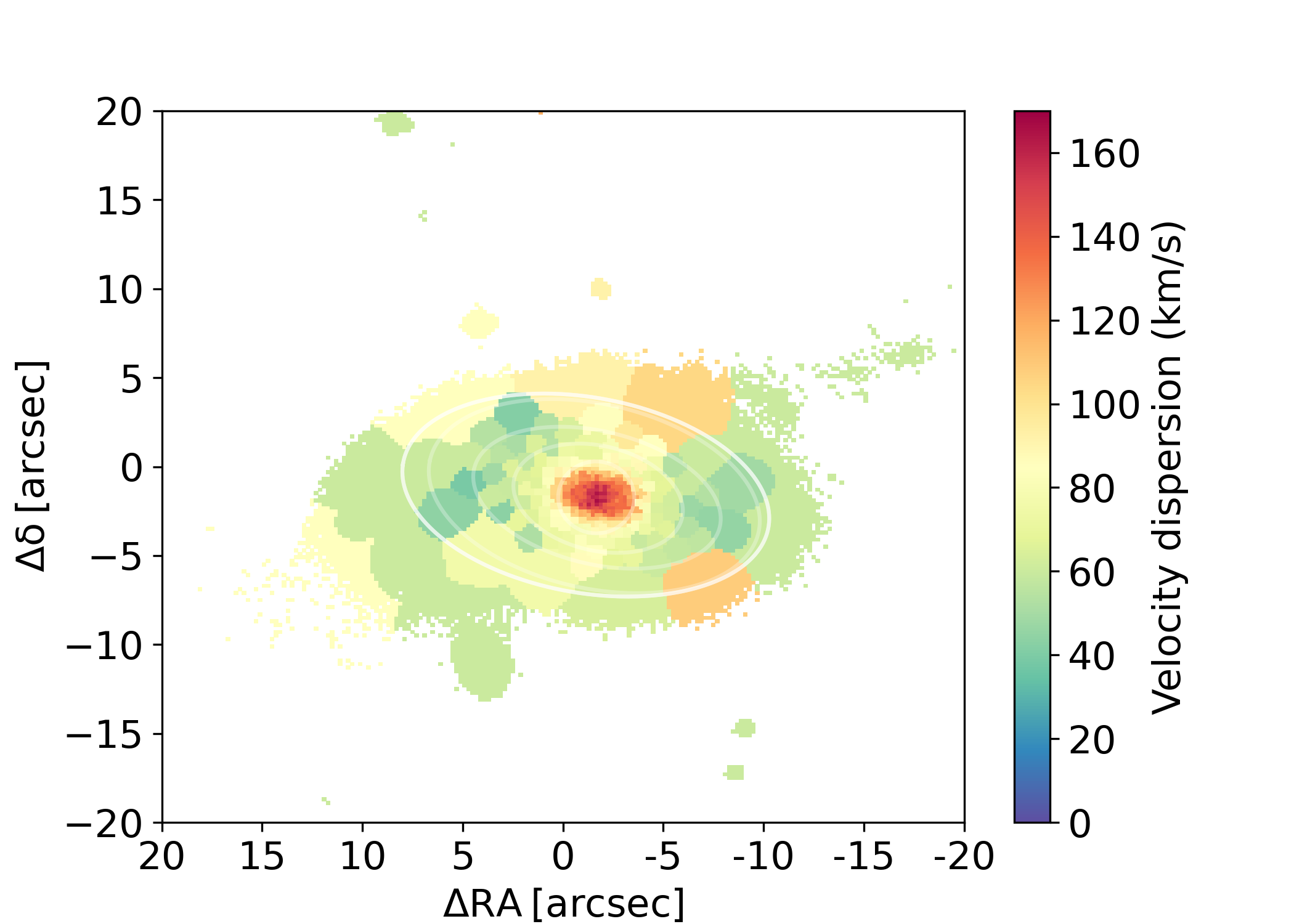}
    }\qquad   
\caption{Flux, velocity, and velocity dispersion maps for the stellar component of SDSS1510+07.
The top panel shows the integrated flux, obtained from collapsing the data cube around 8000 \AA. Overplotted in white are the isophote profile fits, which show the extension and inclination of a stellar component, consistent with a disk. The velocity map (middle panel) shows that the kinematics of the stellar component are rotation-dominated. The velocity dispersion map (bottom map) shows higher velocities concentrated in the center, as expected from a rotation-dominated disk. }
 \label{fig:star_mom_1510}
\end{figure}

\textit{SDSS 1510+07:} In Fig. \ref{fig:star_mom_1510}, we show the optical continuum flux by collapsing the data cube around 8000\AA. We carry out an isophotal profile ellipse fitting, from which we infer an ellipticity ($\epsilon$) and position angle (PA) of 0.1 and 50\degree, respectively, for the central stellar component. A structure reminiscent of a tidal tail is observed to the NW and SE. The velocity and velocity dispersion maps are consistent with a relatively unperturbed rotating component. 

\begin{figure}
    \centering
    \subfloat{
        \includegraphics[width=0.95\columnwidth]{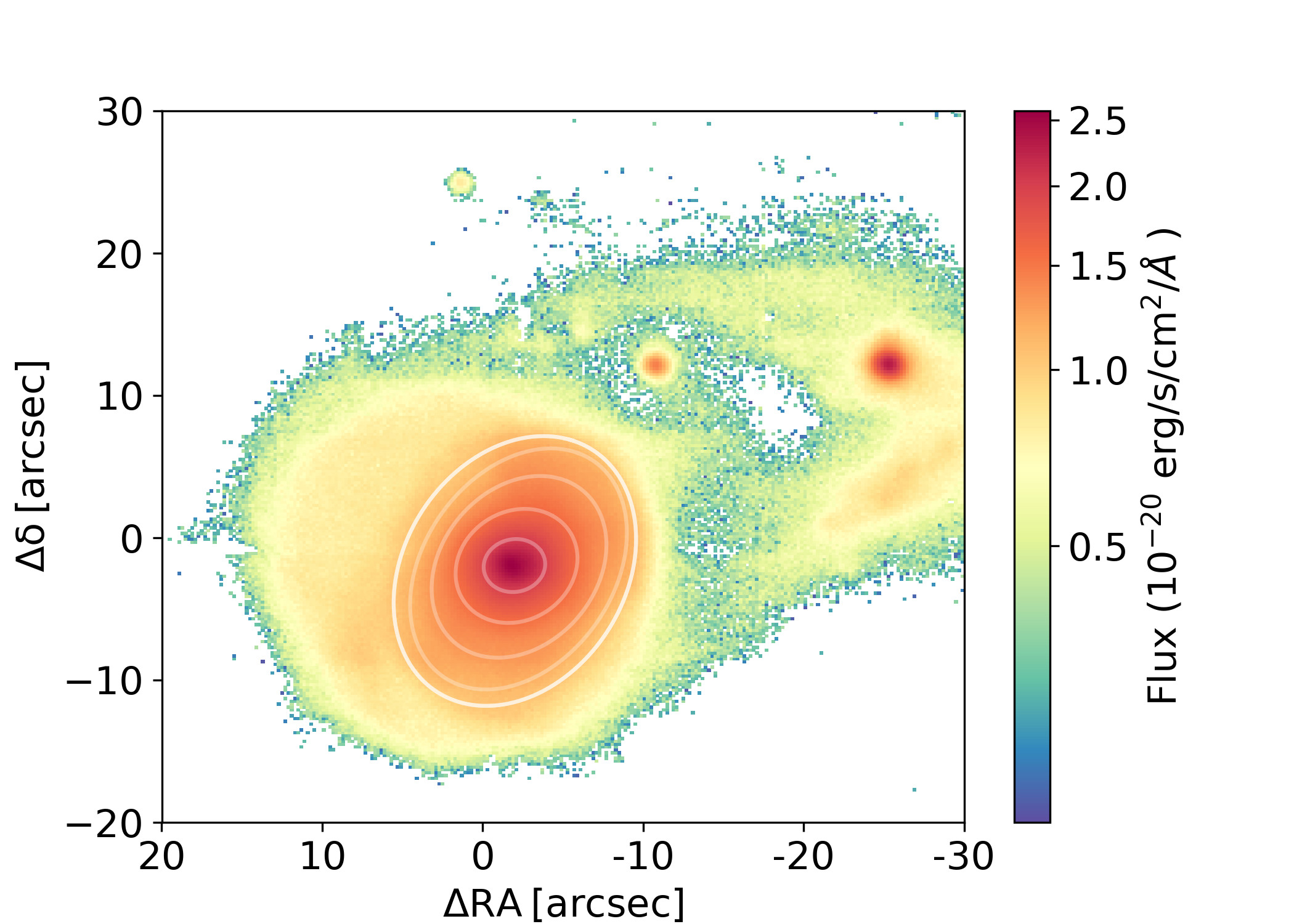}
    }
    
    \subfloat{
        \includegraphics[width=0.95\columnwidth]{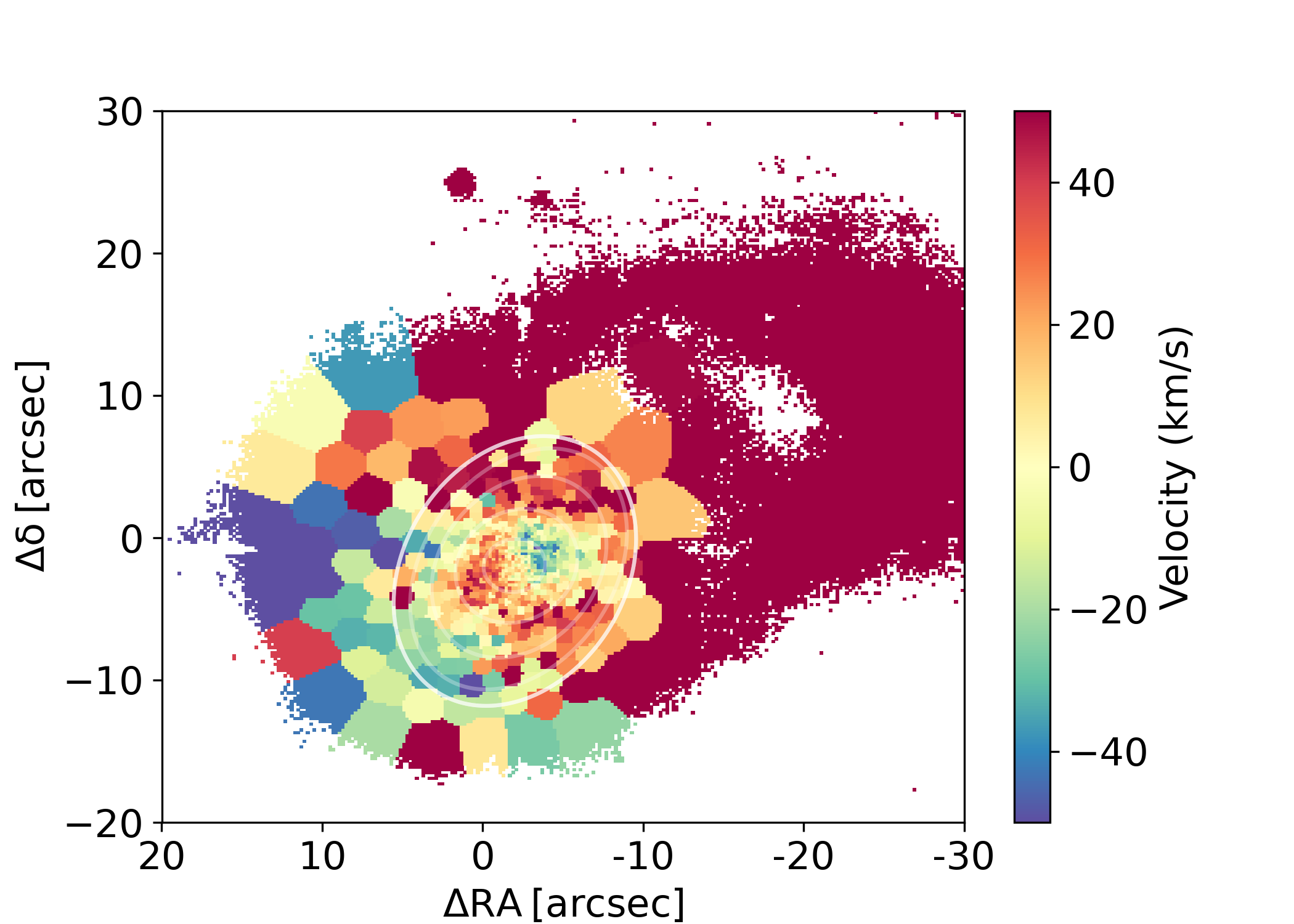}
    }
    
   \subfloat{
        \includegraphics[width=0.95\columnwidth]{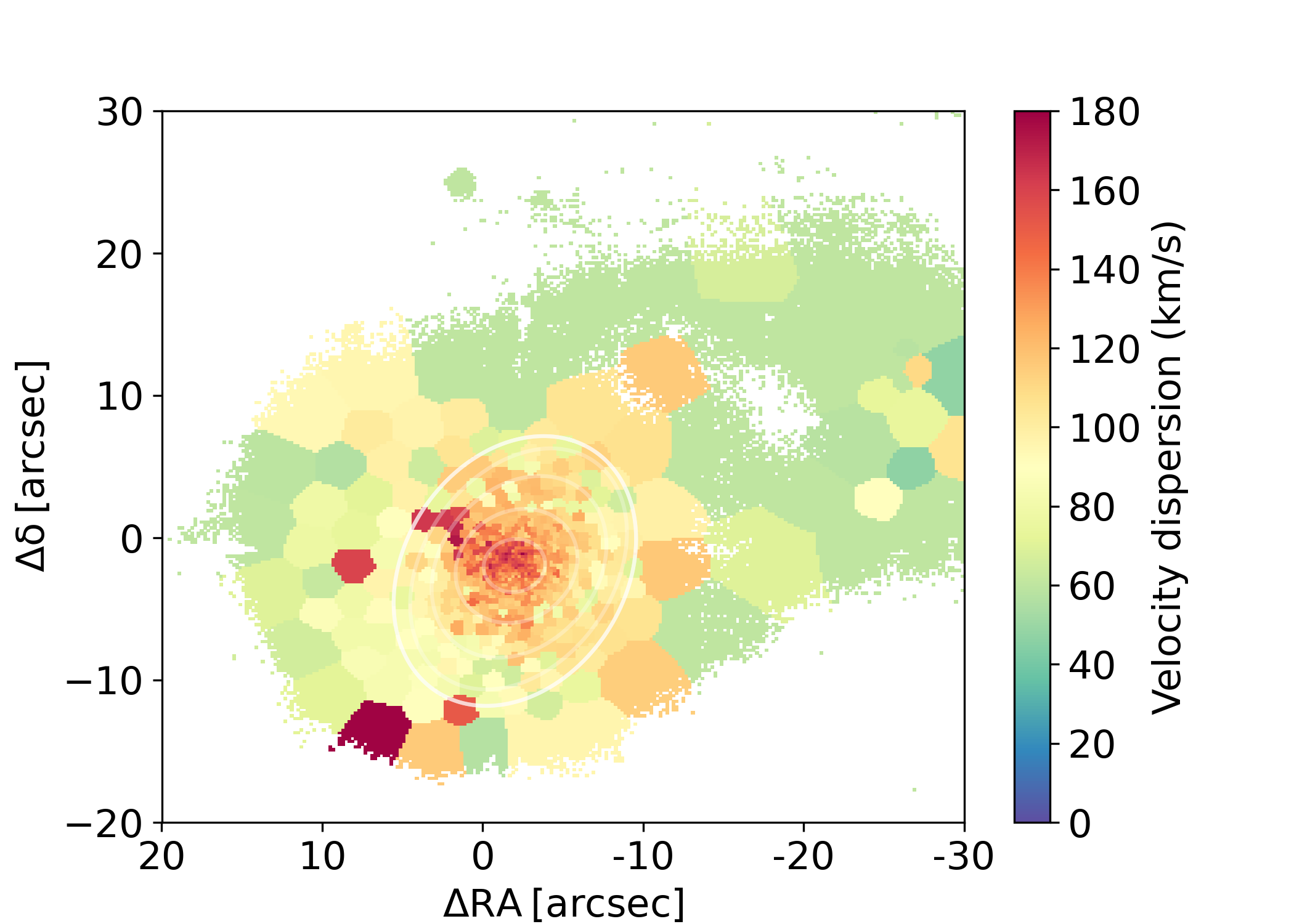}
    }
\caption{Flux (top), velocity (middle), and velocity dispersion bottom) maps for the stellar component of UGC 7342.
The integrated flux maps show the extension of the stellar component, which is highlighted with the ellipses fitted to the isophotal profiles. Fitted ellipses are shown as white contours in all the maps.
Beyond the centrally concentrated high surface brightness stellar structure, tidal arms dominate the distribution of the stellar component. The velocity map in the middle panel shows a gradient that spatially coincides with the centrally concentrated component, indicating the possible presence of rotation. Beyond the centrally concentrated distribution, the kinematics of the stellar component are clearly perturbed due to the streaming motions. The velocity dispersion map shows a higher velocity concentration in the nuclear region, indicating some possible rotational support.
}
 \label{fig:star_mom_7342}
\end{figure}

\textit{UGC 7342:} Fig. \ref{fig:star_mom_7342} shows the flux distribution of the stellar component, which is dominated by a large tidal structure towards the NW, while a fainter structure is also seen to the SE. A component extending 20 \arcsec\ is fitted with isophotal ellipses, which indicates an $\epsilon$ that changes between 0.2 and 0.1 in the inner 10\arcsec\ and a PA that increases steadily from 20\degree\ to 75\degree.
The velocity and velocity dispersion maps reveal a small (10\arcsec\ along PA $\sim 120$\degree) stellar component, reminiscent of a rotating disk, outside which the kinematics are largely non-rotational. 

\begin{figure}
    \centering
    \subfloat{
        \includegraphics[width=0.95\columnwidth]{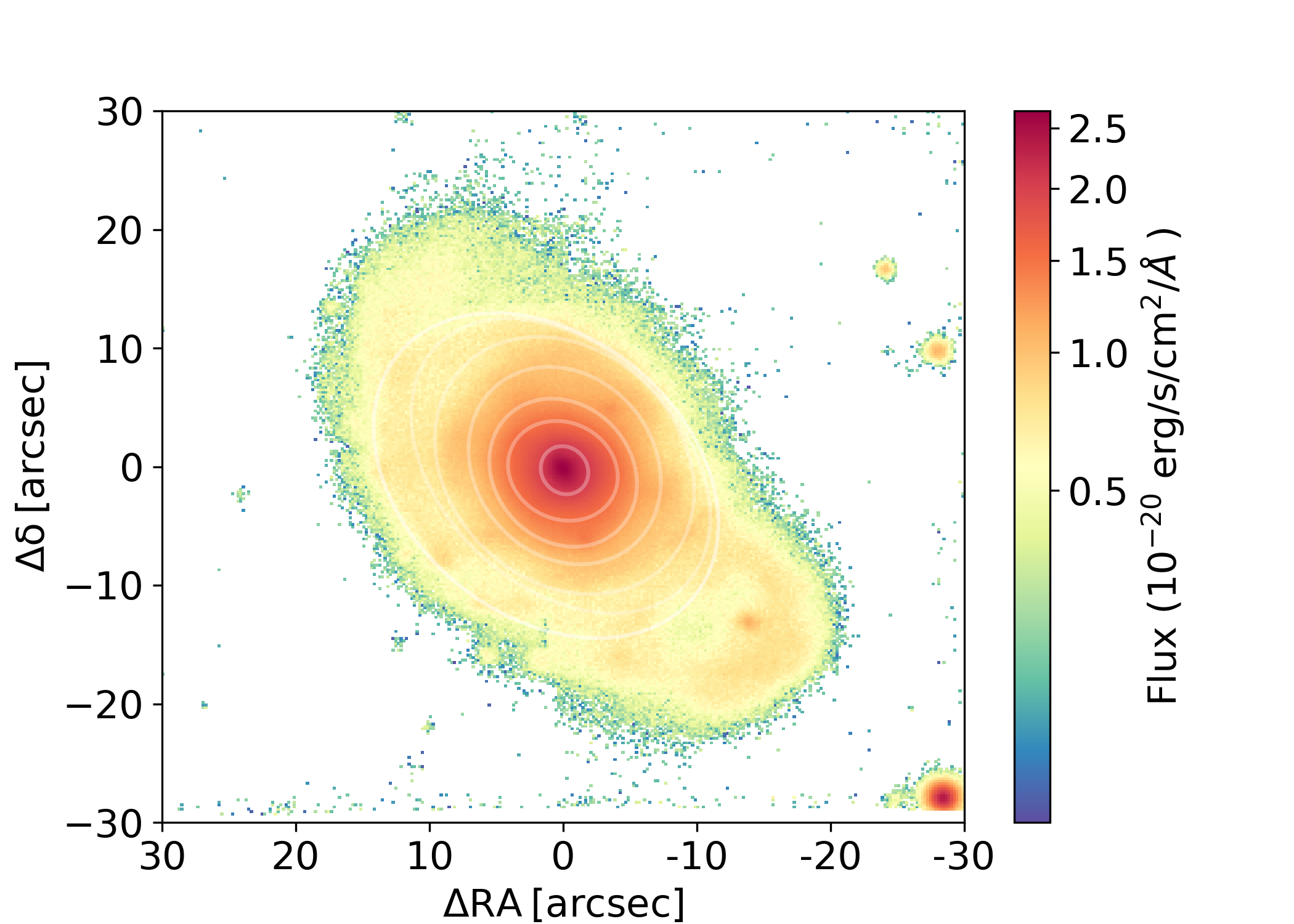}
    }
    
    \subfloat{
        \includegraphics[width=0.95\columnwidth]{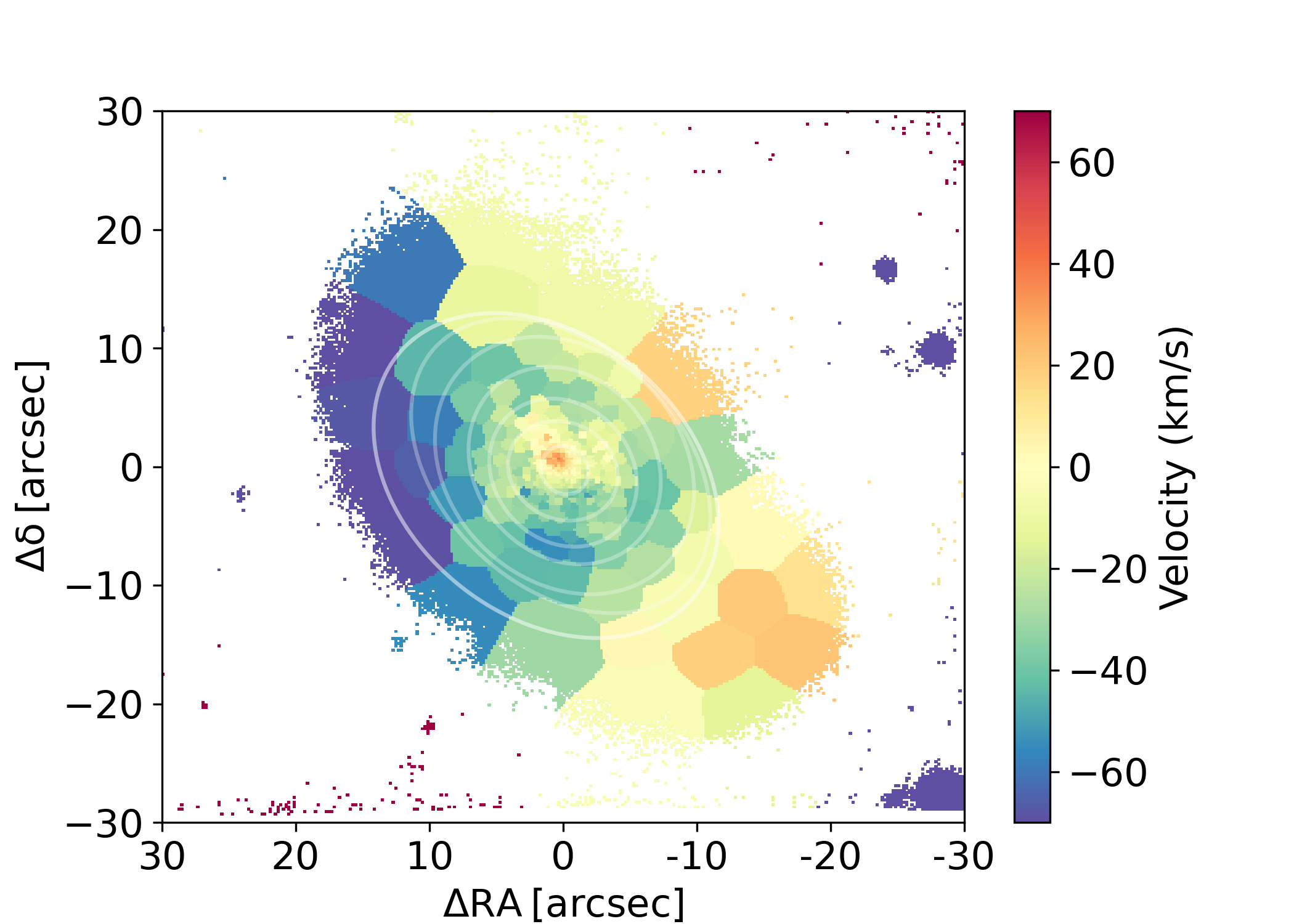}
    }
    
   \subfloat{
        \includegraphics[width=0.95\columnwidth]{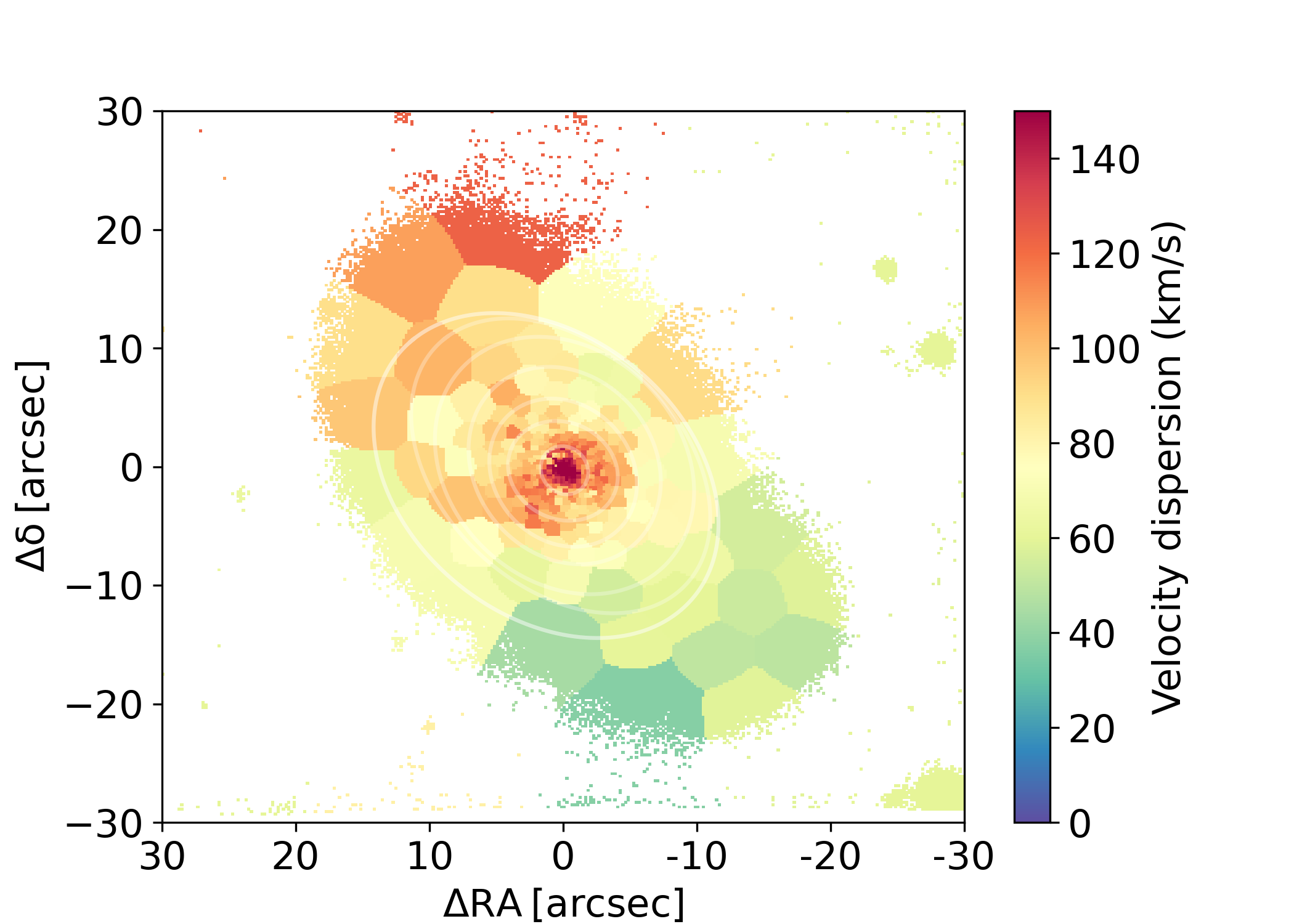}
    }
  \caption{Flux, velocity, and velocity dispersion maps for the stellar component of SDSS1524+08. Overplotted in gray are the ellipses fitted to the isophotal profiles. This fit indicates a close-to-face-on disk. Outside the main centrally-concentrated stellar distribution, there are clear tidal tails that dominate the stellar distribution. The velocity map (middle panel) indicates little to no sign of rotation, with an apparent small gradient in velocity in the central region. The velocity dispersion map (bottom panel) shows a high velocity dispersion in the redshifted region NE of the nucleus and seemingly along the tidal tail that goes E to N. This indicates very low rotational support in the stellar distribution.
}
 \label{fig:star_mom_1524}
\end{figure}

\textit{SDSS 1524+08:} The flux distribution (Fig. \ref{fig:star_mom_1524}) shows a clearly perturbed stellar component, with tidal arms towards the NE and SE, and a loop-like feature in the SW curving towards the E. The isophotal ellipse fitting in the stellar component shows a PA of 24\degree\ and an $\epsilon$ of 0.2, changing into 0.1 when reaching 10\arcsec\ in radius.
The kinematics show a highly complex distribution, as evidenced by the velocity and velocity dispersion maps. There is little evidence of rotational support, as it will be discussed in more detail in Sect. \ref{sec:kinematics}.

\subsection{Ionized gas components}\label{sec:gas}
Using our now continuum-subtracted data cube, we fit the emission lines using Gaussian components for every line. We use a custom \texttt{Python} routine based on Bayesian statistics. This routine creates a model spectrum where every emission line is modeled as either one Gaussian or the sum of two components. The model is then fitted to the observed spectra (per bin) using an implementation of the Markov chain Monte-Carlo (MCMC) adaptive Gibbs sampler. We use 3 chains to avoid local minima, and sample 10000 times per spaxel.
The following emission lines are fitted simultaneously: He II 4685\AA, H$\beta$ 4861\AA, [OIII] 4959,5007\AA, [OI] 6302,6365\AA, [NII] 6548,6583\AA, H$\alpha$ 6563\AA, and [SII] 6717,6731\AA. [OIII] 5007\AA\ is hereafter referred to as [OIII].
To reduce the number of degrees of freedom for the fitting process, we tie the following parameters:  the fluxes from [NII]6548,6583 and  [OIII]4959,5007 with ratios of 2.98:1 and 2.97, respectively, and the width of the Gaussian profiles of H$\alpha$ and H$\beta$. 

We run the fit with one Gaussian component for every emission line and then repeat the fit, adding a second component. 
To identify the regions on every source where a second component is needed while not overfitting the spectrum, we create W80 maps for the [OIII] emission line. This line is used as a probe because it is an isolated bright emission line. This non-parametric descriptor is defined as: $W80 = V_{90} - V_{10}$, where $V_{90}$ and $V_{10}$ are the velocities at which 90\% and 10\% of the line flux accumulates, respectively. Hence, W80 describes the velocity range that encloses the central 80\% of the flux within the spectral window.
We add a second Gaussian component in the areas where W80 $> v_{lim}$ km/s, where $v_{lim}$ is 320, 400, and 350 km/s for SDSS1510+07, SDSS1724+08 and UGC 7342, respectively, based on a fit to a 2D histogram of W80, for each galaxy. The contours of this fit above $v_{lim}$ are shown in the flux map for every galaxy. The second Gaussian component is used to better recover the total flux from every emission line, and not for the kinematical analysis.
From these fits, we create flux, velocity, and velocity dispersion 2D maps. 

\begin{figure*}
    \centering
    \subfloat{
        \includegraphics[width=0.33\textwidth]{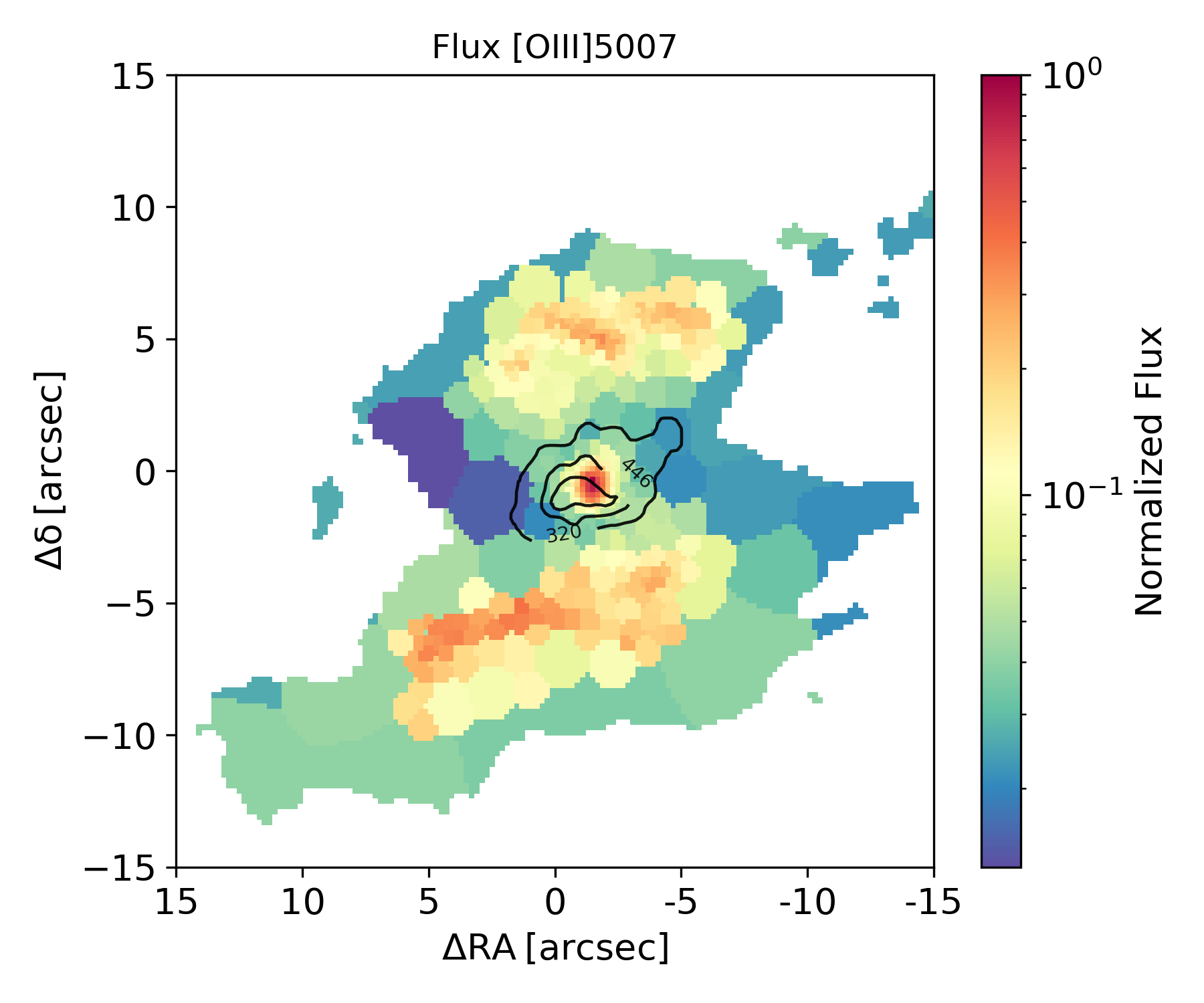}
    }
    \subfloat{
        \includegraphics[width=0.33\textwidth]{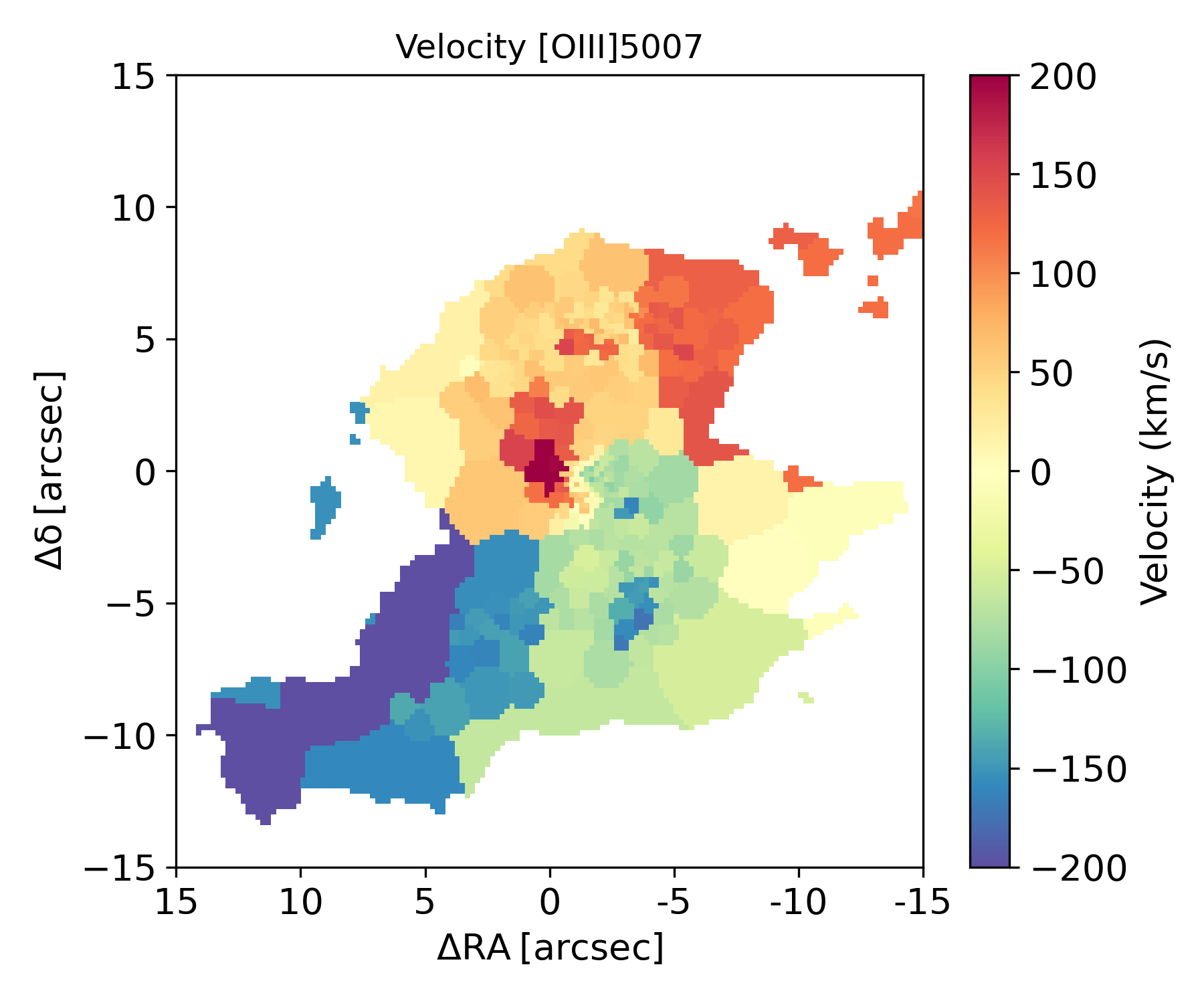}
    }    
    \subfloat{
        \includegraphics[width=0.33\textwidth]{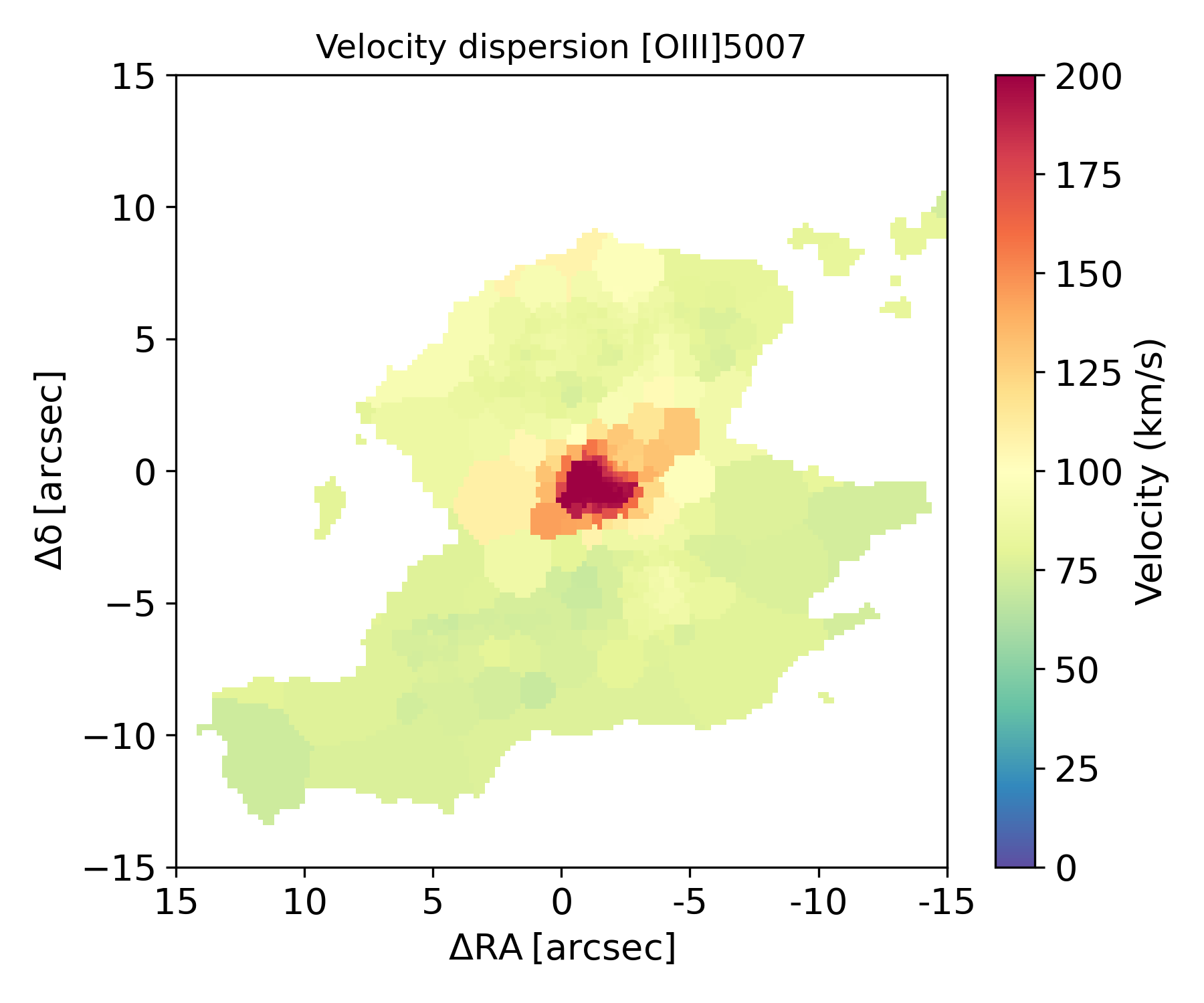}
    }
\caption{ Normalized flux (left), velocity (middle), and velocity dispersion (right) maps for SDSS1510+07 based on the measurements of the [OIII] emission line. The black contours represent the values (in km/s) of the parameter W80, above the velocity limit used to fit a secondary Gaussian component to the emission line fit.
}
 \label{fig:moms_1510}
\end{figure*}

\textit{SDSS1510+07:} In Fig. \ref{fig:moms_1510}, we show the flux, velocity, and velocity dispersion maps for the [OIII] emission line. The flux distribution is largely dominated by the ionized gas, which seems to follow the stellar tidal tails to the NW and SE. These features are clear in [OIII], and a similar distribution is observed in the other emission lines.
The velocity field shows clear rotation within the inner 4\arcsec, reaching +162/-243 km/s. This nuclear component is co-rotating with the stellar component, along a different PA.
Outside this radius, the velocity field's major axis seems to shift from the NE to the NW. The tidal tails show redshifted (blueshifted) emission that coincides with the redshifted (blueshifted) side of the disk, reaching projected velocities of -200/+100 km/s. 
The velocity field for [OIII] shows twisted zero-velocity iso-contours, which can indicate the presence of non-rotating components or an inner kinematically decoupled disk.

\begin{figure*}
    \centering
    \subfloat{
        \includegraphics[width=0.33\textwidth]{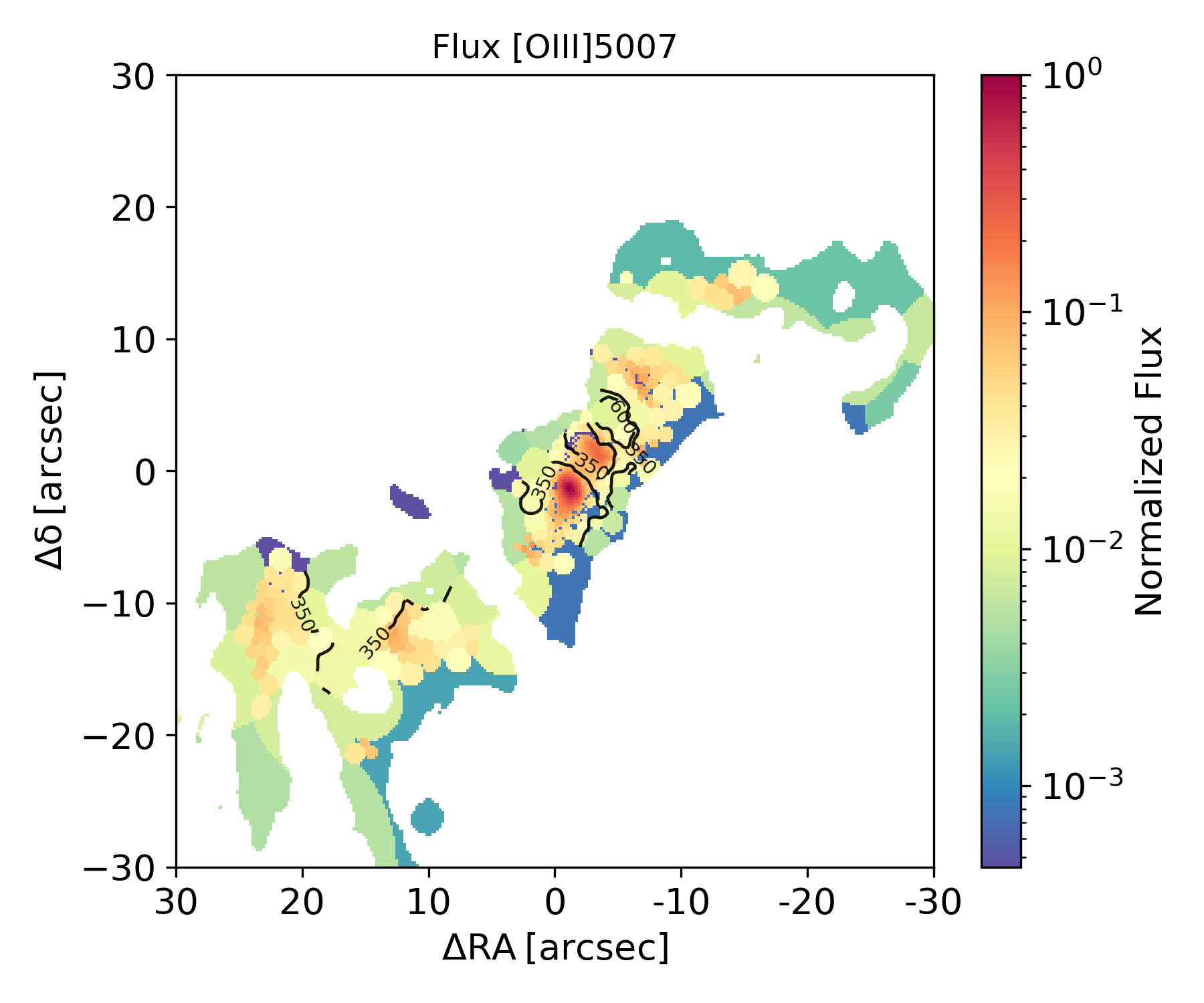}
    }
    \subfloat{
        \includegraphics[width=0.33\textwidth]{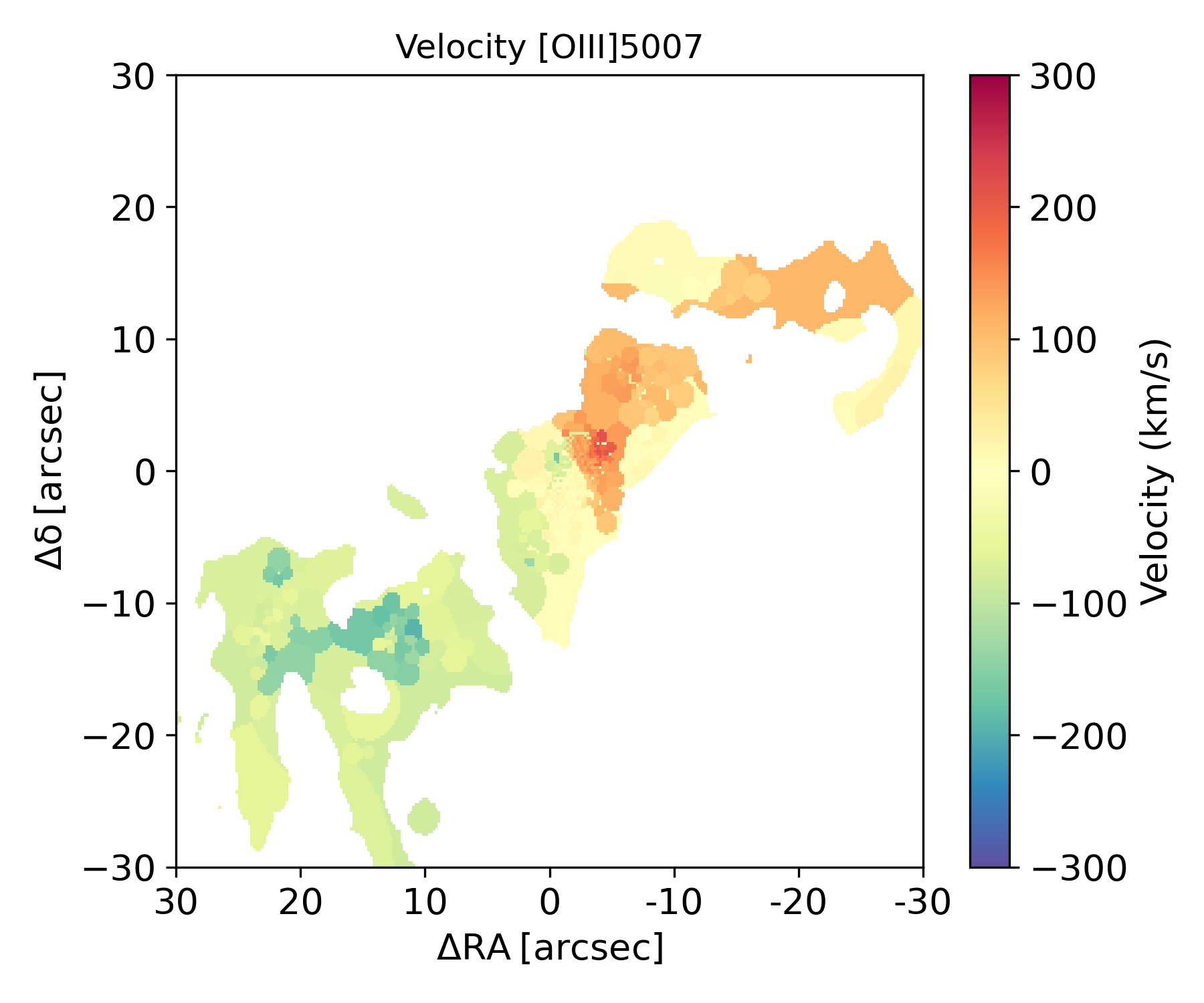}
    }    
    \subfloat{
        \includegraphics[width=0.33\textwidth]{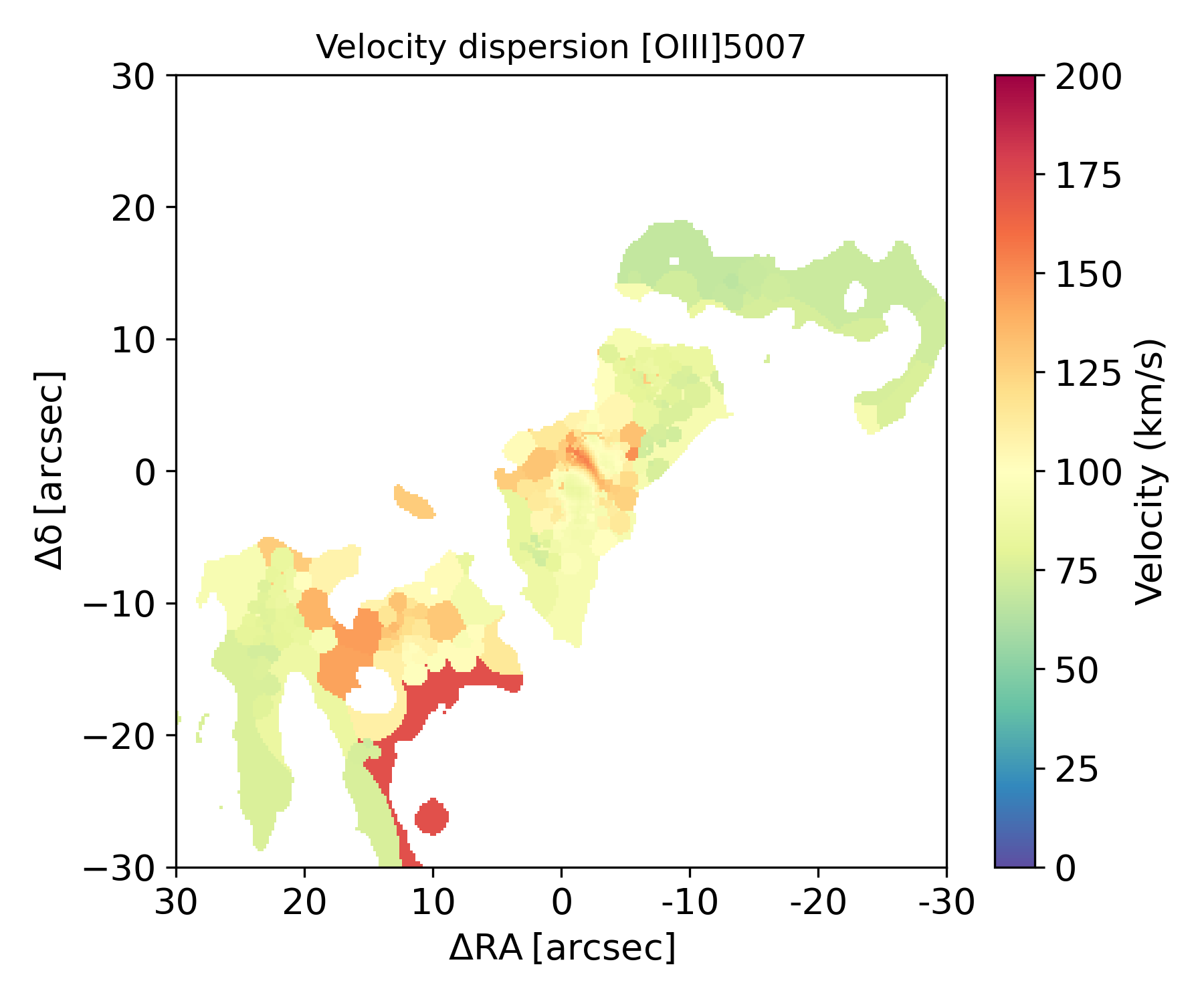}
    }
    
\caption{Normalized flux (left), velocity (middle), and velocity dispersion (right) maps for UGC 7342 based on the [OIII]  emission line. The black contours represent the values (in km/s) of the parameter W80, above the velocity limit used to fit a secondary Gaussian component to the emission line fit.
}
 \label{fig:moms_7342}
\end{figure*}

\textit{UGC 7342:} Fig.\ref{fig:moms_7342} shows the flux, velocity, and velocity dispersion maps for the [OIII] emission line as representative of the other emission lines, which show very similar behavior. The flux distribution shows very extended [OIII] emission from SE to NW, reaching 32\arcsec\ (33 kpc at the redshift of the galaxy) from the center. The distribution is complex and highly filamentary. The ionized gas does not seem to have the same morphology as the tidal tail observed on the NW of the stellar distribution. Furthermore, there is gas to the opposite side (SE) of this tidal tail, where we do not observe a strong stellar counterpart. The velocity field shows a NW-to-SE velocity gradient along PA 310\degree, which could be an indication of rotation. However, it is highly perturbed. We further explore these perturbations in Sect.\ref{sec:kinematics}. 

\begin{figure*}
    \centering
    \subfloat{
        \includegraphics[width=0.33\textwidth]{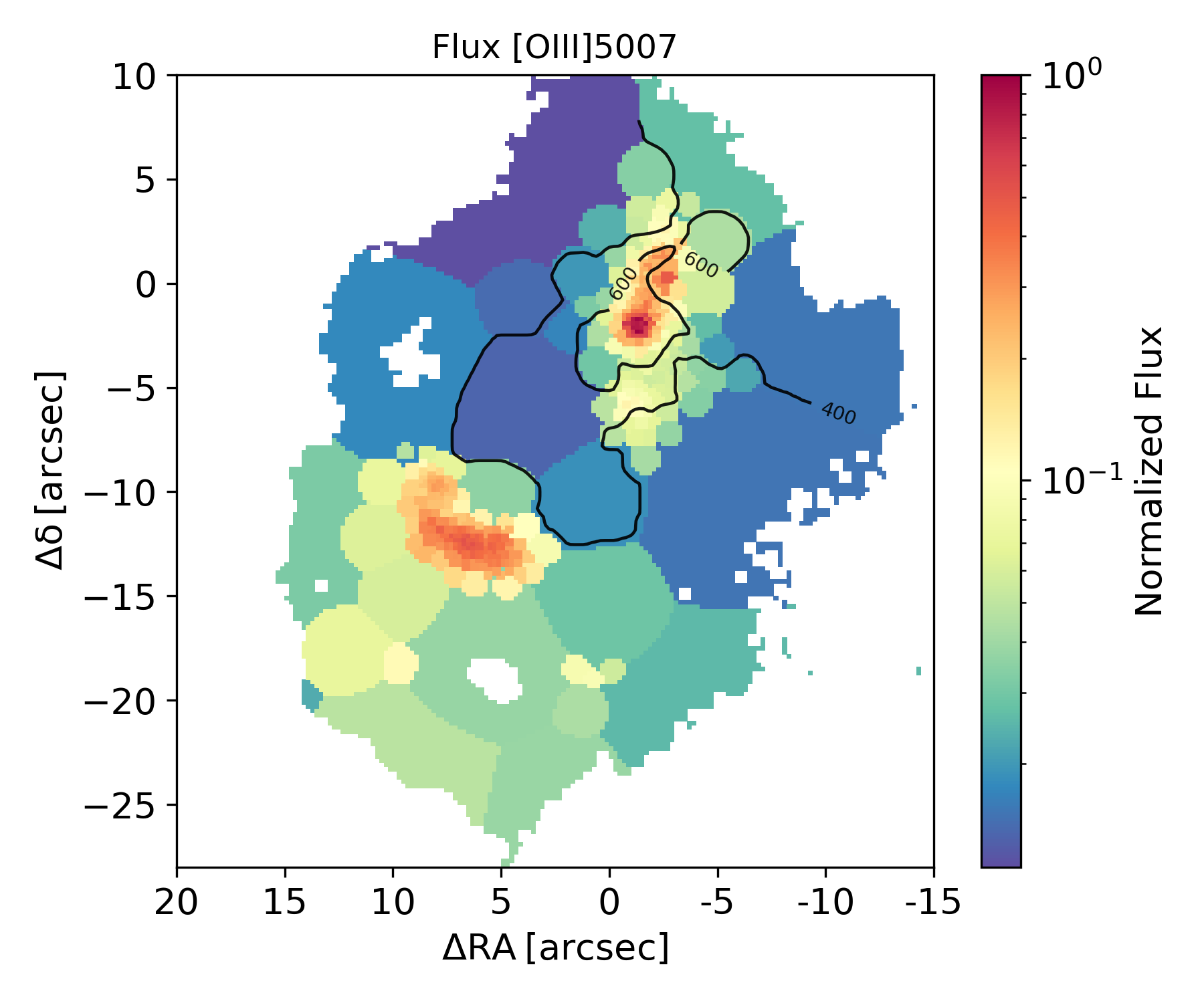}
    }
    \subfloat{
        \includegraphics[width=0.33\textwidth]{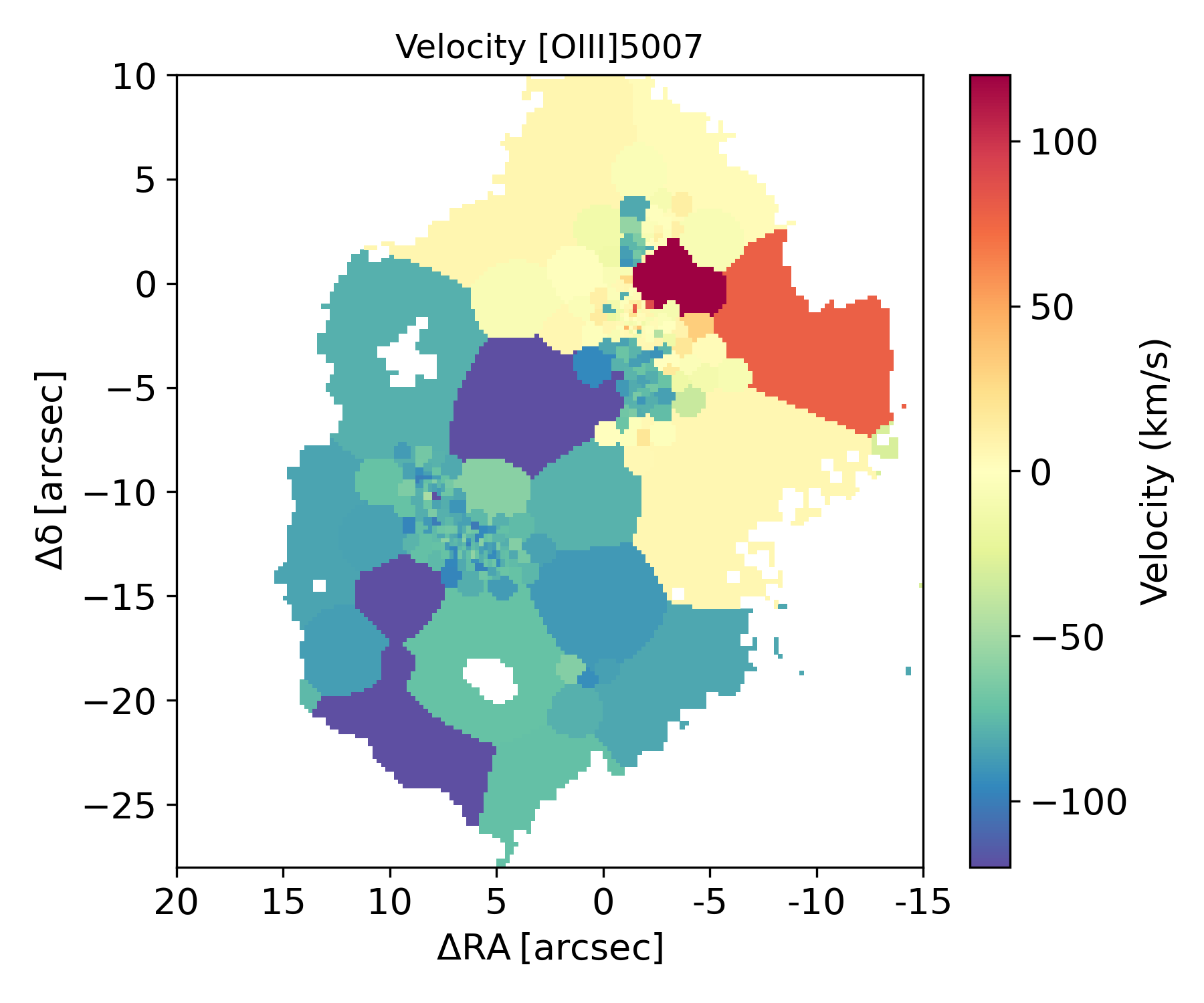}
    }    
    \subfloat{
        \includegraphics[width=0.33\textwidth]{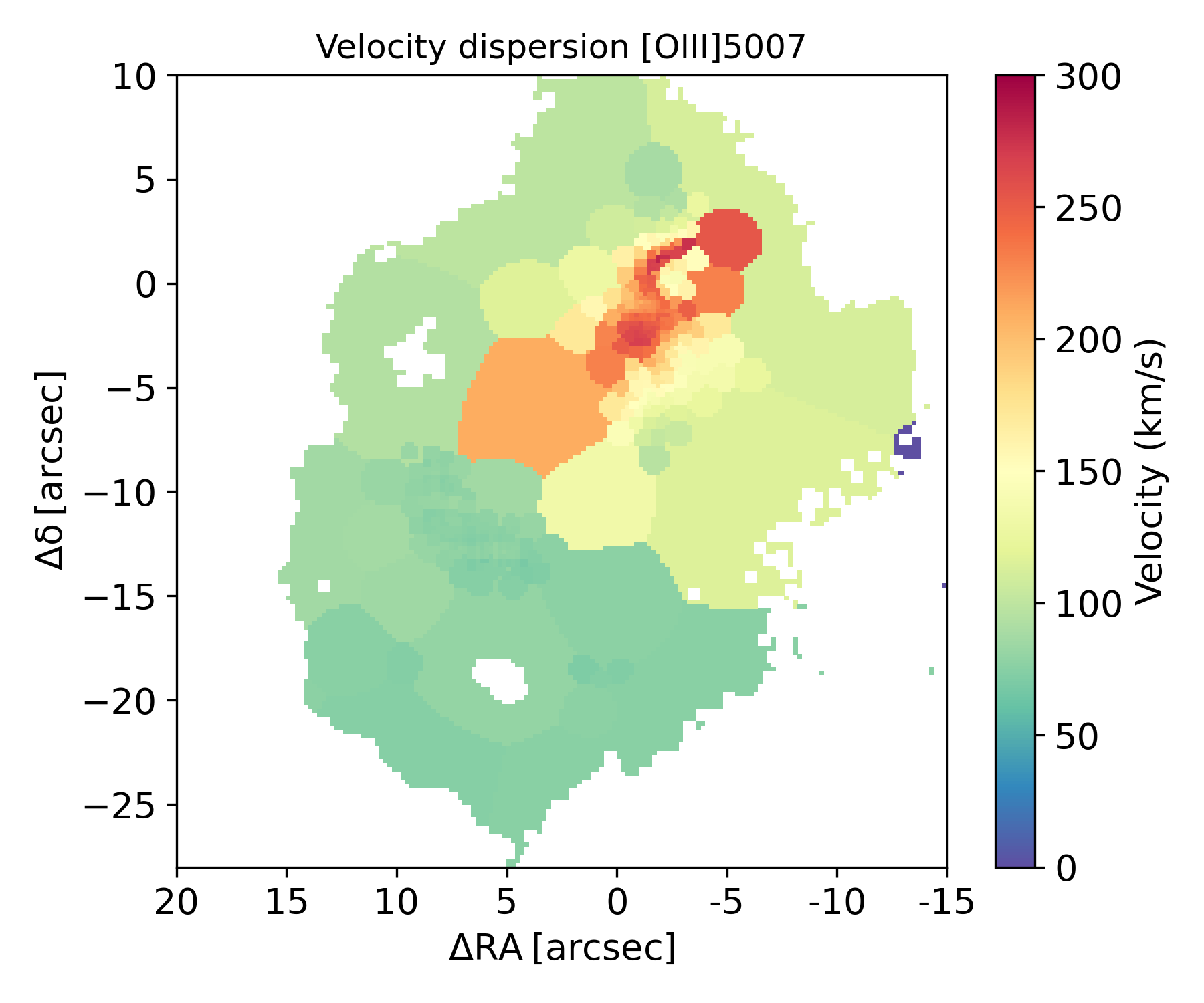}
    }
    
\caption{Normalized flux (left), velocity (middle), and velocity dispersion (right) maps for SDSS1524+08 obtained from measurements of the [OIII]  emission line. The black contours represent the values (in km/s) of the parameter W80, above the velocity limit used to fit a secondary Gaussian component to the emission line fit. 
}
 \label{fig:moms_1524}
\end{figure*}

\textit{SDSS1524+08:} Fig.\ref{fig:moms_1524} shows the flux, velocity, and velocity dispersion maps for the [OIII] emission line, as representative of the behavior of the other emission lines. We observe ionized gas mainly in the region of the centrally-focused stellar feature and 21\arcsec (16 kpc) to the SE of it, perpendicular to the stellar component extension. The ionized gas coincides with the tidal tail observed to the E, indicating that it may have been swept along the tail and then photoionized. Contrary to the other galaxies in this sample, there is no symmetric counterpart to the EELR in this galaxy. The velocity field shows blueshifted velocities for the EELR.
A negative to positive velocity gradient is observed along PA 320 degrees. This is perpendicular to the major axis of the stellar component distribution. The velocity dispersion is also larger along this PA. 
\citet{keel+2024} finds that SDSS1524+08 has a companion galaxy to the SE, along the PA of the ionization cone. This companion shows extended ionized tidal tails, which were photo-ionized by the nuclear source in SDSS1524+08. This phenomenon is referred to as cross-ionization. The distance of the ionized gas in the companion galaxy to the AGN in SDSS1524+08, suggests longer timescales than the one traced in this work (see Sect. \ref{sec:lumhist}), reaching $\sim 10^{5}$ yr.

\subsubsection{Dust extinction}\label{sec:dustext}

We estimate the total extinction in the V band (A$_{V}$) from the Balmer decrement, using the H$\alpha$/H$\beta$ ratio. We use the PyNeb \citep[][]{luridiana+2012} software, assuming an H$\alpha$/H$\beta = 2.86$ intrinsic ratio, with an electron temperature of T$_{e} = 10^{4}$ K \citep[][]{osterbrock+2006}. With this, we derive a reddening map using the extinction curve of \citet[][]{cardelli+1989}, for a galactic diffuse ISM (R$_{V} = 3.12$). We then correct the flux maps of our target galaxies for each bin using the derived extinction values.

\subsubsection{BPT classification maps}\label{sect:bpts}

\begin{figure*}
    \centering
    \subfloat{
        \includegraphics[width=0.33\textwidth]{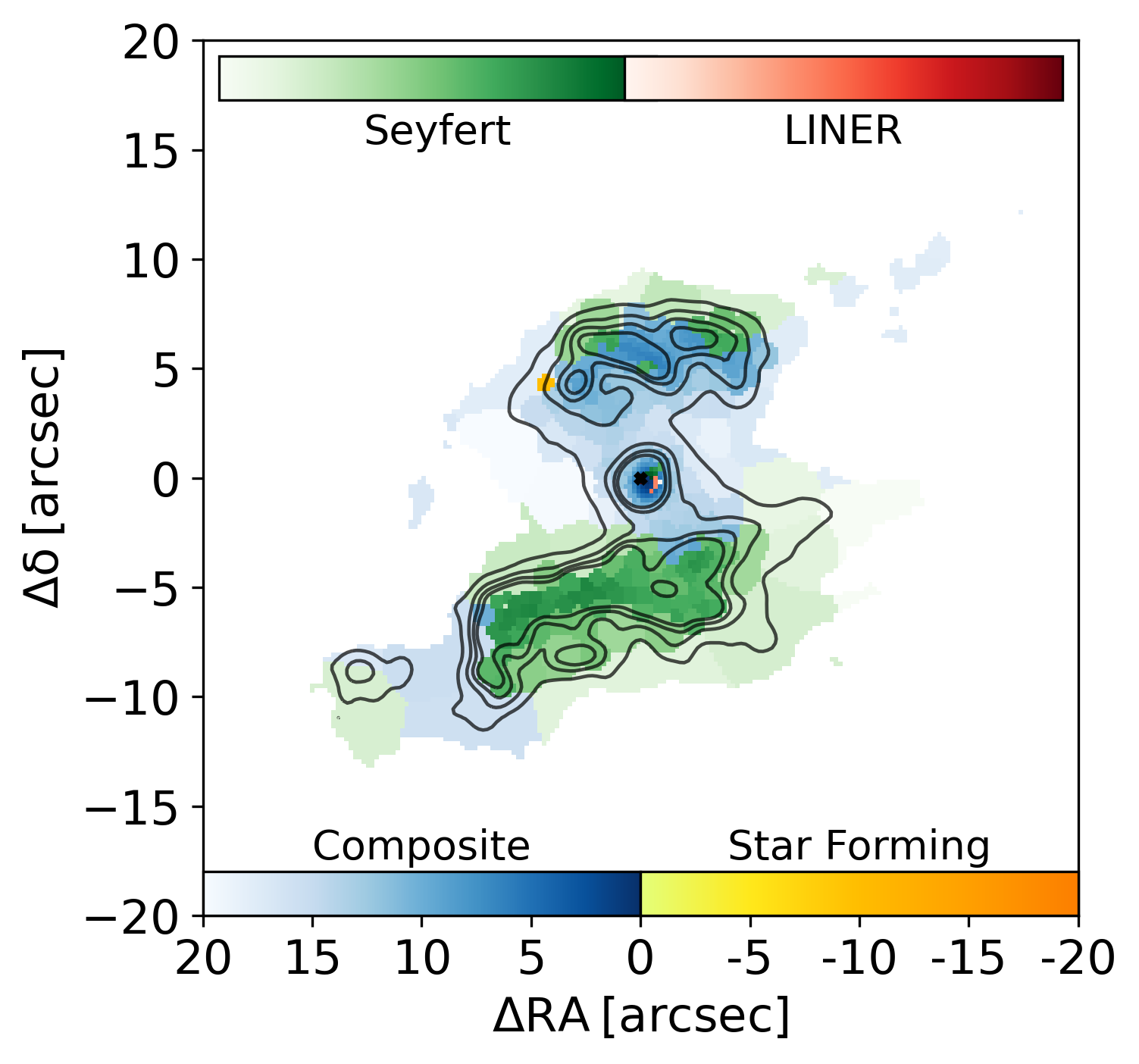}
    }
    \subfloat{
        \includegraphics[width=0.33\textwidth]{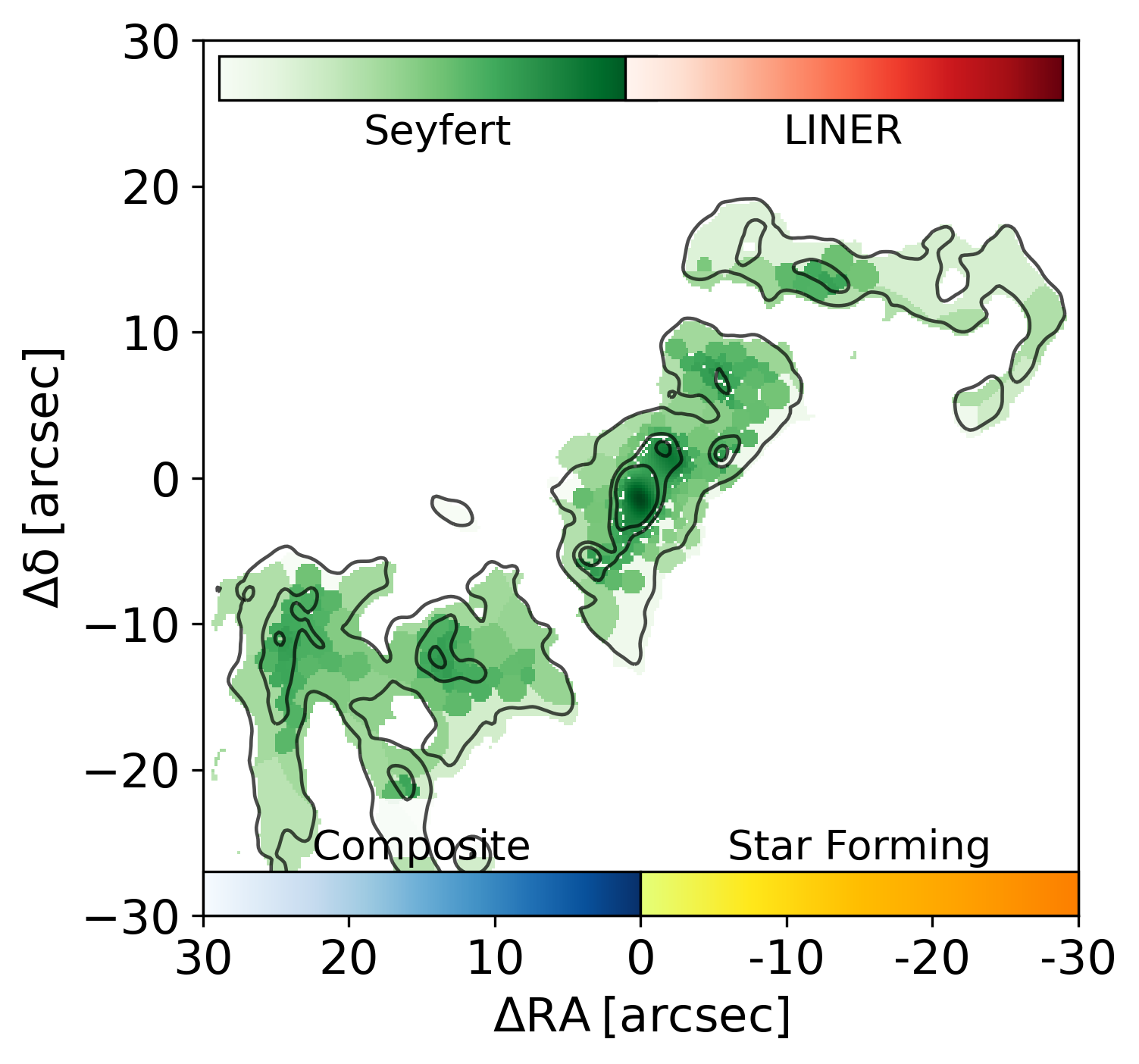}
    }    
    \subfloat{
        \includegraphics[width=0.33\textwidth]{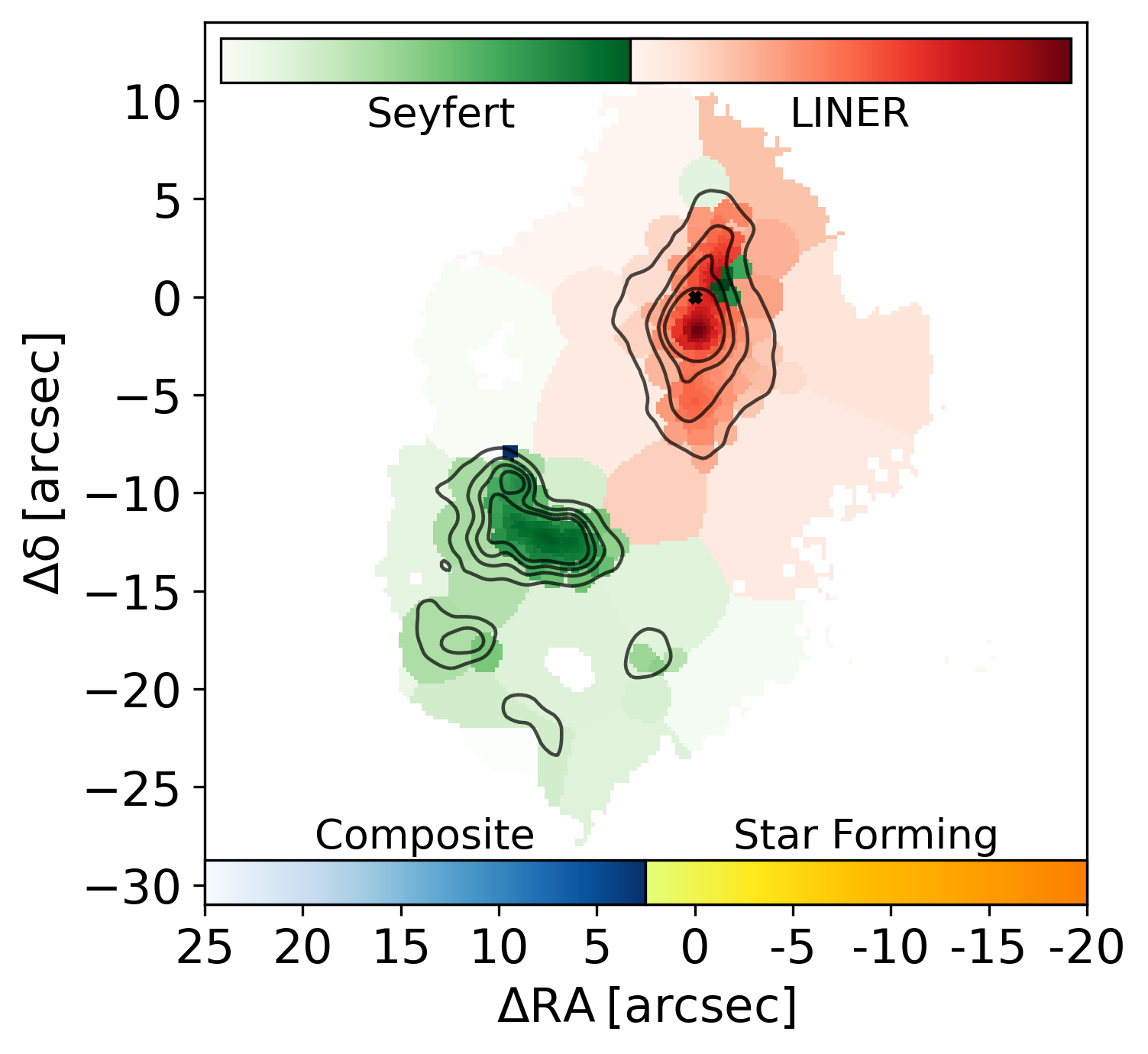}
    }
    
    \subfloat{
        \includegraphics[width=0.33\textwidth]{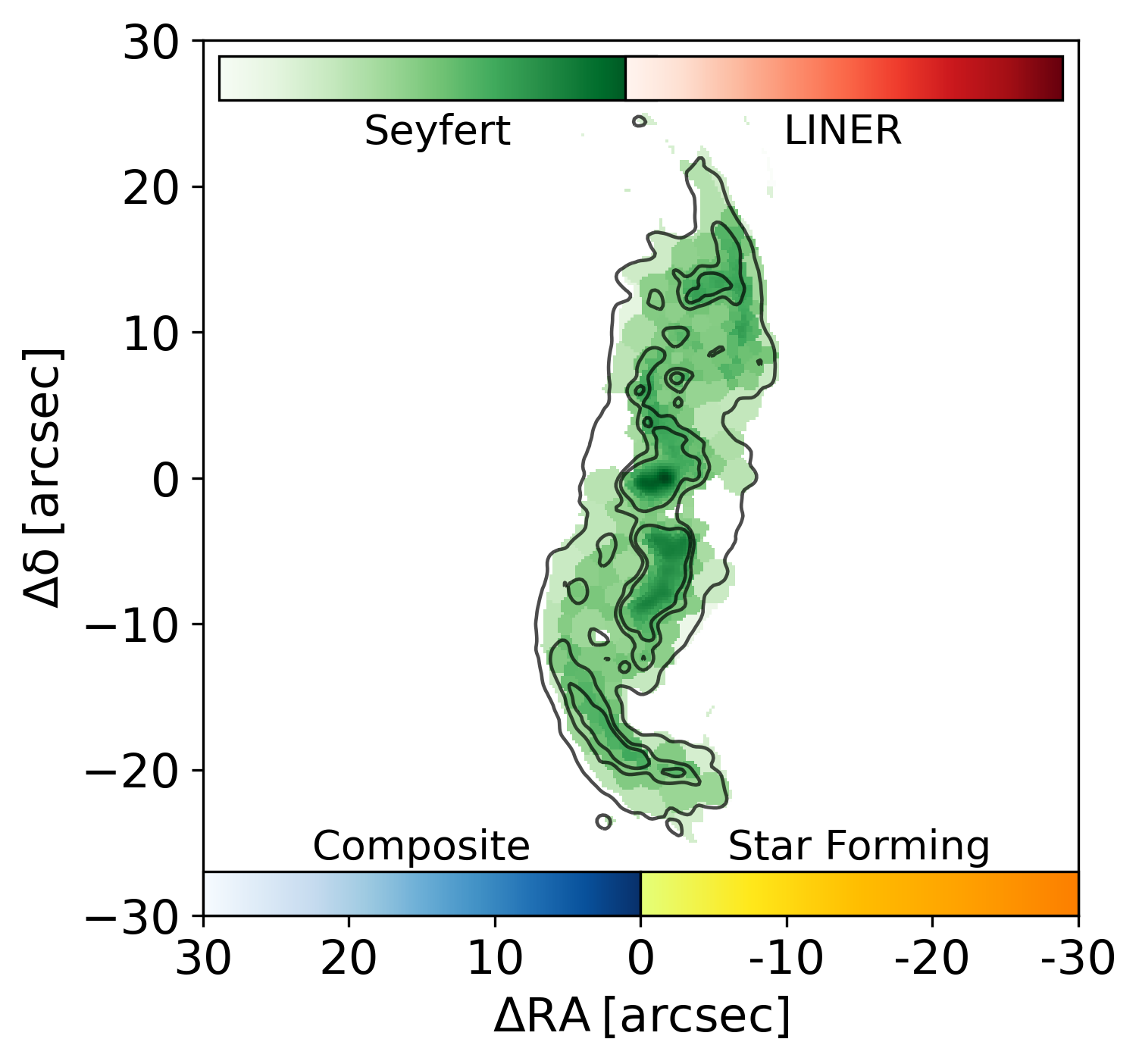}
    }
    \subfloat{
        \includegraphics[width=0.33\textwidth]{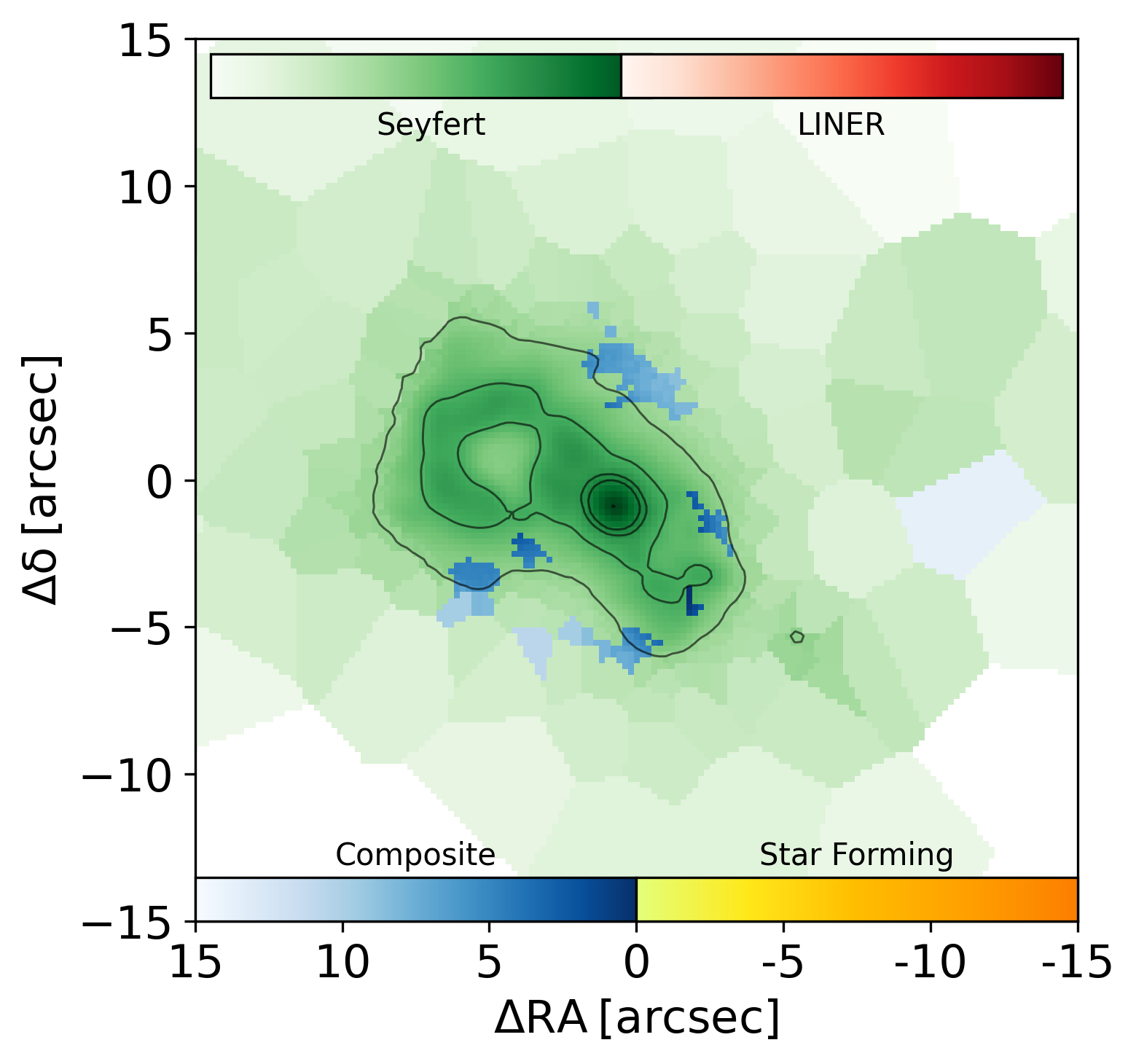}        
    }
\caption{ Spatially-resolved BPT classification maps based on the  [OIII]/H$\beta$ vs [NII]/H$\alpha$ BPT diagrams for SDSS1510+07, UGC 7342, SDSS1524+08 (top from left to right), NGC 5972, and The Teacup (bottom from left to right). The color bars indicate the different BPT regions, where green indicates the spaxels fall in the Seyfert section of the diagram, red for LINER, blue for composite, and yellow for star-forming. Every color is weighted with the [OIII] emission line flux to more clearly show the bright photo-ionized regions. }
 \label{fig:bpts_2d}
\end{figure*}

Using the reddening-corrected flux maps for the emission lines, we calculate the spatially-resolved [OIII]/H$\beta$ vs [NII]/H$\alpha$  Baldwin-Phillips-Terevich \citep[BPT;][]{baldwin+1981} diagrams, in order to identify the dominant contribution to the gas ionization on every bin, for every galaxy. This diagram is used to classify the main ionization mechanism into Star Forming \citep[][]{kauffmann+2003}, Seyfert, LINER \citep[][]{kewley+2001} or Composite \citep[][]{schawinski+2007}, depending on the position of the spaxel's emission line ratios in the diagram. In Fig. \ref{fig:bpts_2d}, we show the BPT classification maps based on the [OIII]/H$\beta$ vs [NII]/H$\alpha$ BPT diagram for all the galaxies in the sample. All three EELRs indicate that they have been mainly photoionized by the central AGN.

\textit{SDSS 1510+07} shows emission classified as composite along PA $\sim$ 90\degree, covering most of the centrally-concentrated stellar component. In the inner 2\arcsec\, there is Seyfert-like photoionization along PA $\sim$ -20\degree, and outside 5\arcsec\, we can observe Seyfert photoionization along the tidal arms towards the NW and SE. 

\textit{UGC 7342} shows only Seyfert-like emission along the entire EELR, with some Voronoi bins perpendicular to the ionization cone indicating LINER and composite-like photoionization. 

In \textit{SDSS 1524+08}, we observe that in the inner 5\arcsec\ region, extending in the direction of the stellar component, there is LINER-like photoionization. Some Seyfert-like photoionization is observed along PA $\sim$140\degree\ from the center. While the entire EELR towards the SE indicates that it is mainly AGN-photoionized. The counterpart towards the NW shows a similar behavior, with a couple of Voronoi bins indicating composite emission.

All galaxies in the sample show that most of the gas is consistent with Seyfert-like ionization. In the cases of SDSS1510+07 and SDSS1524+08, the AGN-like ionization presents evidence for bi-conical structures.

\subsection{Kinematics}\label{sec:kinematics}

To analyze the kinematics of these galaxies, we generate position-velocity diagrams (PVDs) using the 3D-Barolo package \citep[][]{teodoro+2015}. This code fits a rotation model using rings propagating from the center of the map. Based on the 3D-Barolo 2Dfit routine, we fit tilted rings with different PA, inclination, and rotation velocities to the stellar component velocity map of each galaxy. With this information,  we constructed models of the line-of-sight velocity (V$_{LOS}$), which at first order, not considering non-circular motions, can be expressed as 
$$ V_{LOS} (R) = V_{sys} + V_{rot}(R) \cos\theta \sin i $$
where V$_{sys}$ corresponds to the systemic velocity, $\theta$ to the azimuthal angle measured in the plane of the galaxy, and $i$ to the galaxy inclination. The disk rotation model is then subtracted from the velocity map in order to reveal non-circular motions. 
To analyze if the stellar component is supported by rotation, we use the $\lambda_{R}$ parameter, defined by \citet{emsellem+2007} as a proxy of the observed projected angular momentum of the stellar component, which can indicate whether the galaxy kinematics exhibit clear large scale rotation or not. This parameter is defined as, 

$$ \lambda_{R}= \frac{\Sigma_{i=1}^{N} F_{i}R_{i}|V_{i}| }  {\Sigma_{i=1}^{N} F_{i}R_{i} \sqrt{V_{i}^{2} + \sigma_{i}^{2}   } }$$

to be measured in 2-dimensional spectroscopy. Where $F_{i}$, $R_{i}$, $V_{i}$, $\sigma_{i}$ are the flux, distance to the center, velocity, and velocity dispersion of the stellar component in each $i$th spaxel (of a total of $N$ in each galaxy), respectively. 

\begin{figure*}
	\includegraphics[width=\textwidth]{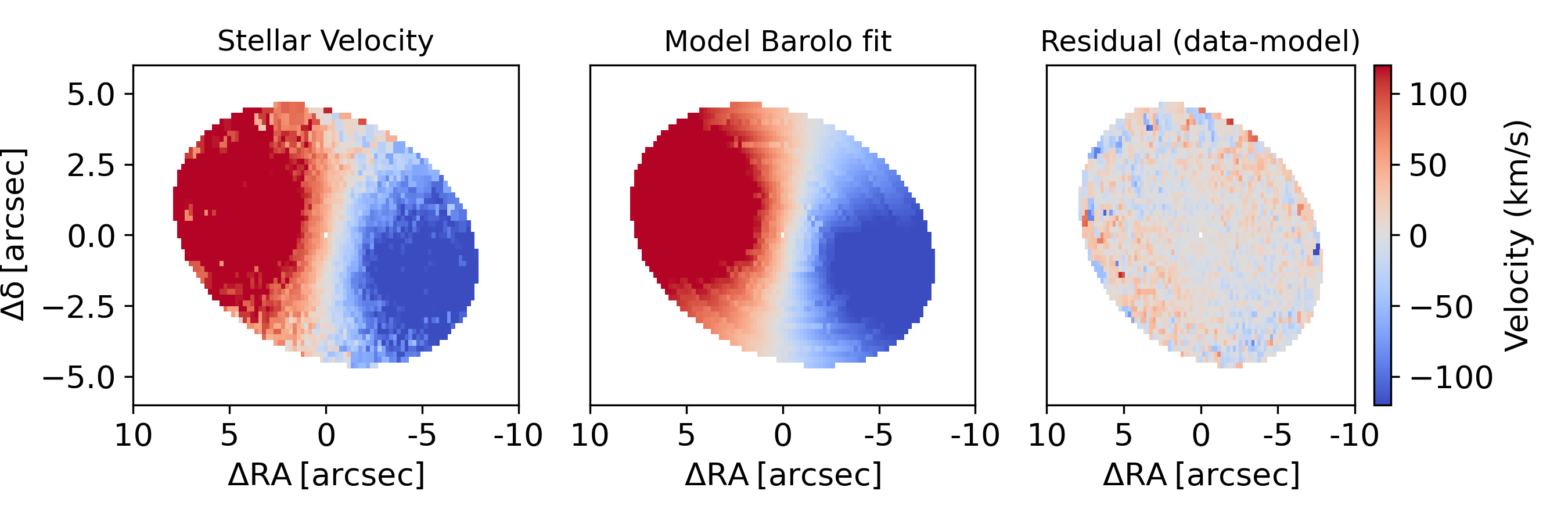}
    \caption{Velocity map for the stellar component of SDSS1510+07 (left). Best fit kinematic Barolo model (middle). Residual map, i.e., velocity map subtracted from the model map (right)}
    \label{fig:stellarmod_1510}
\end{figure*}

\begin{figure*}
    \centering
    \subfloat{
        \includegraphics[width=\textwidth]{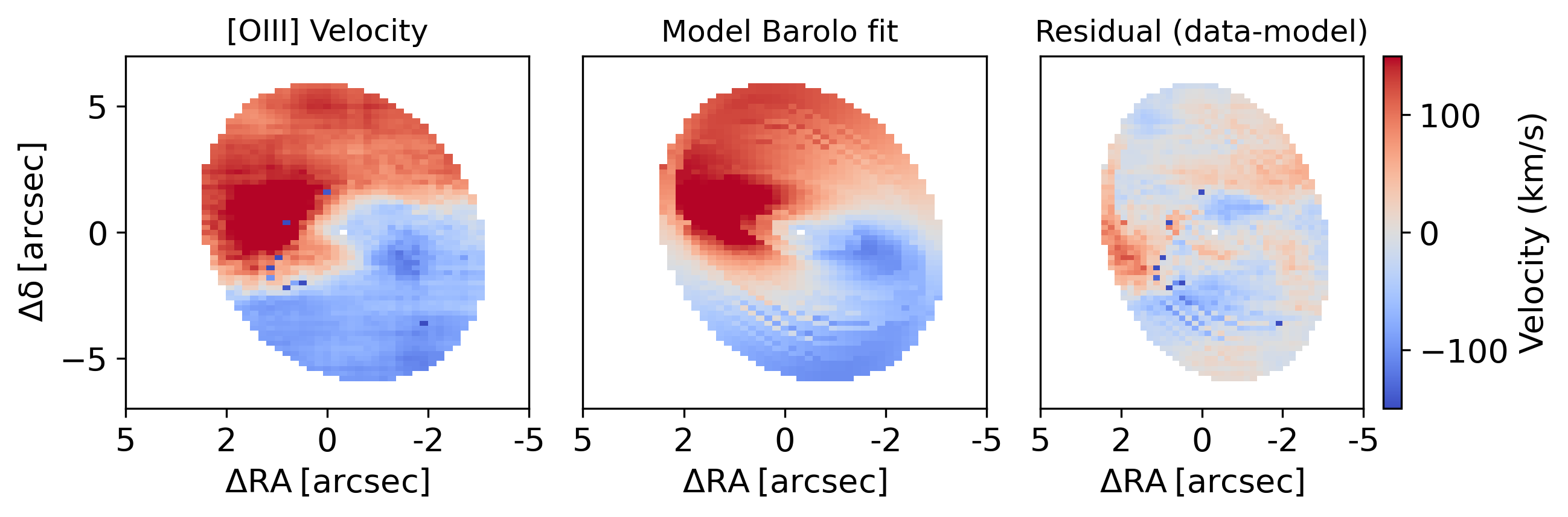}
    }
    
    \subfloat{
        \includegraphics[width=\textwidth]{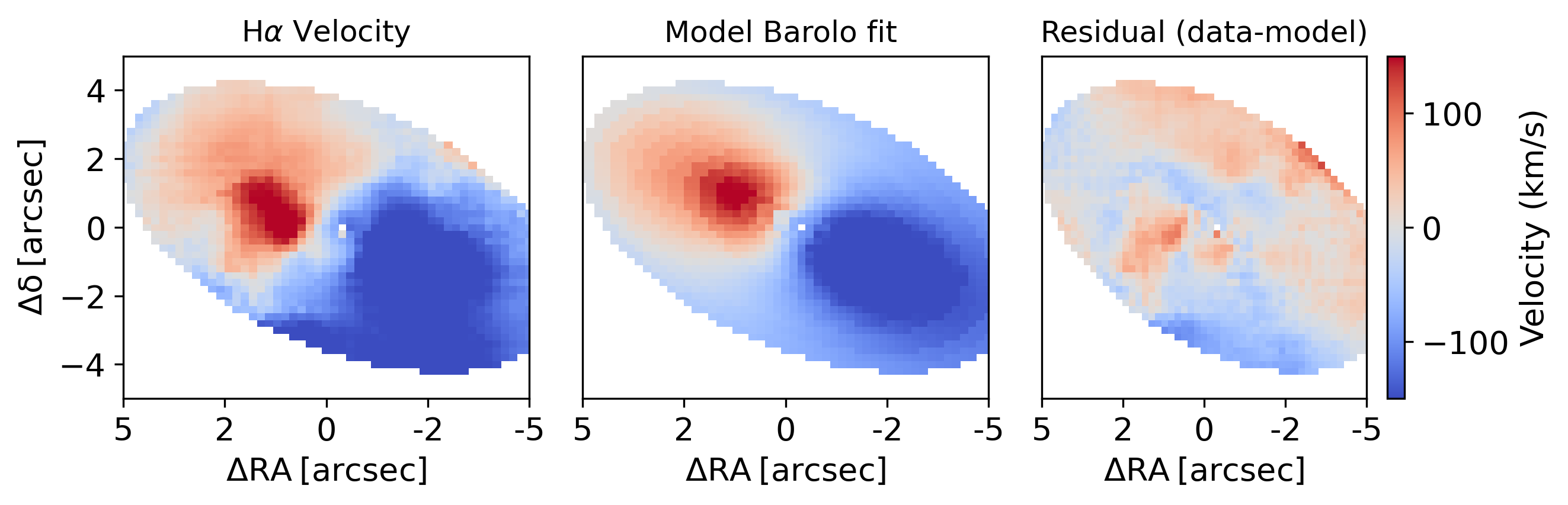}
    }    
\caption{ Velocity maps for the emission lines [OIII] (top row) and H$\alpha$ (bottom row). Velocity map for the ionized gas component of SDSS1510+07 (left). Best fit kinematic 3D-Barolo model (middle). Residual map, that is, velocity map subtracted from the model map (right) }
 \label{fig:gasmod_1510}
\end{figure*}

\textit{SDSS 1510+07:} In Fig. \ref{fig:stellarmod_1510}, we show the rotating disk model fitted to the stellar velocity map with the 2Dfit routine. The best-fit model shows mean values of 73\degree and 55\degree\ for the PA and inclination, respectively. These values remain almost constant along the different radii. The kinematics are consistent with a rotating stellar disk. 
For the emission lines, however, we use the 3Dfit task, which fits the 3D data cube centered and trimmed to contain only the [OIII] line. For the emission lines, we opt for the 3Dfit, considering that information from complex or multiple component kinematics can be lost in a velocity map. Therefore, a fit directly to the emission line and not just its median velocity can provide more accurate results.
We perform this fit for [OIII] and H$\alpha$ emission lines, using the values obtained from the posited stellar disk as first guesses. In Fig. \ref{fig:gasmod_1510} we show the resulting model subtracted from the corresponding velocity maps. We find that the PA for both lines changes noticeably on every ring after 0\farcs7 radius. The PA inside 4.5\arcsec\ radius is fitted to be 65\degree, resulting in an 8\degree\ offset from the stellar disk. The continuously changing PA outside this radius indicates that the disk is likely warped. As it can be seen from the BPT classification maps (Fig. \ref{fig:bpts_2d}) and velocity maps (Fig. \ref{fig:moms_1510}), at  5.5\arcsec\ radius, we find the EELR towards the NW and SE. The kinematics in these tidal arms appear to be non-rotating. This implies that if they remain in the plane of the galaxy, there would be outflowing gas, reaching 220 km/s. The second scenario corresponds to the tidal tail extending outside the plane of the disk; therefore, depending on the inclination angle, this could reach up to $\sim 490$ km/s deprojected velocity.
In the inner 1.2\arcsec\, we find a component with a blueshift-to-redshift velocity gradient along PA 135\degree, reaching velocities of 100 km/s. This feature can correspond to a kinematically decoupled component (KDC), rotating on a major axis perpendicular to the rotation of the large-scale disk.
The PA offset between the ionized gas and the stellar disk, the tidal arms, and the KDC can all be indicators of past interactions, implying SDSS 1510+07 is likely a post-merger.

\begin{figure*}
	\includegraphics[width=\textwidth]{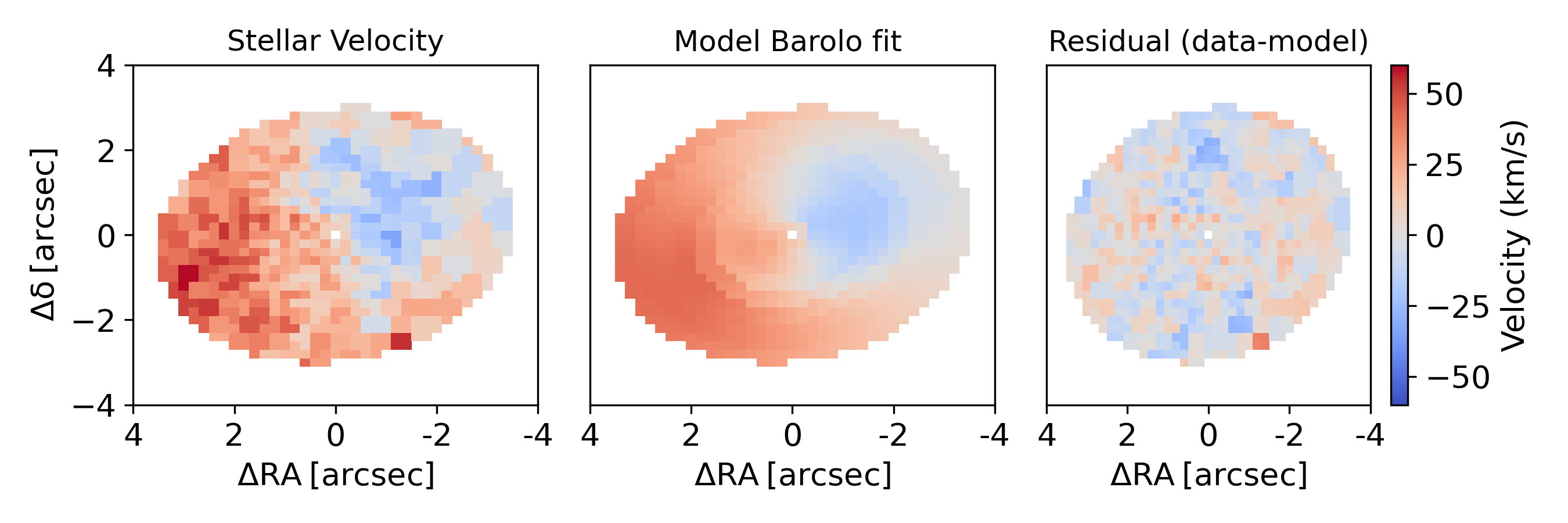}
    \caption{Velocity map for the stellar component of UGC 7342 (left). Best fit kinematic Barolo model (middle). Residual map, that is, velocity map subtracted from the model map (right)}
    \label{fig:stellarmod_7342}
\end{figure*}

\begin{figure*}
	\includegraphics[width=\textwidth]{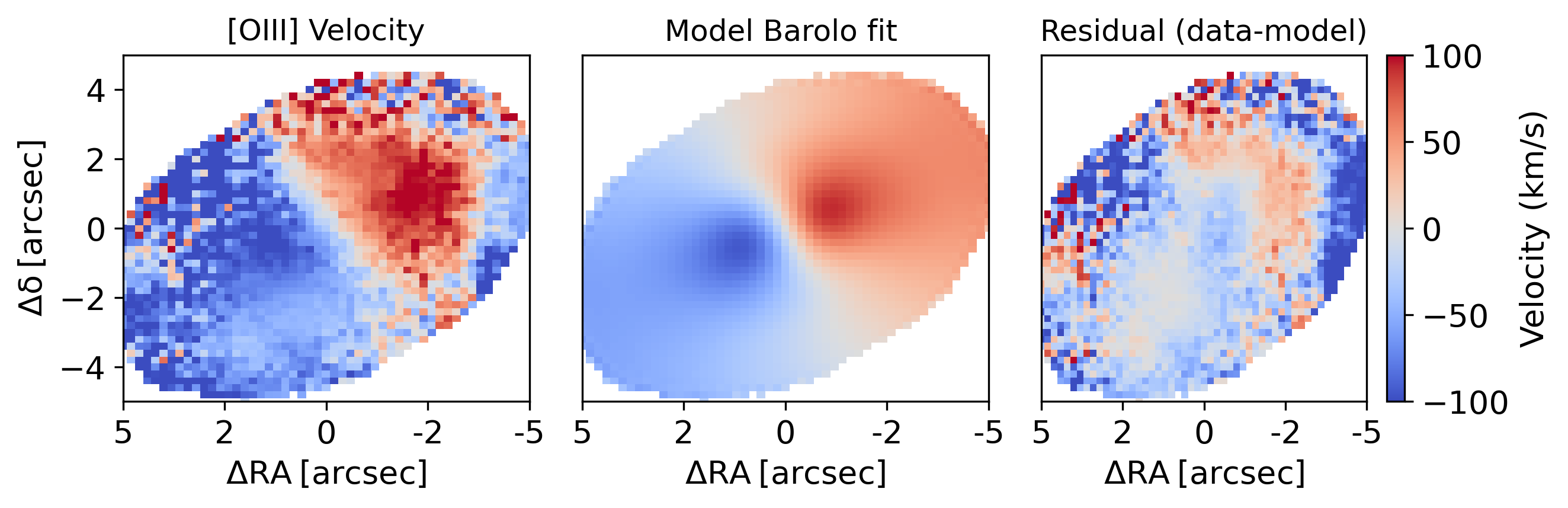}
    \caption{Velocity map for the ionized gas component of UGC 7342 (left). Best fit kinematic 3D-Barolo model (middle). Residual map, that is, velocity map subtracted from the model map (right)}
    \label{fig:gasmod_7342}
\end{figure*}

\textit{UGC 7342:} 
The stellar component shows an inner component that extends over 10\arcsec, with two tidal tails extending from it, forming loops towards the NW and the NE (Fig. \ref{fig:star_mom_7342}). The NE tail is fainter and closer to the centrally-concentrated component.
The main stellar component shows largely non-rotating kinematics. A gradient of $\pm 50$ km/s is observed in the inner 4\arcsec\ along PA $\sim$ 115\degree\ (anti-clockwise from N, see Fig. \ref{fig:star_mom_7342}). Outside this nuclear region, there are signs of rotation (blueshift on the SE and redshift on the NW)  along PA $\sim 290$\degree, i.e., misaligned 180\degree\ from the nuclear component. 
This could indicate a nuclear disk that is counter-rotating and kinematically decoupled from the remainder of the stellar component. This type of feature can often be observed on disturbed galaxies \citep[e.g.,][]{emsellem+2007}, as seems to be also suggested by the presence of two tidal tails. We fitted the possible rotation on a disk up to 4\arcsec\ radius, following the same method described above. We obtain mean values of 115\degree\ and 35\degree\ for the PA and the inclination, respectively (Fig. \ref{fig:stellarmod_7342}).
The velocity maps for the ionized (Fig. \ref{fig:moms_7342}) gas show a highly disturbed gradient from blue to redshifted velocities along a similar PA (310\degree) to the large-scale stellar distribution. 
We fitted a rotating model to the inner 6\arcsec, to cover the most undisturbed section of the gaseous disk (Fig. \ref{fig:gasmod_7342}). We find that part of the observed velocities can be fitted by a rotating disk. The PA varies between 300-320\degree\ indicating a warping of the disk. The residuals denote an important contribution of non-rotating components. 

The highest-velocity residuals are redshifted towards the E and blueshifted towards the W. The spectra in these regions are complex. In the former, we find a broad wing, while in the latter, we see a separated blueshifted component, as well as a broad wing. The velocity deviation from systemic reaches 100 km/s for the W feature, and up to $\sim$400 km/s for the E. These non-rotating components suggest the presence of an outflow if the SW side of the galaxy corresponds to the near side.
Beyond the central kpcs, the ionized gas retains the same sense of rotation along the tidal tails. In the blueshifted region (SE), there are areas with higher velocity dispersion (Fig. \ref{fig:moms_7342}) and double-peaked profiles. This implies that kinematics are more complex than pure rotation. We probe this region, extracting flux along different slits. The associated PVDs show the presence of an additional -fainter- component in the areas where the velocity dispersion reaches $\sim 200$ km/s. This component could be interpreted as outflowing gas. There is no observable counterpart to this possible outflow in the redshifted (NW) side of the galaxy; this can be due to more entrained gas or an asymmetric morphology. Another possibility is that it corresponds to the presence of multiple components along the line of sight.
Considering that the gas in the tidal tails has likely left the plane of the disk, the velocities observed are harder to interpret. However, the velocity dispersion is mostly around $\sim$ 100--150 km/s. Kinematic studies on interacting galaxies with nuclear activity indicate that tidally induced forces can produce dynamically hot systems characterized by velocity dispersions of $\sim 100 $ km/s, while systems with $\sigma > 300$ km/s show signs of gas flows \citep[e.g.][]{rich+2015,bellocchi+2013}. Therefore, it is likely that in this source, the tidal features display kinematic turbulence that can be attributed to past interactions.

\textit{SDSS 1524+08:} We calculated the $\lambda_{R}$ parameter along the major axis and found a resulting profile consistent with a slow rotator classification. This denotes that the kinematics of the large-scale galactic distribution are not dominated by rotation. However, in the inner 10\arcsec\ of the stellar velocity field, there are possible signs of rotation along PA $\sim$20\degree. The velocity field of the large-scale ionized gas component (Fig. \ref{fig:moms_1524}) shows a strong velocity gradient at PA $\sim$321\degree. This gradient reaches velocities of -220 km/s towards the SE and 360 km/s to the NW. A second gradient can be observed along PA $\sim$20-30\degree, similar to the stellar component. This gradient reaches velocities of -60 km/s from systemic towards the NE and 78 km/s to the SW.

\begin{figure*}
	\includegraphics[width=\textwidth]{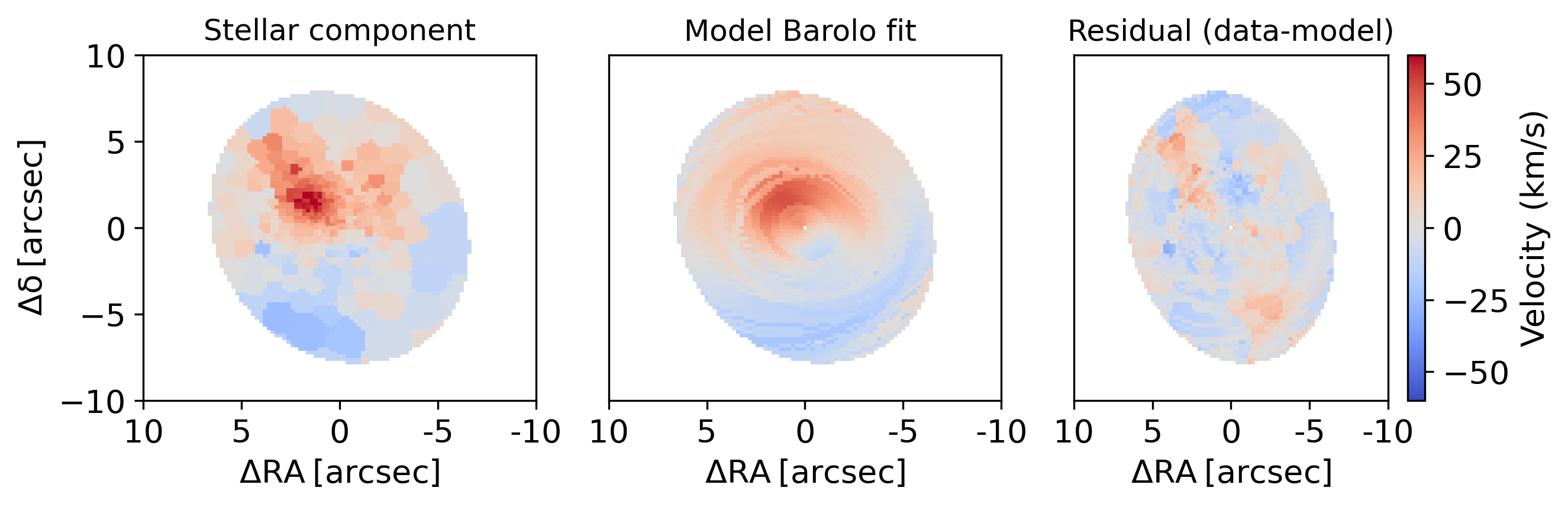}
    \caption{Velocity map for the stellar component of SDSS1524+08 (left). Best fit kinematic 3D-Barolo model (middle). Residual map (right). The residual map indicates possible rotation along PA 22\degree in the inner 10\arcsec.
    }
    \label{fig:stellarmod_1524}
\end{figure*}

\begin{figure*}
	\includegraphics[width=\textwidth]{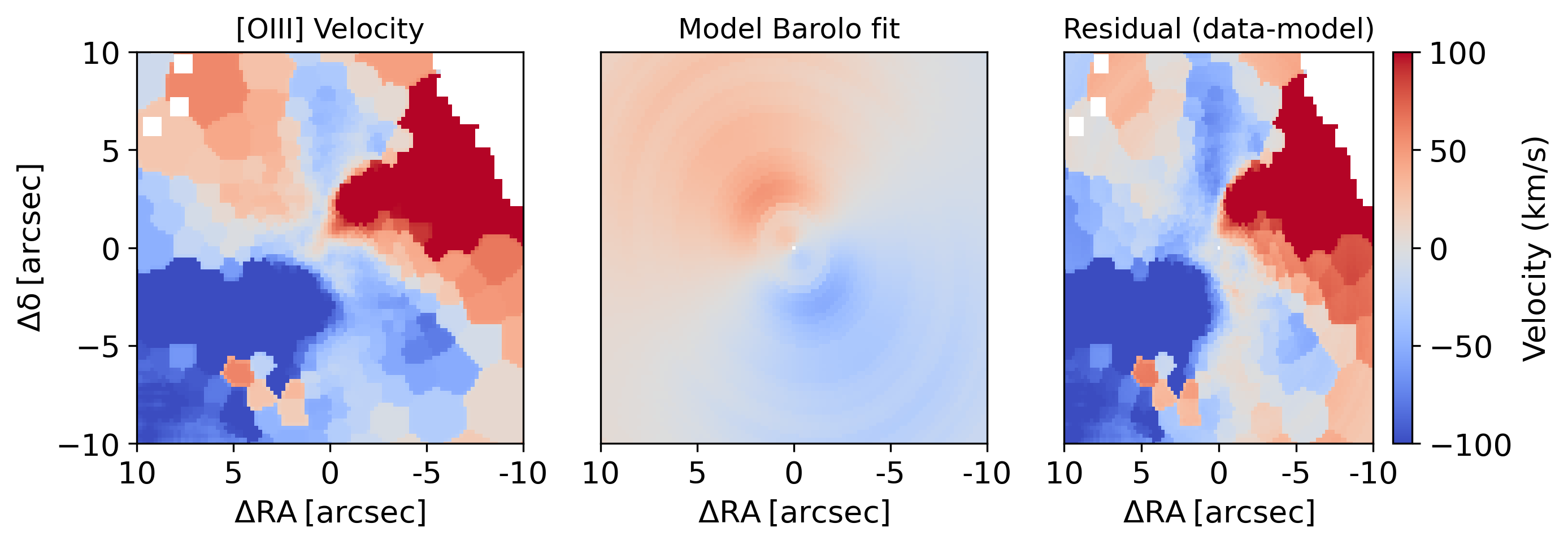}
    \caption{Velocity map for the ionized gas component of SDSS1524+08 (left). Best fit kinematic 3D-Barolo model (middle). Residual map (right). The residual map shows that rotation along PA 22\degree\ can explain some of the features along this PA, indicating there could be a rotating component in the same PA of the stellar component. However, the strong components along PA 320\degree clearly dominate the fit.
    }
    \label{fig:gasmod_1524}
\end{figure*}

\begin{figure}
	\includegraphics[width=\columnwidth]{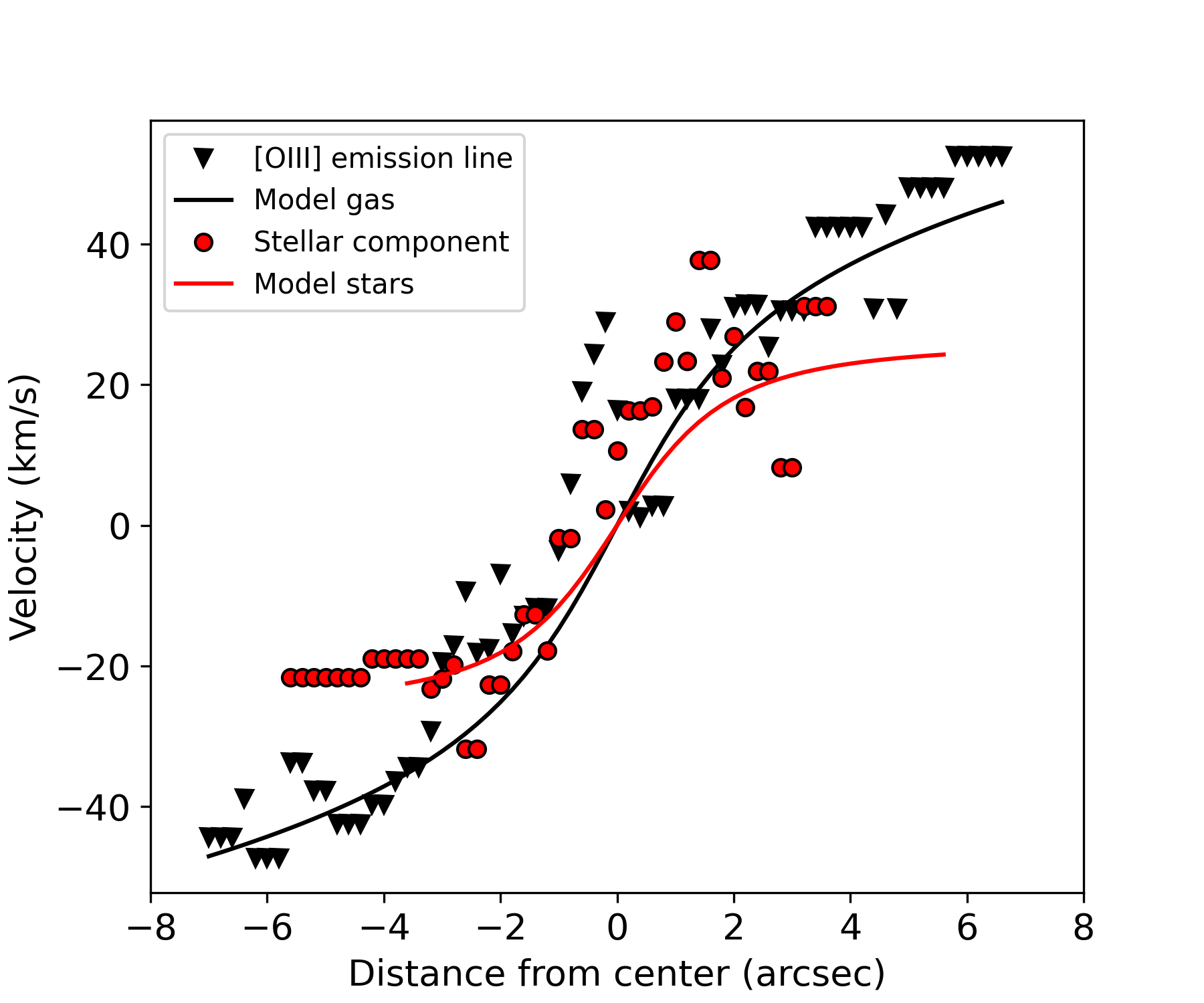}
    \caption{Rotation curve for SDSS1524+08, extracted along the major axis for the stellar component (red) and the [OIII] emission velocity map (black). Following the same color scheme, we show the fitted rotation curves for the stars and the gas.   }
    \label{fig:rotcurvefit_1524}
\end{figure}

Considering that the ionization cone aligns with the stronger velocity gradient, as shown by the spatially resolved BPT classification maps, and that the perpendicular velocity gradient aligns with the stellar rotation major axis, it is clear that, at large scales, the stellar and the ionized gas components are offset.
We consider a scenario in which both the stellar component and the ionized gas remain partially rotating along PA $\sim$ 22\degree, in the inner 15\arcsec\ and 25\arcsec\ for the stellar and gas components, respectively. 
To test this scenario, we use the 2D routine of Barolo to fit the velocity maps of the stellar component and the [OIII] emission line. For the latter, we use an [OIII] velocity map derived from a new Gaussian fit, which was carried out in a data cube binned considering the SNR in a spectral window that contains only the [OIII] line. This was done to increase the spatial resolution of this map.
We fit the inner ($<10$\arcsec) of the stellar component leaving PA, V$_{sys}$ and V$_{rot}$ as free parameters (Fig. \ref{fig:stellarmod_1524}). The residual map of this shows some agreement with a gradient caused by rotation in a plane along PA $\sim 22$\degree. Considering the possibility that this rotation corresponds to a component of the ionized gas, we use the same PA to fit the velocity map of the [OIII] map, leaving only V$_{sys}$ and V$_{rot}$ as free parameters (Fig. \ref{fig:gasmod_1524}). Considering that the gradient along PA $\sim 321$\degree heavily dominates the velocity map, the fit at all radii will be strongly biased by this component. Therefore, the residual map shows there was not a clean subtraction of the gradient along PA 22\degree. To better visualize this, we extract the velocities of both components along this PA and
fit a rotation curve (Fig. \ref{fig:rotcurvefit_1524}) that shows the rotation gradient is similar in the inner $\sim 8$\arcsec, but the ionized gas reaches higher velocities. 
We interpret the strong gradient along PA $\sim$ 321\degree as gas outflowing along the ionization bicone if we consider the NW side as the far side of the galaxy. The velocity dispersion along the bicone axis (Fig. \ref{fig:moms_1524}) reaches values of $\sim$400  km/s. Additionally, the emission line profiles along this axis reveal the presence of a wing. These features, combined with the high velocities observed, suggest the presence of an outflow, likely along the ionization bicone.

Finally, a feature of bright ionized gas towards the SE also shows high blueshifted velocities. Considering that this feature is mostly disconnected from the main galactic distribution and physically coincides with the tidal arms, we infer that this gas was located along the tidal arm and was ionized by the AGN and likely pushed by the outflow along the bicone. However, given the low velocity dispersion of this feature and the unknown 3-dimensional position of the tidal arms, it is not possible to confirm that the gas is indeed outflowing as well. 

The kinematics of the other two galaxies from this sample, NGC 5972 and the Teacup, were already described in detail by \citet[][]{finlez+2022} and \citet{venturi+2023} respectively.

\subsection{Extended emission line regions}\label{sec:eelrs}

The spatially-resolved BPT classification maps established that the EELRs observed in this sample have been mainly ionized by the nuclear AGN emission (See Sect. \ref{sect:bpts}). It follows that the central source has illuminated and ionized the entire extension of the observed EELRs. Considering the light travel time from the nuclear source to the extension of the EELR, we can trace the luminosity evolution over this time range, thus generating the AGN light curve over timescales of $10^{4-5}$ yr. To probe the AGN luminosity at different radii, we compare our observations with photoionization models, following the procedure described in detail by \citet[][]{finlez+2022} and briefly summarized below.

\subsection{Description of the photoionization models}\label{sec:lbolmodels}

\begin{figure*}
    \centering
    \subfloat{
        \includegraphics[width=0.33\textwidth]{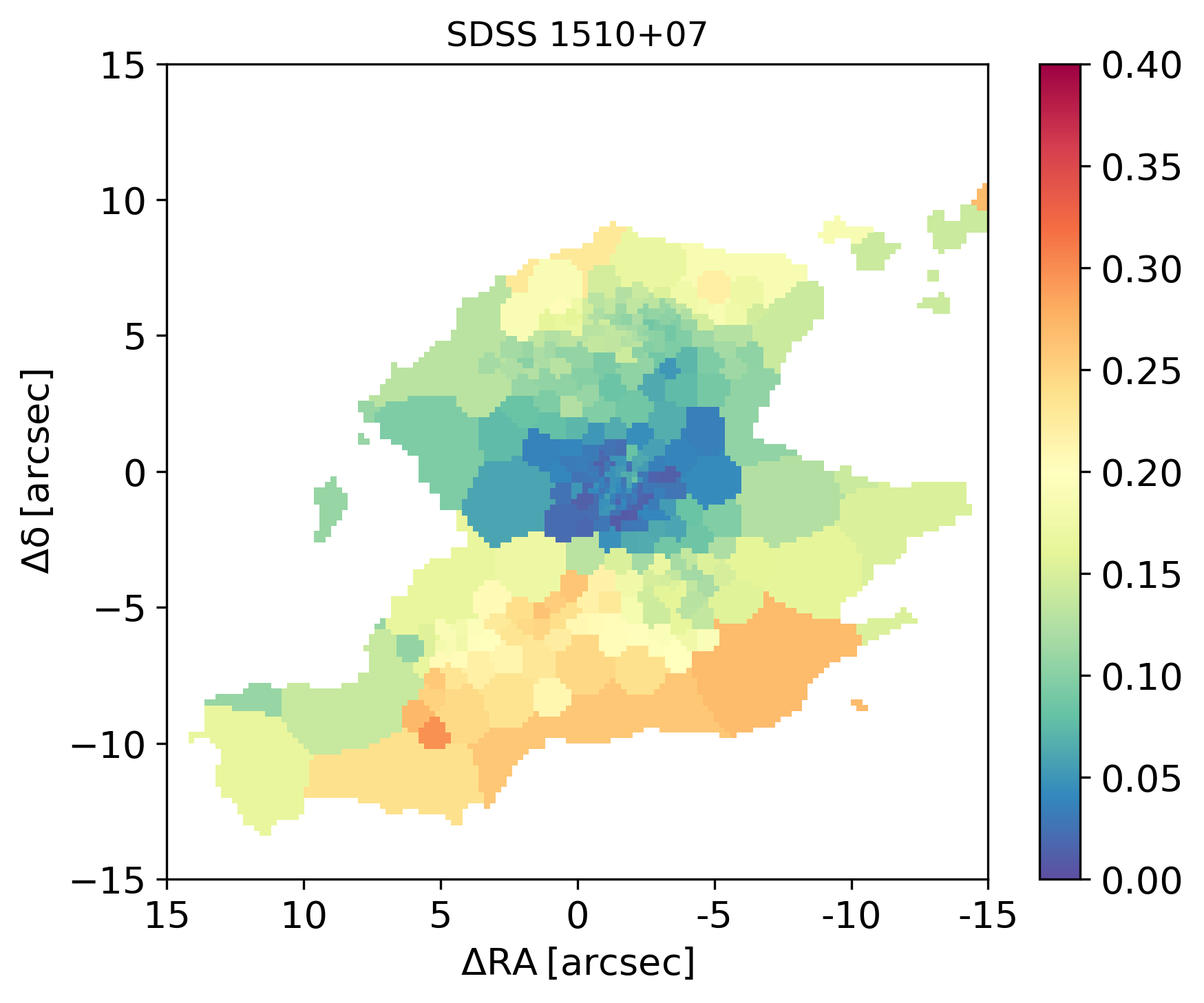}
    }
    \subfloat{
        \includegraphics[width=0.33\textwidth]{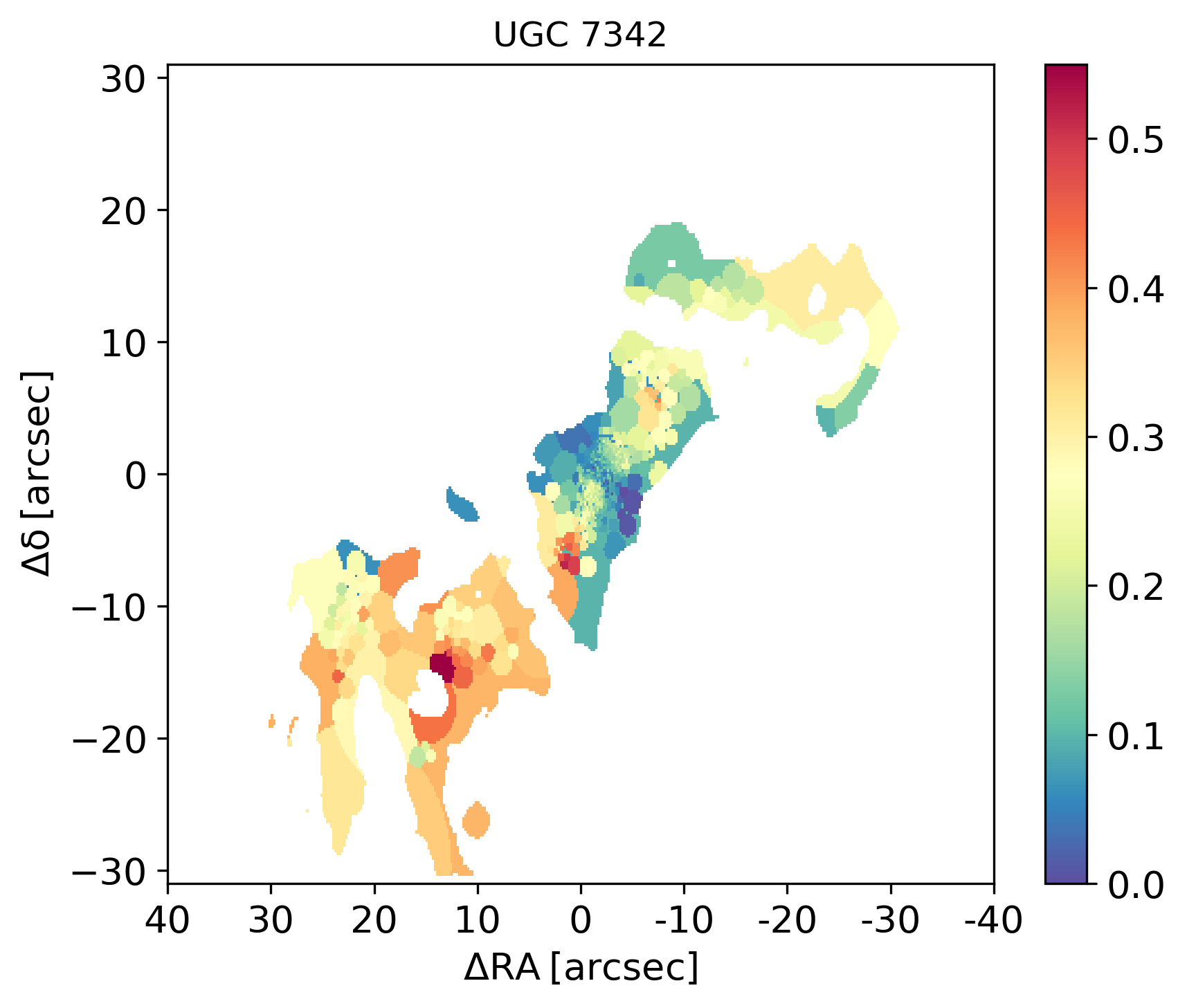}
    }    
    \subfloat{
        \includegraphics[width=0.33\textwidth]{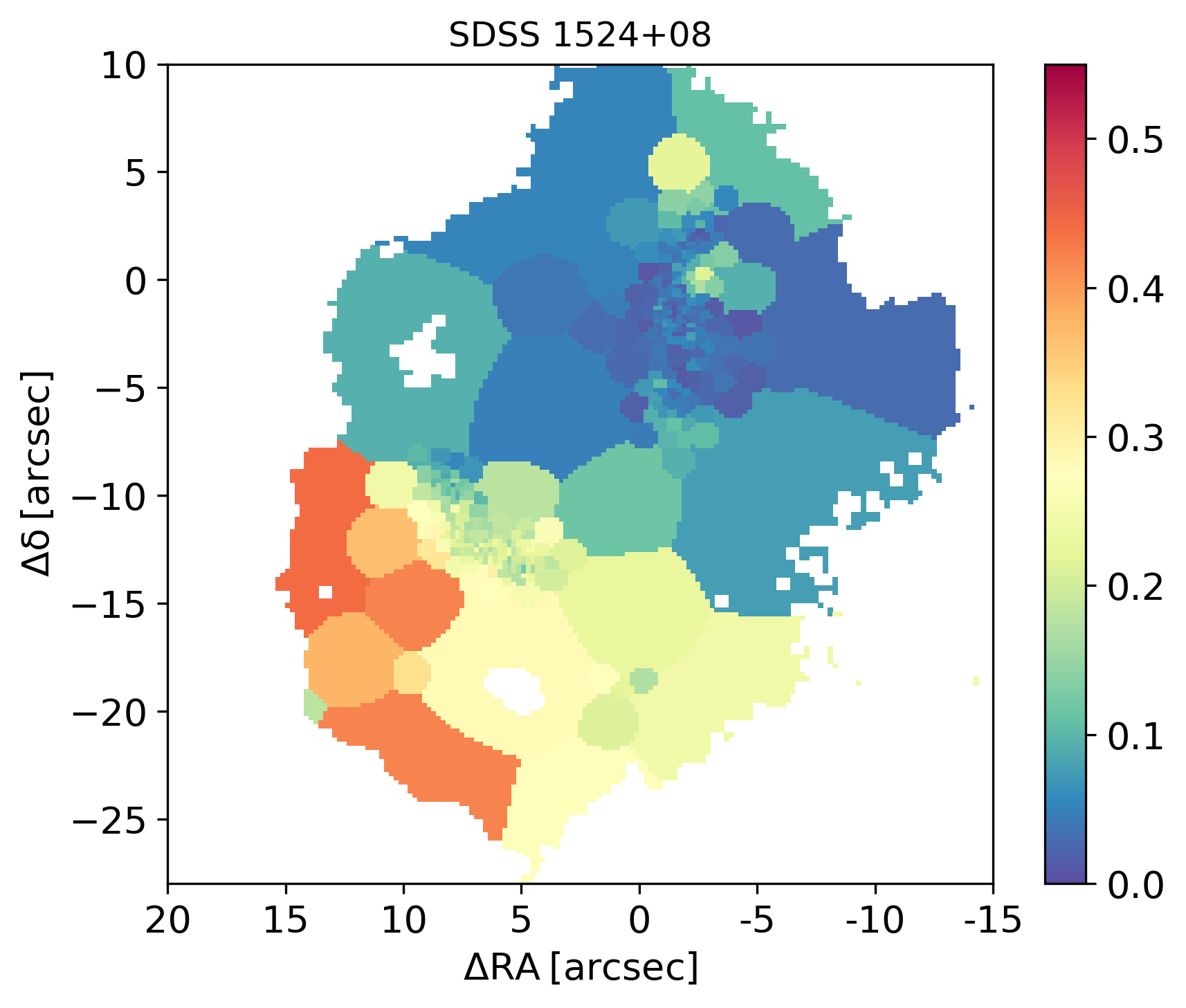}
    }
    
    \subfloat{
        \includegraphics[width=0.33\textwidth]{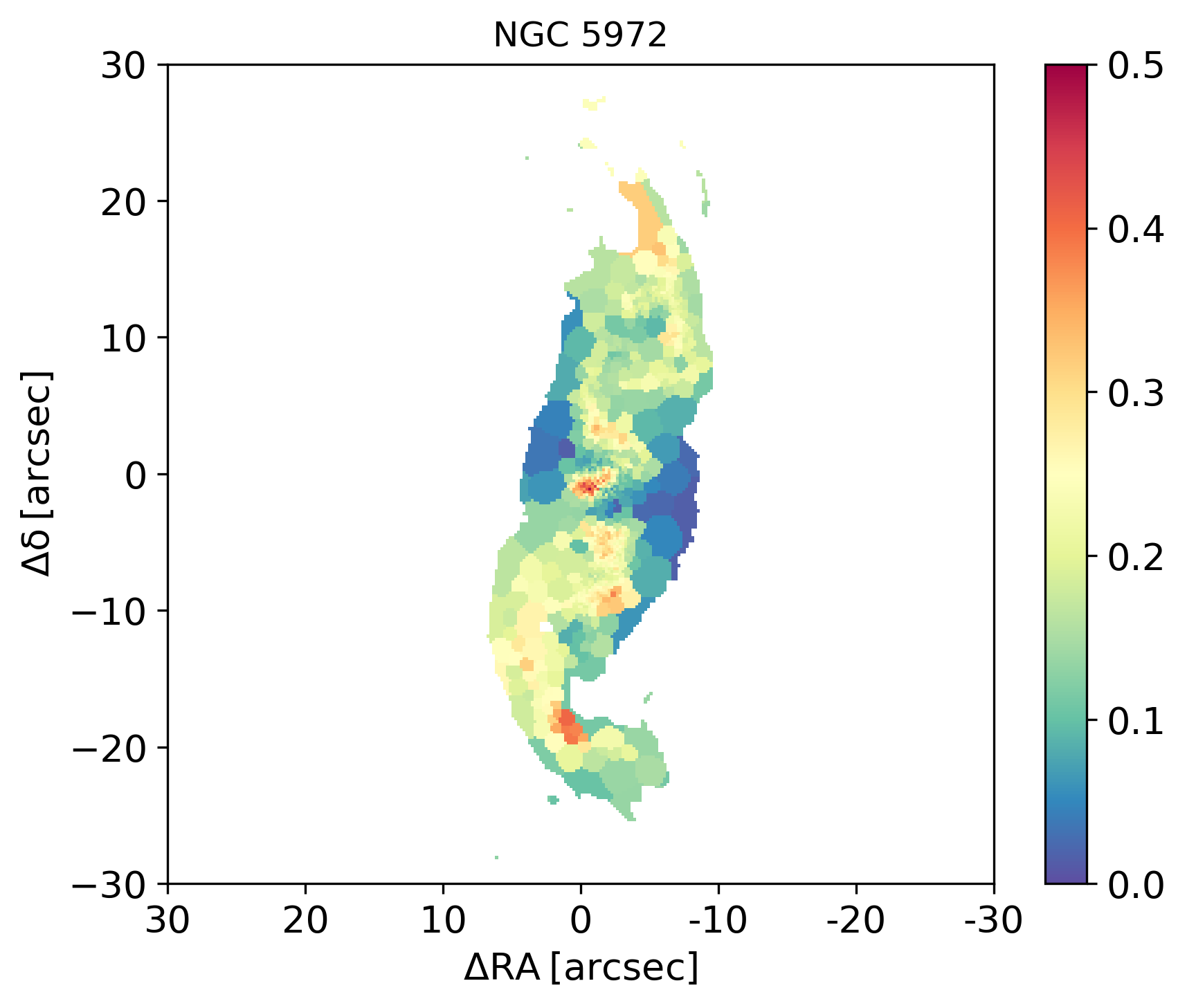}
    }
    \subfloat{
        \includegraphics[width=0.33\textwidth]{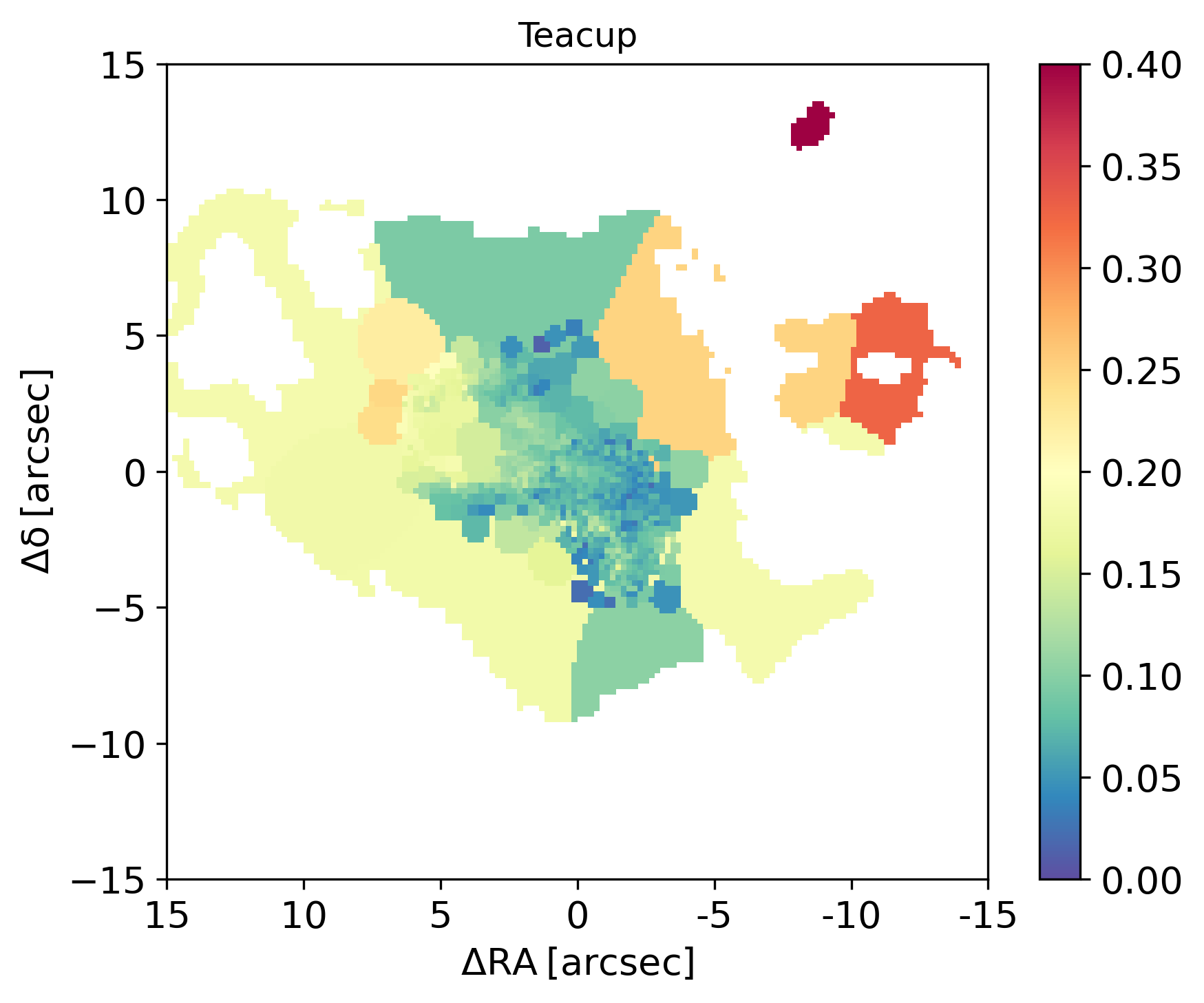}        
    }
\caption{Spatially resolved He II/H$\beta$ ratio maps. The ratio remains largely constant along the EELRs for all sources, with lower values in the central region. This area is excluded from our luminosity history analysis. Since this ratio reflects the continuum shape and is mostly constant along the ionized clouds, we assume a fixed spectral shape of the continuum with radius.}
 \label{fig:hehbratio}
\end{figure*}

These models were constructed using the photoionization code CLOUDY \citep[version 17.02,][]{ferland+2017}. This software solves the thermal and statistical equilibrium equations on a plane-parallel slab of gas that is being ionized by a central source. 
To model the ionization source we consider observations of local AGN \citep[as described in][]{elvis+1994,kraemer+2000} and define a consistent spectral shape of the central source by a broken power law ($L_{\nu} \propto \nu^{\alpha}$), with $\alpha = -0.5$ for E $<$ 13.6 eV, $\alpha = -1.5$ for 13.6 eV $<$ E $<$ 0.5 keV, and $\alpha = -0.8$ for E $>$ 0.5 keV, where E$=h\nu$ corresponding to the photon energy.
The He II/H$\beta$ ratio is sensitive to the shape of the ionizing continuum \citep[e.g.][]{penston+1978,kraemer+1994,gagne+2014}. More specifically, it is related to the ionizing UV power law index $\alpha$ in the range 228-912 \AA. In Fig. \ref{fig:hehbratio}, we show the spatially resolved ratio for our objects. The ratio is mainly constant along the EELRs in each galaxy, albeit different from source to source, and it is lower (suggesting a flatter ionizing spectrum) in the nuclear region of all galaxies. This region is not considered in the following analysis. Considering that the ratio does not change with radius along the EELR, we keep the spectral shape of the continuum fixed for all models, as further discussed in Section \ref{Sec:caveats}.

The ionization state of the gas depends on the intensity of the ionizing photon flux relative to the gas density. This can be encompassed in the dimensionless ionization parameter, U, which at the illuminated face of the cloud is defined as:

$$ U = \frac{Q(H)}{4\pi r_{0}^{2}n(H)c} $$
where $r_{0}$ is the distance between the nuclear source and the cloud, $n_{H}$ is the hydrogen number density in cm$^{-3}$, and c is the speed of light. 

The number of ionizing photons per unit of time (s$^{-1}$) is then defined as:

$$ Q(H) = \int_{\nu_{0}}^{\infty} \frac{L_{\nu}}{h\nu} d\nu $$
with $L_{\nu}$ representing the AGN luminosity as a function of frequency, $h$ the Planck constant, and $\nu_{0}=13.6$ eV/$h$  the frequency corresponding to the hydrogen ionization potential \citep{osterbrock+2006}.

We then create a grid of models varying the value of the ionization parameter, while assuming a solar metallicity. Following \citet{keel+2012a}, we consider a fully-ionized gas scenario. Hence, we assume that the hydrogen density (n$_{H}$) is the same as the electron density (n$_{H}$ = n$_{e}$). We then estimate the electron density using the [SII]$\lambda6716/6730$ \AA\ emission line ratio (hereafter referred to as the [SII] ratio), with the help of the PyNeB Python package. This software can compute the emission line emissivities \citep{luridiana+2012}. We further assume a temperature of 10$^{4}$ K \citep{osterbrock+2006}, and finally, obtain an electron density from the [SII] ratio for every aperture used. The models were run with five different values of column density, N(H), ranging from $10^{19}$cm$^{-2}$ to $10^{23}$cm$^{-2}$. We describe in more detail the effects of the N(H) changes in Appendix \ref{sec:appendixA}. In Sect. \ref{sec:lumhist} we show the case with a lower mean chi-square value for all objects, where N(H)$=10^{20}$cm$^{-2}$.

\begin{figure*}
    \centering
    \subfloat{
        \includegraphics[width=0.33\textwidth]{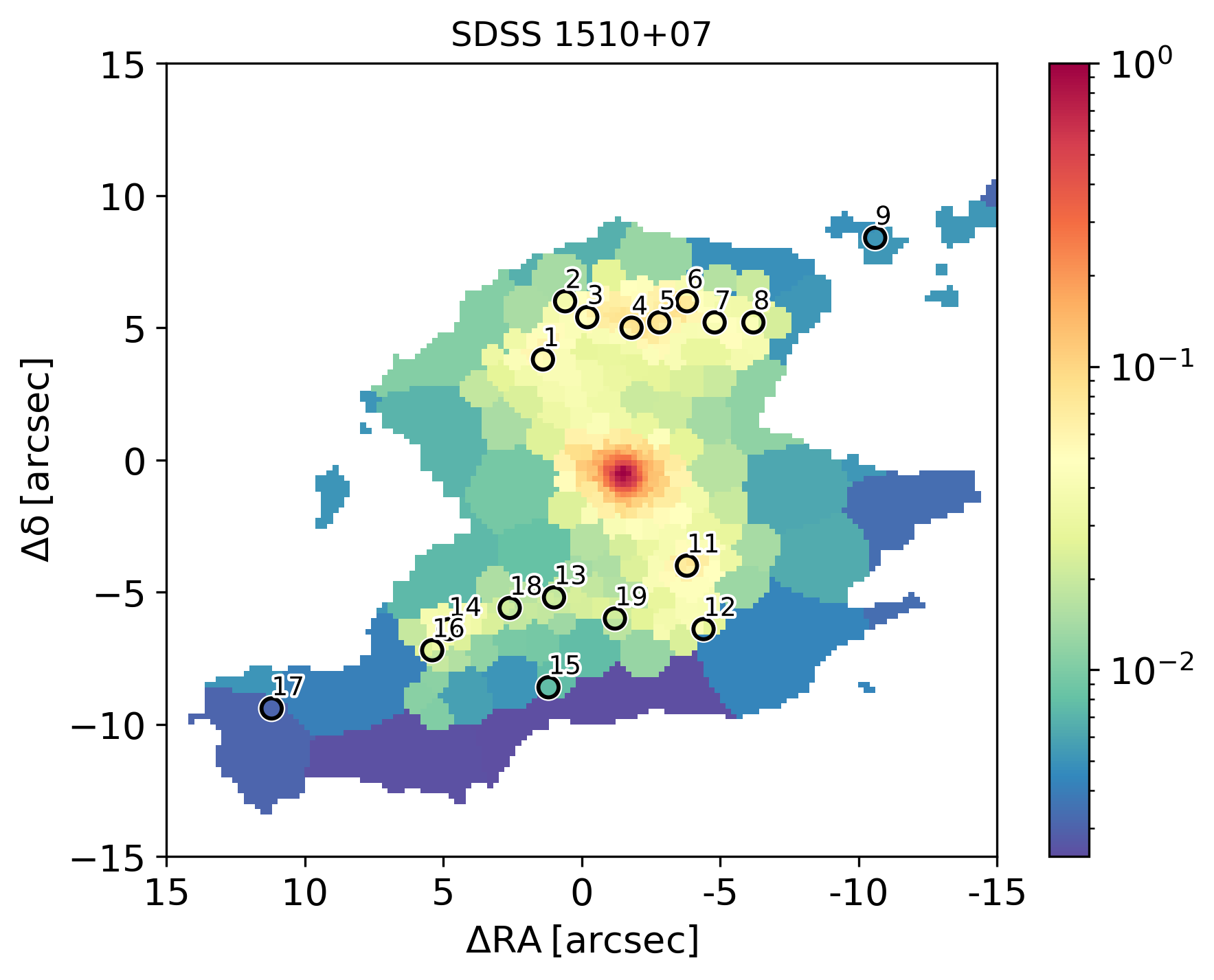}
    }
    \subfloat{
        \includegraphics[width=0.33\textwidth]{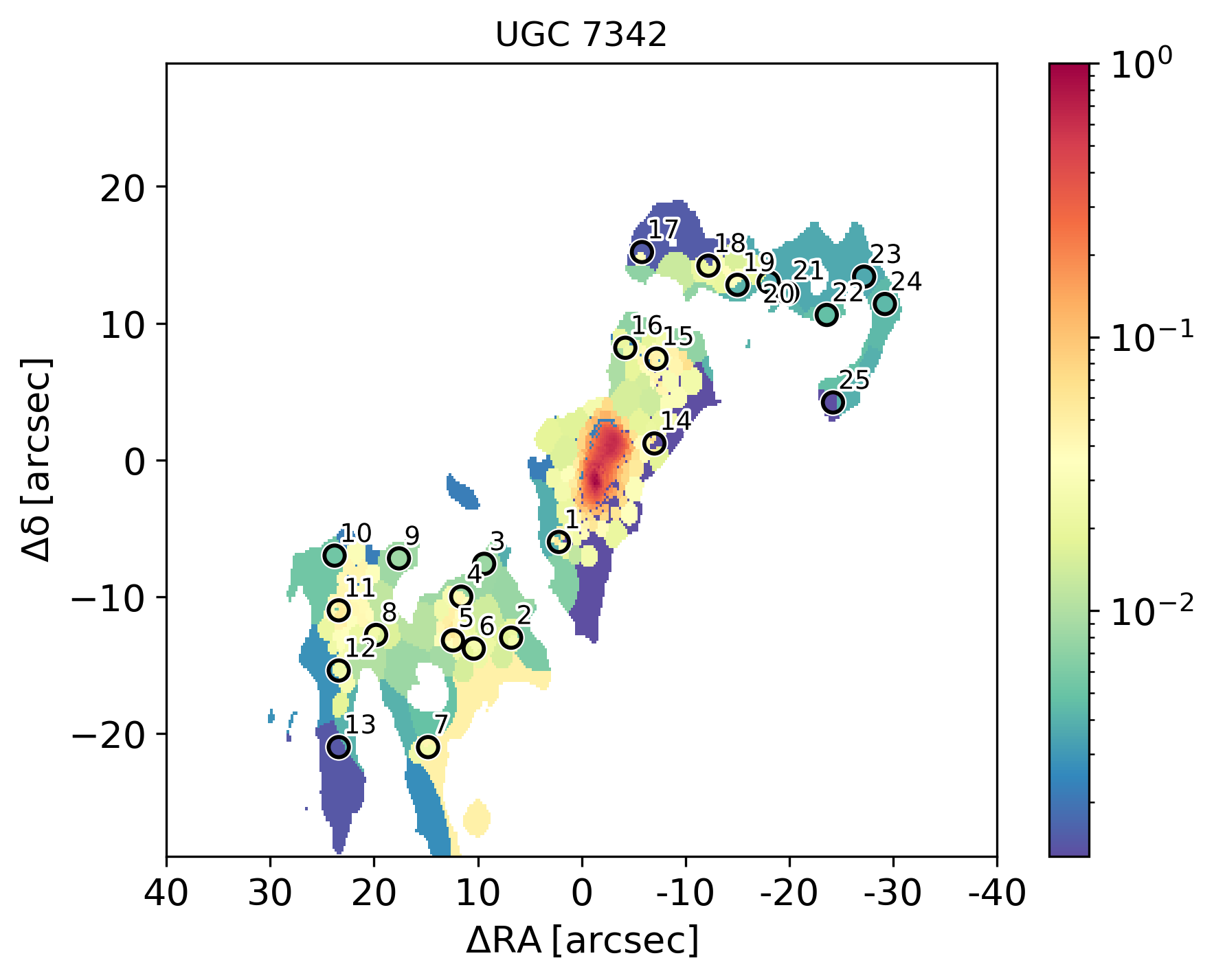}
    }    
    \subfloat{
        \includegraphics[width=0.33\textwidth]{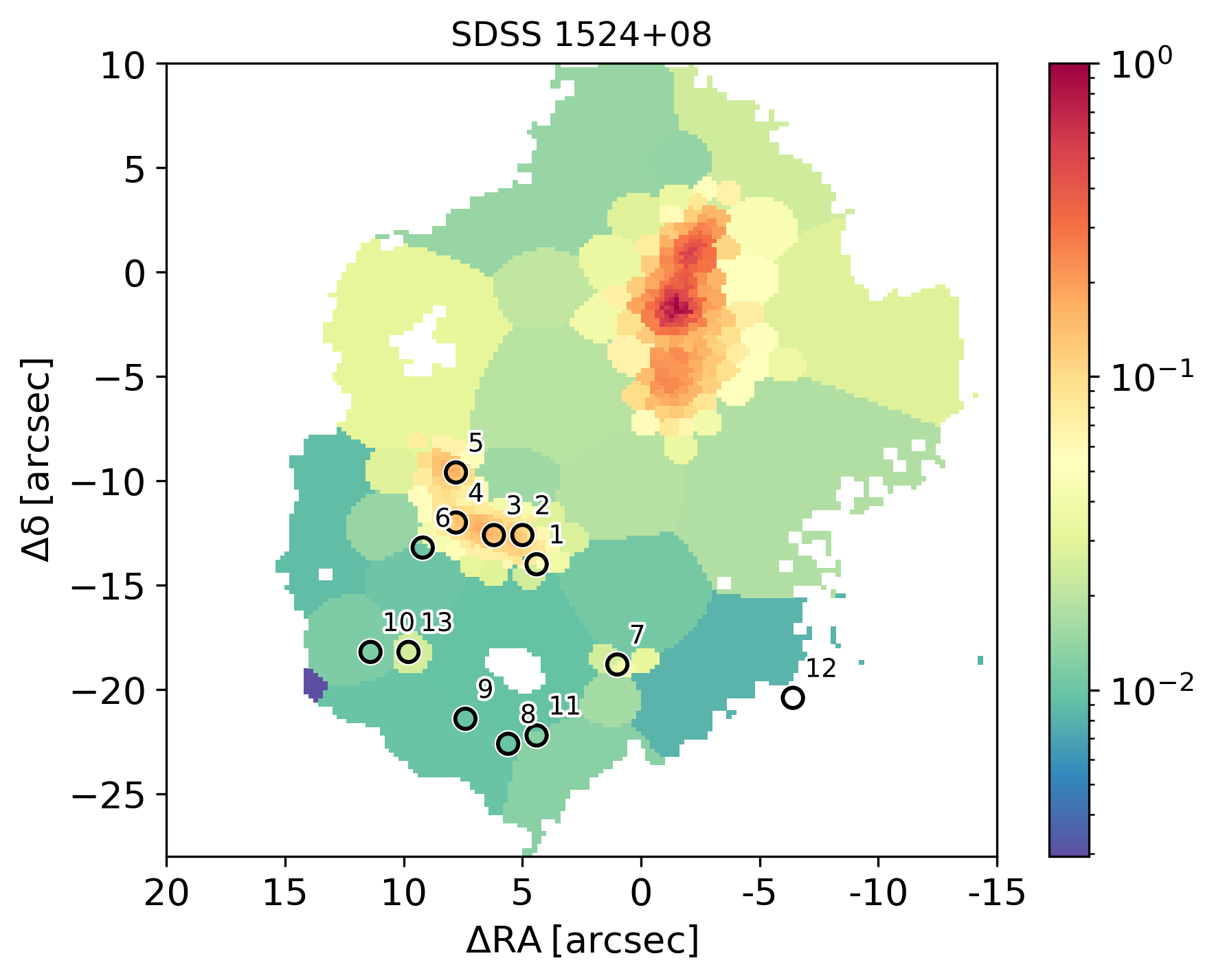}
    }
\caption{ Apertures used for the luminosity history analysis marked as black circles over [SII] emission line normalized flux maps. Left to right: SDSS1510+07, UGC 7342, and SDSS1524+08. }
 \label{fig:apertures_map}
\end{figure*}

In every source, we define a sequence of apertures, starting from the nucleus and reaching the largest extension of the EELR (shown in Fig. \ref{fig:apertures_map}). We fill the vicinity of each aperture, centered up to 1.5 times the radius of the aperture, with 30 new apertures of the same size. From these new apertures, we choose the one with the largest SNR to carry out the analysis.  We extract the integrated spectrum of each aperture and fit two Gaussian components to every emission line. The fitted emission lines are the following: H$\beta\lambda$4681, [OIII]$\lambda$5007, He II$\lambda$4685, [OI]$\lambda$6365, [NII]$\lambda$6548,6583, H$\alpha\lambda$6563, and [SII]$\lambda$6716,6730. We compute the integrated emission line fluxes and their ratios relative to H$\beta \lambda$4861 in order to compare them to the line ratios produced by our photo-ionization models, as CLOUDY delivers the emission line fluxes as ratios relative to H$\beta \lambda$4861. 

We consider a model to be acceptable when the difference between the observations and model ratios is less than a factor of two for every line ratio. However, considering that [OIII] is the brightest emission line, we require its ratio to match within 50\%. The models that do not follow these requirements are rejected before the $\chi^{2}$ calculation. In this manner, we ensure that all the lines fall within an acceptable range. We compute the reduced  $\chi^{2}$ for every model and choose as our best fit the model that matches our criteria and has the smallest reduced  $\chi^{2}$.

We then adopt the obtained best-fit ionization parameter, the electron density from the [SII] ratio, and the distance to the cloud as the projected distance from the nucleus to each aperture. With this, we are able to obtain a value of $Q(H)$ for each aperture. We then integrate our SED, together with the value for $Q(H) $, in order to estimate the bolometric luminosity required to ionize the gas to its observed state at different radii from the galaxy nucleus. 

We estimate uncertainty levels for these luminosity estimations using MonteCarlo (MC) simulations, where random noise is added to the observed emission line ratios. The simulated ratios are then fitted, as described above, and the standard deviation of the best-fit values obtained are considered as the associated uncertainties.

We consider the projected distance between the nucleus and the cloud as an approximation of the true distance between the AGN and the VP. However, the true spatial distribution of the clouds is unknown. Considering the angle between the cloud and the line-of-sight ($\theta$), the true distance is given by $r_{proj}/\sin \theta$, where $r_{proj}$ is the projected in-the-sky distance. Applying this correction, the estimated light-travel time difference is given by 

$$\Delta t = \frac{r_{proj}}{c \sin \theta} (1-\cos \theta) $$

which should be considered in our estimate of the light travel time for every cloud. 

\subsection{Luminosity history}\label{sec:lumhist}

\begin{figure*}
	\includegraphics[width=\textwidth]{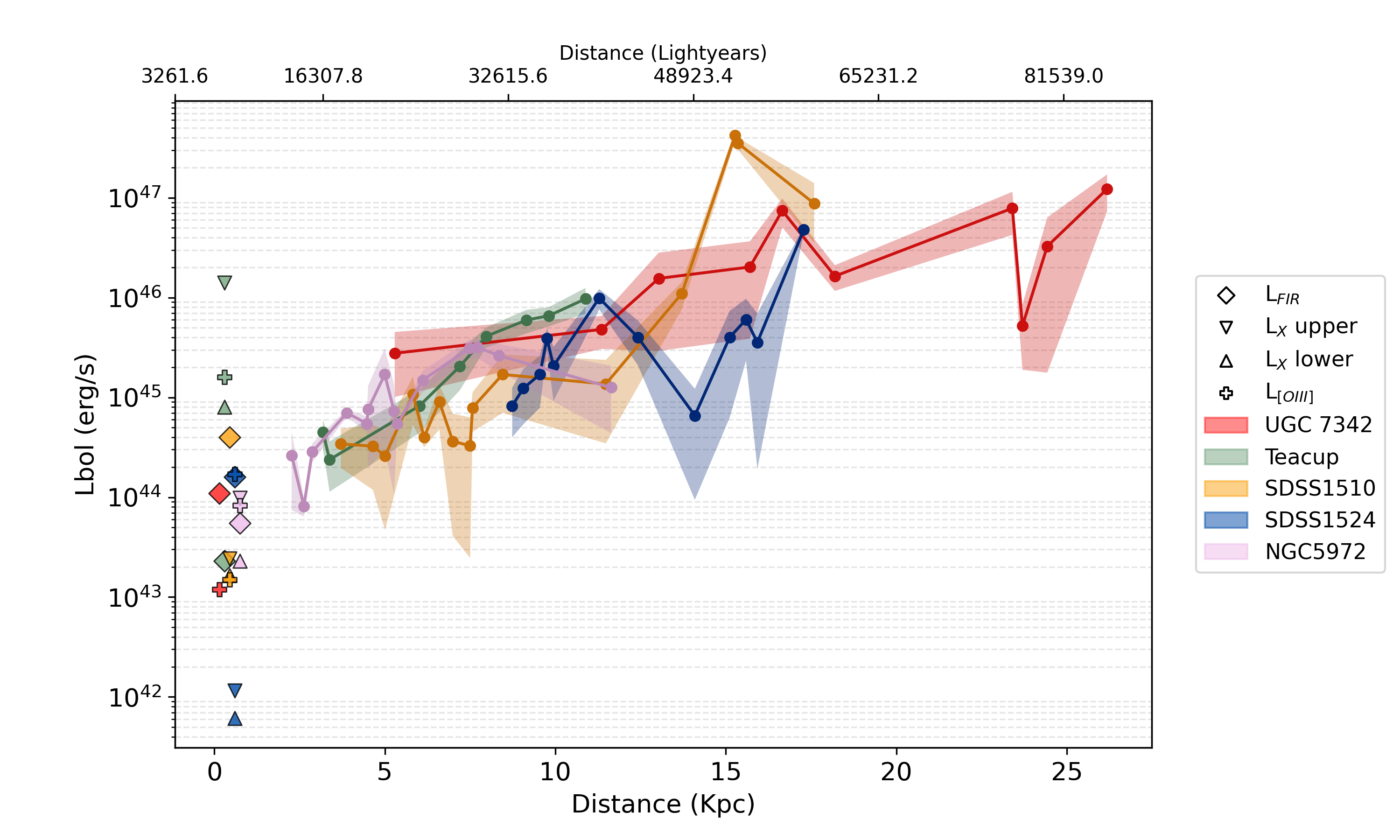}
    \caption{Derived bolometric luminosity as a function of the projected distance from the galaxy center. Color circles correspond to the derived luminosity for each aperture along a source, using different colors for each system, as presented in the figure. The different estimates of current L$_{Bol}$ are marked in the corresponding color for every galaxy; a small offset from the nucleus is added to improve the clarity of the different symbols.
    Shaded areas of corresponding color represent the error for each luminosity value derived from MC simulations, as described in the text. There is a clear trend of increasing luminosity with increasing projected distance. This implies a significant change between $\sim$ 1--3 dex over $\sim$$10^{4}$ yr timescales.}
    \label{fig:lbol_all}
\end{figure*}

We present in Fig. \ref{fig:lbol_all} the results of the process described above. For all the sources analyzed here, the bolometric luminosity increases with increasing radius from the nucleus. Considering the projected light travel time associated with each radius, we convert the distance from the nucleus to a derived timescale. 

\textit{SDSS 1510+07:} We created 16 apertures along the extension of the arm-like structures formed by the ionized gas in the N-S direction. The derived AGN luminosity from 2 to 14 kpc ranges from $2\times 10^{44}$ to $10^{46} $ erg/s. Considering that the projected distance corresponds to the distance traveled by the light from the nucleus, this implies a change of AGN luminosity of $\sim$1.7 dex on a timescale of $5\times 10^{4}$ yrs.

\textit{UGC 7342:} We defined 18 apertures along the ionized gas, SE to NW. The regions further from the nucleus ($\sim$ 25-27 kpc) are consistent with a bolometric luminosity between $2-6 \times 10^{47}$ erg/s. At $\sim$5 kpc, the luminosity is 5 $\times 10^{45}$ erg/s. The physical distance from the nucleus reached by the ionized gas (25 kpc) in this galaxy is the largest in our sample and corresponds to $\sim$8$\times 10^{4}$ yrs, in which the nucleus has faded 2 dex in luminosity.

\textit{SDSS 1524+08:} We generated 8 apertures, most associated with the SE ionized cloud, covering distances from  2 to 13 kpc. The bolometric luminosity at these limits corresponds to 1$\times 10^{44}$ and 3$\times 10^{46}$ erg/s, respectively. This implies that the AGN luminosity has faded 2.5 dex over $4\times 10^{4}$ yrs.

\section{Discussion}
\label{sec:discussion}

The spatial resolution provided by our VLT/MUSE observations allows us to spatially resolve the EELRs of these sources and study in detail their luminosity histories together with their morphologies and kinematics. 

\subsection{Origin of the EELRs}

The presence of tidal tails in the stellar component of all three objects, as well as the kinematic perturbations observed in the ionized gas, all point towards a scenario where recent mergers or galaxy interactions have been part of the sources' histories. Likely, these interactions have hauled gas from the galactic disk along the tidal arms, where it has been illuminated and photoionized by the AGN. 
The presence of this gas can cover a larger volume, as seen from the AGN, in contrast with a galaxy where all, or most, of the gas remains within the confines of the disk. This volume filling percentage increases the likelihood of the AGN opening angle to intersect with gas, and consequently ionize it. This could explain the high incidence of merger or interaction features within the known VPs.

We explore the environmental properties of the sources in our sample using the group and cluster catalog of \cite{tempel+2017}. This catalog was obtained using the friends-of-friends algorithm.  We found that SDSS1510+07 is part of a group containing 16 sources. UGC 7342 lies within a lower-density environment in a small group of 3 galaxies, and SDSS1524+08 is part of a group containing 107 galaxies. Both SDSS1510+07 and SDSS1524+08 can be found at lower radius and velocity in the line-of-sight (LOS) on a phase-space diagram, which combines information on clustocentric velocity and clustocentric radius \citep[see][for a detailed description]{rhee+2017}. This could indicate that these sources are in the ‘’infalling’’ region of the phase-space diagram and, therefore, are recently arriving in the more dense region of their respective groups.

\subsection{Luminosity history}

To derive estimates of the current AGN bolometric luminosity, we employ the values obtained from the luminosities measured from the [OIII] emission line, and the IR and X-ray emission. We then compare these current luminosity estimates with the bolometric luminosity derived from CLOUDY in the aperture closest to the center.

 In Table \ref{table:current_lums}, we show the current AGN luminosity indicators in comparison to the values previously obtained from the CLOUDY computations. For the [OIII]-derived luminosity, the flux is taken from a fit to an aperture in the central region. We assume a correction factor of $L_{bol}/L_{[OIII]}  \simeq 3500$, with a variance of 0.38 dex, to estimate the bolometric luminosity, following \cite{heckman+2004}. For the bolometric luminosities estimated from L (2-10 keV), we use an X-ray bolometric correction $\kappa_{bol} = L_{bol}/L_{2-10} $ assuming a range of $10 < \kappa_{bol} < 100 $, based on typical AGN values \citep[e.g.][]{vasudevan+2009,lusso+2012}.  The far-infrared (FIR) luminosity from \cite{keel+2012a} is used as a conservatively high estimate bolometric luminosity indicator, assuming potentially obscured AGNs, and thus, for these cases, they can be considered as upper limits. 

For the sources UGC 7342, SDSS1510+07, and SDSS1524+08, we use available archival X-ray data to estimate the intrinsic AGN flux. UGC 7342 was observed with NuSTAR and XMM-Newton, but in both cases, the galaxy was only marginally detected. The spectrum can be well fitted by an absorbed power law, with a very high value of photon index ($4.4_{-1.3}^{+2.1}$), and a low column density ($2.0_{-1.5}^{+2.5} \times 10^{21}$ cm$^{-2}$). Uncertainties on the parameters are at the 90\% confidence level. The emission may be consistent with star formation, primarily reflecting host galaxy X-ray emission.
SDSS1510+07 has been observed with NuSTAR and XMM-Newton, and its spectrum can be well fit by an absorbed power-law plus host galaxy emission contributing to the soft X-rays. The derived column density is $8.9^{+4.1}_{-2.8}\times 10^{22}$ cm$^{-2}$, while the photon index is $1.5_{-0.2}^{+0.3}$. SDSS1524+08 was observed by Chandra, and its spectrum can be well fitted with a simple power law (no absorption besides Galactic) model, with $\Gamma=1.8_{-0.7}^{+0.8}$. The bolometric luminosity estimated from the intrinsic 2-10 keV fluxes are listed in Table \ref{table:current_lums}.

\begin{table*}
\caption{Current bolometric luminosity estimators}             
\label{table:current_lums}      
\centering          
\begin{tabular}{lcccccl}   
\hline\hline      
Source & L$_{[OIII]}$ & L$_{nearest}$ & L$_{FIR}$ & L$_{X}$ & Ref \\
(1) & (2) & (3) & (4) & (5) & (6) \\
\hline                    
SDSS1510+07  & $1.5\times 10^{43}$   & $2.1\times 10^{44}$   & $4\times 10^{44}$     & $1.6-2.4\times 10^{43}$      & \cite{keel+2012a} \\
UGC 7342     & $1.2\times 10^{43}$   & $4.7\times 10^{45}$   & $1.1\times 10^{44}$   &      & \cite{keel+2012a} \\
SDSS1524+08  & $1.7\times 10^{44}$   & $1.3\times 10^{44}$   & $1.6\times 10^{44}$   & $6.12-11.5\times 10^{41}$      & \cite{keel+2012a} \\
NGC 5972     & $8.3\times 10^{43}$   & $7.2\times 10^{44}$   & $ 5.5\times 10^{43}$ & $2.3-9.8 \times 10^{43}$ & \makecell{\cite{keel+2012a, zhao+21}; \\  \cite{harvey+23}} \\
The Teacup   & $1.6\times 10^{45}$   & $2.2\times 10^{45}$   & $2.3\times 10^{43}$   & $8-140 \times 10^{44}$   & \makecell{ \cite{lansbury+2018}; \cite{harrison+2014}; \\ \cite{keel+2017} } \\
\hline                  
\end{tabular}
\tablefoot{Column (1) Target name. (2) Bolometric luminosity derived from [OIII] luminosity in the nucleus. (3) Bolometric luminosity derived from the CLOUDY simulations in the nearest aperture to the nucleus. (4) FIR luminosity. (5) Bolometric luminosity derived from X-ray (2-10 keV). (6) References. All luminosity values are in erg/s. (2) and (3) are calculated in this work, as well as (5) for SDSS1510+07 and SDSS1524+08.}
\end{table*}

We compare the estimated bolometric luminosity with the highest values (L$_{Bol}^{max}$), reached at the largest distances from the nucleus, with the different current AGN luminosity estimators. 
For SDSS1510+07, L$_{Bol}^{max}$ reaches $10^{47}$ erg s$^{-1}$. Comparing this value with the current estimators derived from L$_{FIR}$ \citep[from][]{keel+2012a}, L$_{X}$, and L$_{[OIII]}$ suggests fading factors of 2.6, 3.9, and 4 dex, respectively. This occurs over a timescale of $4.6\times 10^{4}$ yr.
For UGC 7342, L$_{Bol}^{max}$ corresponds to $4\times 10^{46}$ erg s$^{-1}$, indicating a fading of 2.5, and 3.5 dex from the FIR, and [OIII] L$_{Bol}$ estimators, respectively, over a timescale of $8\times 10^{4}$ yr. For this source, the X-ray luminosity estimator is likely reflecting the host galaxy X-ray emission, rather than the AGN, which may be heavily obscured.
For SDSS1524+08, L$_{Bol}^{max} = 4 \times 10^{46}$ erg s$^{-1}$, comparing this luminosity with the FIR, X-ray and [OIII] indicators, we find fading factors of 2.4, 4.7, and 2.4 dex, respectively, in timescales of $4.5\times 10^{4}$ yr.
For NGC 5972,  L$_{Bol}^{max} = 2 \times 10^{46}$ erg s$^{-1}$, indicating a fading of 2.6, 2.5, and 2.4 dex fading over timescales of $5.5\times 10^{4}$ yr, as derived from the FIR, X-ray, and [OIII] L$_{Bol}$ indicators.
For The Teacup, L$_{Bol}^{max}$ corresponds to  $9 \times 10^{45}$ erg s$^{-1}$, at a distance corresponding to $5.4 \times 10^{4}$ yr. The current AGN indicators derived from L$_{FIR}$ \citep[from][]{keel+2012a}, L$_{X}$ and L$_{[OIII]}$ indicate a difference with L$_{Bol}^{max}$ of 2.6, 1.1 and 0.7 dex. This source has the largest dispersion on the derived fading factor, ranging from 1.5 to 1000 times the current L$_{Bol}$. A significant contribution to this scatter comes from the uncertainty on the L$_{X}$, as can be seen in \ref{fig:lbol_all}. 

The derived variability on timescales of $\sim 10^{4}$ yr for our sample is consistent with previous studies of other similar sources \citep[e.g.][]{lintott+2009,gagne+2014,sartori+2016, keel+2017}. These results could be interpreted in the context of a scenario in which the AGN duty cycle is broken down to shorter phases of $10^{4-5}$ yr, resulting in a total $10^{7-9}$ yr "on" lifetime \citep[][]{schawinski+2015}.

\subsection{Structure function}

\begin{figure*}
	\includegraphics[width=\textwidth]{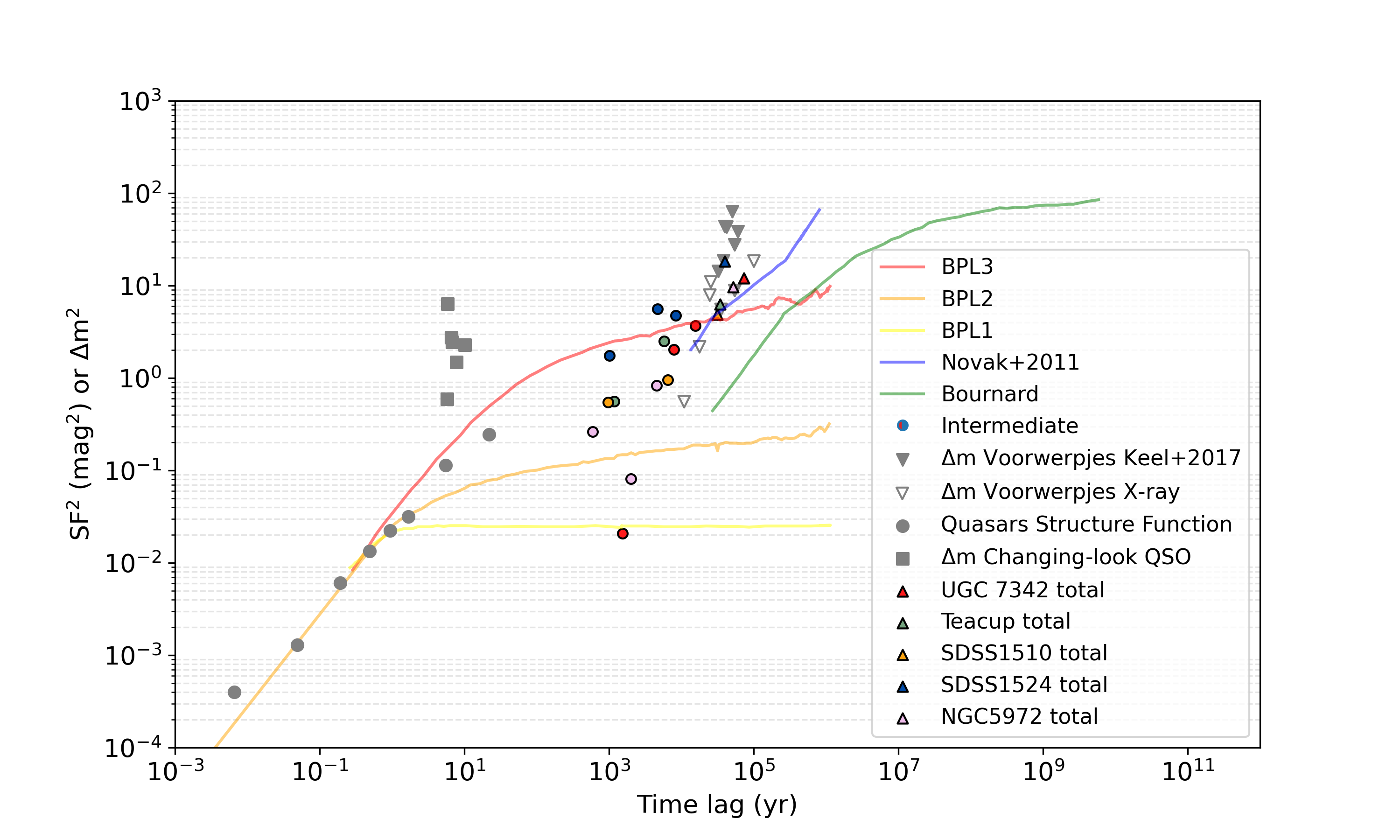}
    \caption{SF for the magnitude difference at every measured radius (colored circles as described in the legend) and the total magnitude difference (colored triangles). We add in gray the variability data summarized in Fig. 2 of \cite{sartori+2018} for comparison. Gray circles represent the quasars' structure function. Squares represent the $\delta m$ for changing-look quasars. Downward triangles show the data for Voorwerpjes from \citet[][filled triangles]{keel+2017} and from X-ray data (empty triangles). The solid lines represent the BPL1, BPL2, and BPL3 models in yellow, orange, and red, respectively. While the green and blue solid lines indicate SF from hydrodynamic simulations from \citet[][]{novak+2011} and Bournaud \citep[adapted from][]{sartori+2018}, respectively. See \citet{sartori+2018} for more details.}
    \label{fig:SF_all}
\end{figure*}

We can interpret our results in the context of the framework for AGN variability presented by \citet{sartori+2018}. In this work, the variability of AGN is linked over a wide range of timescales, from days to $10^{9}$ yr, as derived from simulations \cite{novak+2011}.
The total variability of every source is estimated as the luminosity difference between the closest and the farthest aperture from the nuclear source. This is then translated into a magnitude difference at a given time lag \citep[$\tau$; as defined in eq. 1 of][]{sartori+2018}. 
We repeat this computation for the entire light-time travel along the EELR for every source. We obtain a value of $\delta m$ (mean luminosity difference) for every bin in a time lag, which is computed from the intermediate points in the recreated luminosity curves (Fig. \ref{fig:lbol_all}). 
These values are shown in Fig. \ref{fig:SF_all}, where they can be contrasted with observations and the structure function (SF) from hydrodynamics simulations. Furthermore, we included models BPL1, BPL2, and BPL3 from Fig. 2 in \citet{sartori+2018}. These models represent the SF from light curve simulations, derived from different power spectral densities (PSDs) of the Eddington ratio distribution. 
The total luminosity variability for the sources in the sample is consistent with the values obtained for the Voorwerpjes in the SF \citep[as derived by][]{keel+2017,sartori+2018}. Our estimated $\delta m$ values cover the time lag gap between changing-look QSOs and Voorwerpjes. The observed distribution is more consistent with the SF model BPL3, which considers a broken power-law for the PSD, with $\alpha_{low} = -1$, $\alpha_{high} = -2$, $\nu_{break} = 2\times 10^{-10}$ Hz $\sim 160$ yr. Adding to the analysis carried out in \cite{sartori+18}, it can help to further constrain the underlying PSD and the ability to link AGN variability on different timescales.

\subsection{Physical timescales}
Our analysis indicates that all the AGN in this sample have experienced a luminosity decline of 1--4 dex over the past $7-8\sim 10^{4}$ yr. Similar results were observed by \cite{keel+2017} for SDSS1510+07, UGC 7342, NGC 5972, and the Teacup galaxy. Their technique made use of the HST H$\alpha$ surface brightness of the brightest pixel at each radius. Both procedures track each other within a constant, reconstructing similar luminosity histories. Naturally, these results raise the question of the possible physical mechanisms connected to this decline, which we explore below.

\subsubsection{The role of mergers:}
The presence of merger or galaxy interaction signatures in all the galaxies of this sample suggests the possibility of a link between the strong AGN activity and the merger. In this scenario, the galaxy goes through a merger or interaction, which disturbs the galactic potential, hauling gas outside the disk and possibly funneling gas towards the central region, triggering a powerful AGN phase, which then ionizes the clouds in its path, reaching tens of kpc. Following this phase, the AGN would fade over time until its current luminosity. 
However, it is estimated that a major merger could accumulate a large fraction of galactic gas in the inner region in roughly a dynamical timescale ($\sim 10^{8}$ yr). Considering the discrepancy between the merger timescale and the one derived for the AGN fading (i.e., $10^{4-5}$ yrs), the connection between the merger and the AGN triggering is not evident, and they could be unrelated phenomena. Therefore, the merger increases the volume filling fraction for the AGN to illuminate distant clouds, which may translate into a selection bias.

\subsubsection{Accretion disk instabilities:}

Accretion disk instabilities play a crucial role in regulating the flow of matter onto the SMBH, influencing the luminosity and activity of the AGN. The time it takes for the accretion disc to reach hydrostatic equilibrium, referred to as the dynamical timescale, can be expressed as follows

$$ t_{dyn} \sim \frac{1}{\Omega}   $$
where $\Omega = (G M_{BH}/r^{3})^{1/2}$ corresponds to the angular frequency for a Keplerian orbit. 

The timescale for the cooling of the disk, known as the thermal timescale, is expressed as 

$$ t_{th} \sim \mu^{-1} t_{dyn} $$
Viscous disturbances propagate over a distance $l$ through the disk on timescales of $l^2/v_{k}$, where $v_{k}$ is the kinematic viscosity, and $\mu$ is the viscosity parameter.
\cite[][]{lamassa+2015} parameterized this timescale as

$$ t_{vis} \sim 
\left( \frac{\mu}{0.1}\right)^{-1}  
\left(\frac{\lambda_{Edd}}{0.03}\right)^{-2}  
  \left(\frac{\eta}{0.1}\right)^{2} 
  \left(\frac{r}{10r_{g}}\right)^{7/2} 
  \left(\frac{M}{10^{8} M_{\odot}} \right) $$
where $\eta$ corresponds to the radiation efficiency, $r$ to the radius, and $M$ to the black hole mass. 

To obtain representative timescales for these processes, we assume $\mu \sim 0.03$ \citep{davis+2010}, $\eta \sim 0.1$ \citep{soltan+1982}, and consider representatives values for our sample of $M \sim 10^{8} M_{\odot}$, and $\lambda_{Edd} \sim 0.17$. For the gravitational radius we consider  $r \sim 100 r_{g}$ for a UV emitting region \citep{kato+2008, lamassa+2015}, and obtain the following:

$$t_{dyn} (r/r_{g}=100) \sim 16\: days $$

$$t_{th} (r/r_{g}=100) \sim  1.5\:  yr$$

$$t_{vis} (r/r_{g}=100) \sim 2\times 10^{4}\: yr $$
While for optical emission we consider  $r \sim 200 r_{g}$, which translates into the timescales

$$t_{dyn} (r/r_{g}=200) \sim 45\: days $$

$$t_{th} (r/r_{g}=200) \sim  4\: yr$$

$$t_{vis} (r/r_{g}=200) \sim 2\times 10^{5}\: yr $$

This estimate indicates that both the dynamical and thermal timescales are too short to account for the suggested change in luminosity timescale from our results. Instead, the viscous timescale is more consistent with our inferred timescales. This suggests an accretion rate change in the accretion disk, following the inflow timescale of gas accretion.

\subsection{Potential biases and selection effects}
\label{Sec:caveats}

In \cite{finlez+2022}, we discussed some caveats to take into consideration when the bolometric luminosity is derived from the best-fit CLOUDY models. These include, among others, the geometry, as we assume the real distance traveled by the light from the nuclear source corresponds to the projected distance ($r_{proj}$) from the center to the ionized cloud. A correction for this will include the inclination angle ($i$) of the cloud with respect to the galaxy plane as $r_{proj}/sin(i)$. 
Additionally, it is important to consider that the density, as derived from the  [SII]$\lambda\lambda$6716, 6731 \AA\ doublet ratio, can easily fluctuate in the range 100-1000 cm$^{-3}$. This is mostly explained by the steepness of the slope in the conversion curve between [SII] line ratio and derived electron density \citep{osterbrock+2006}. Furthermore, this tracer can saturate at high densities or in gas with a large matter-bounded fraction.

To test the impact of changing the shape of the continuum based on the He II/H$\beta$ ratio, we test a series of CLOUDY models focused on the median of this ratio along the EELR of every galaxy. We convert the He II/H$\beta$ to an $\alpha$ in the 13.6--500 eV range, following \cite{penston+1978}. Higher values of this ratio translate into higher $\alpha$ values. We run CLOUDY models following the method described in \ref{sec:lumhist}. At higher values of $\alpha$, the grid of simulations moves towards lower values of [NII]/H$\alpha$ and [OIII]/H$\beta$ in the BPT diagram. In this direction, the value of U remains mainly constant. Along this same direction, the density changes. Considering that the ionization parameter stays mainly invariable with changes in the shape of the continuum in the range 13.6--500 eV, and that we use the [SII] emission line ratio to derive the density of the gas, we choose to keep the same shape of the continuum with radius and for every object, to avoid adding more degeneracies to the model fit. The suite of models that consider the dispersion of the He II/H$\beta$ ratio along the EELRs indicates that the variations in the shape of the continuum do not translate into a large difference in the luminosity curve. Therefore, based on the He II/H$\beta$ ratio, the corresponding changes in the shape of the continuum do not largely affect our results. More details can be found in Appendix \ref{sec:appendixA}.

The sources analyzed were selected from the sample described in \cite{keel+2012a}. This sample was obtained from the Galaxy Zoo project, combining targeted and serendipitous approaches based on the visual identification of extended ionized clouds that were near or connected to a galaxy. These were later filtered to be consistent with AGN ionization based on BPT classification.
The main bias of this selection is driven by the extension and brightness of the EELRs to be flagged as VP candidates. This would cause our method to be preferentially sensitive to the most extreme cases, as a powerful AGN in the past would illuminate up to the greatest distances from the nuclear source and strongly ionize large regions of the available gas. Furthermore, the sample selection is not based on confirmed AGN to discover large AGN-ionized clouds with no AGN presence, which can be due to strong obscuration or acute variability during the light travel time from the nucleus to the cloud extension. 

All the objects in our sample indicate fading AGN. Considering the sample selection characteristics described above, the sources in this work may represent very extreme cases of VPs. However, in principle, both fading and rising AGN could be identified from their EELR ionization state. Rising AGN would show large-scale signatures of a weak nuclear source, while the small-scale (within $\sim$10 pc) signatures would denote more powerful activity. To further understand if the detection of fading AGN reflects AGN evolution or the result of a bias in our sample selection, the study of a larger, less biased sample is required. We will explore this in future work (Finlez et al., in prep).

\begin{figure*}
    \centering
    \subfloat{
        \includegraphics[width=0.33\textwidth]{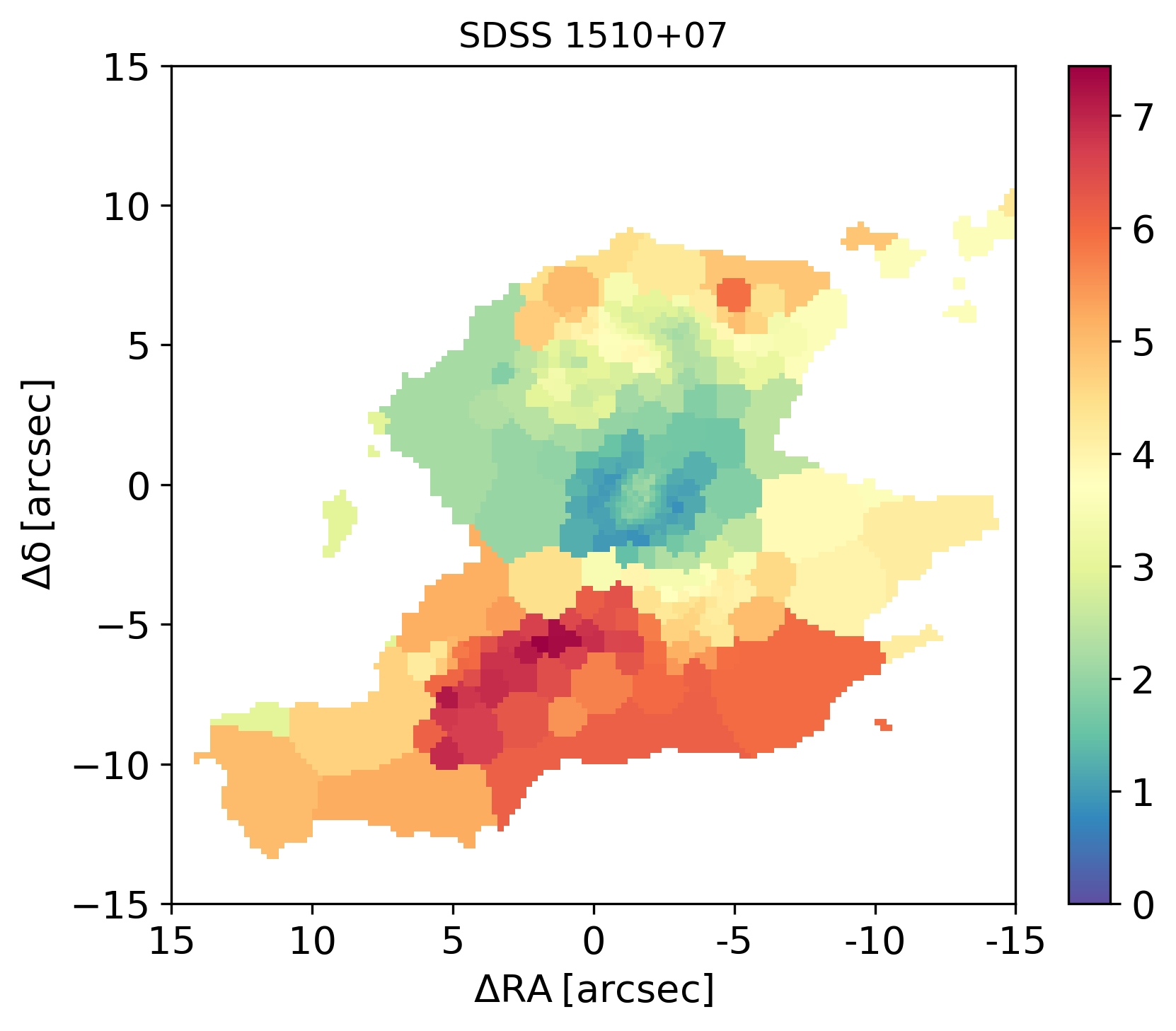}
    }
    \subfloat{
        \includegraphics[width=0.33\textwidth]{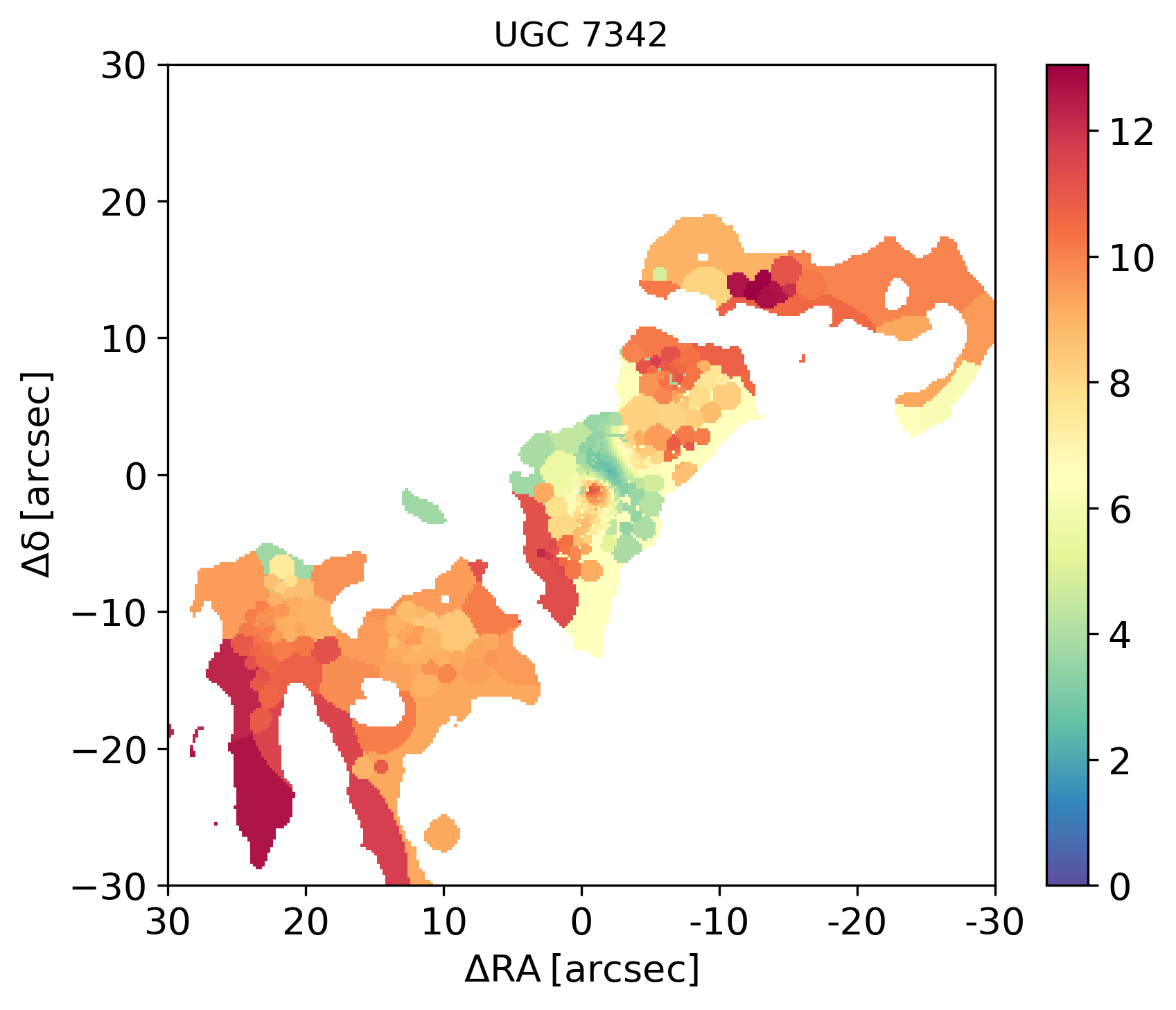}
    }    
    \subfloat{
        \includegraphics[width=0.33\textwidth]{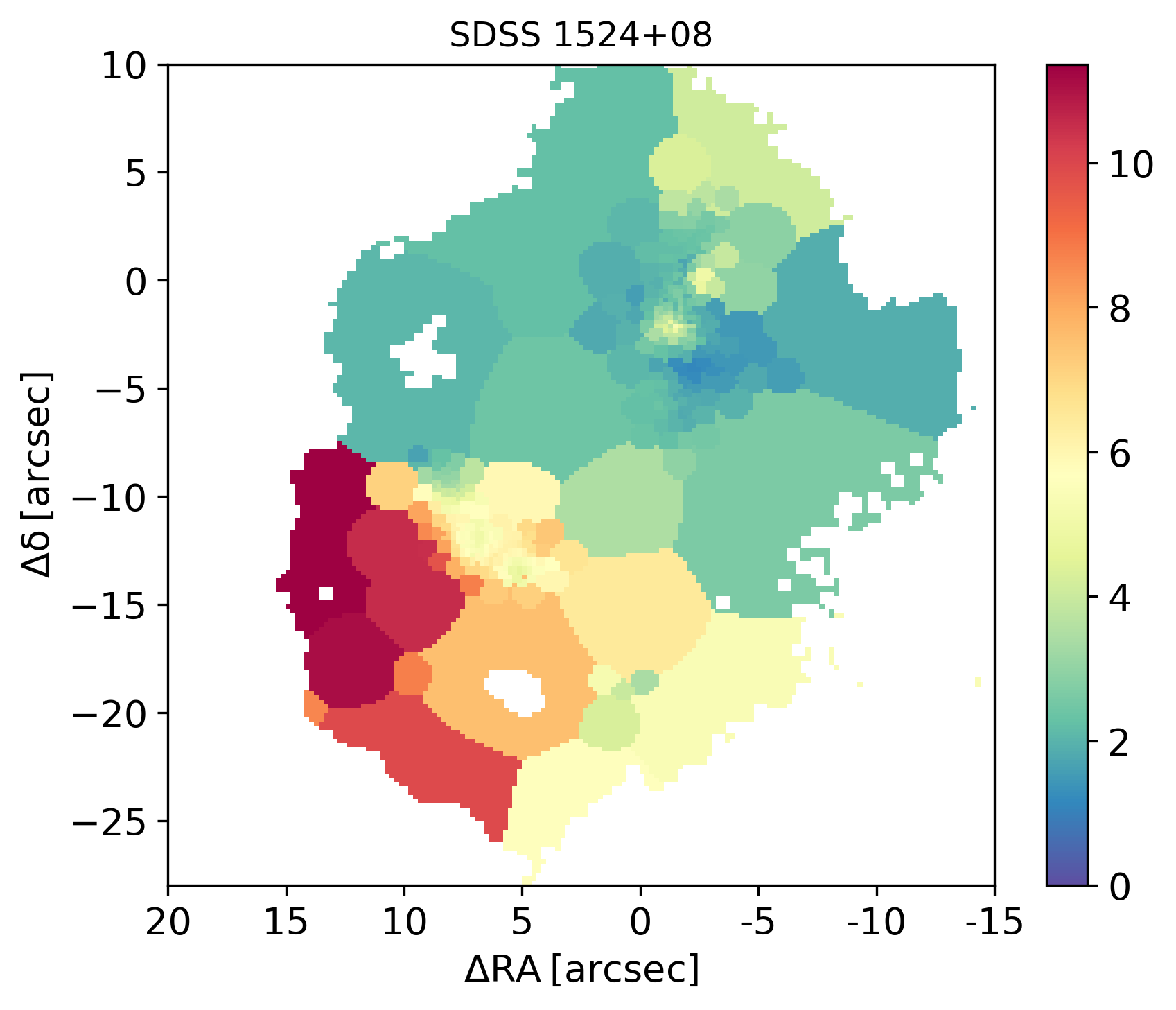}
    }
    
    \subfloat{
        \includegraphics[width=0.33\textwidth]{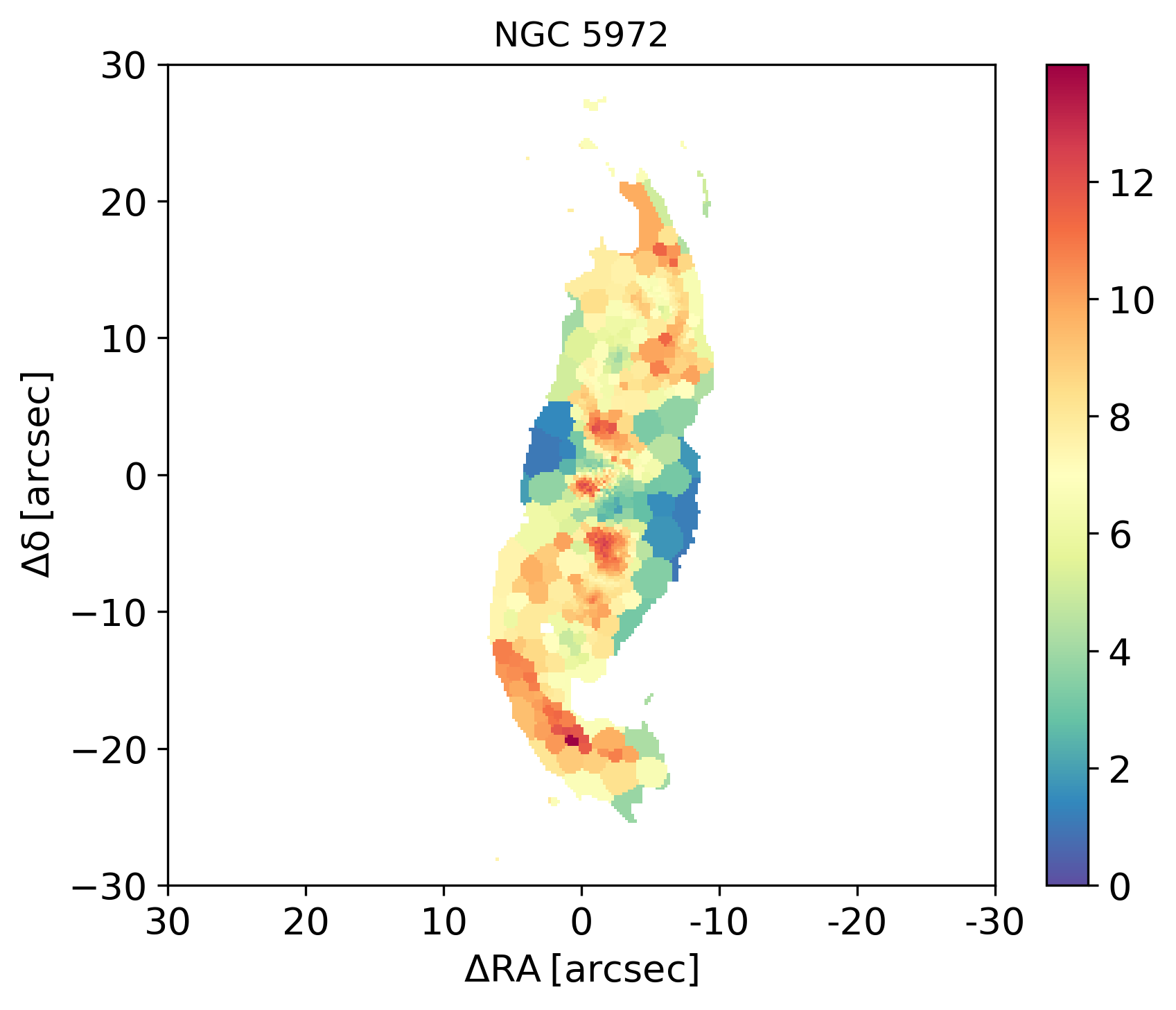}
    }
    \subfloat{
        \includegraphics[width=0.33\textwidth]{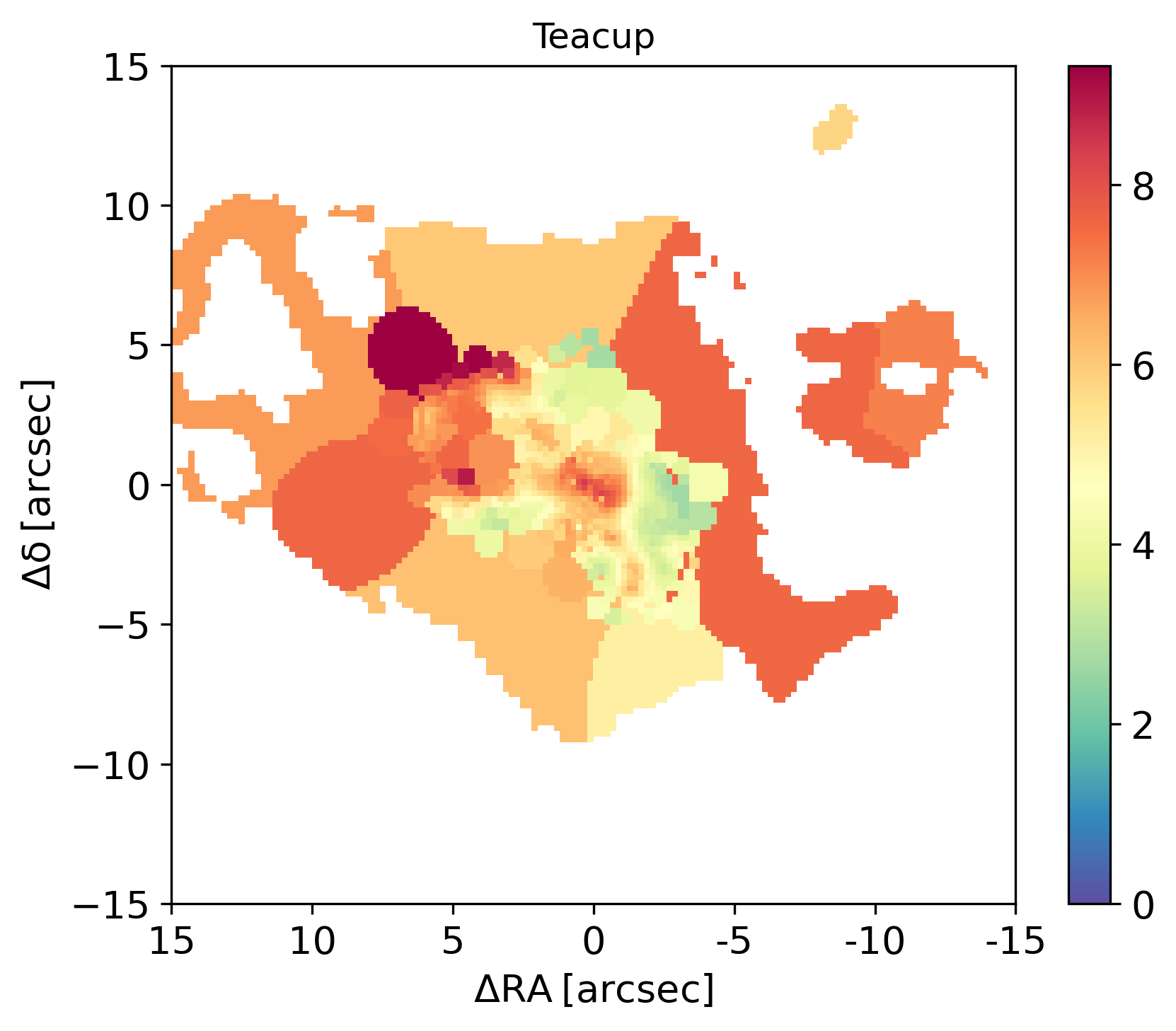}        
    }
\caption{ Spatially resolved [OIII]/H$\beta$ ratio maps. A general trend of higher values of this ratio along the EELR is observed, while closer to the nucleus, the ratio is lower. This suggests that near the nucleus, the [OIII] emission has already declined, but this fading has not yet reached the outer regions because of the large extension of the clouds. This is consistent with the hypothesis that EELRs are ``ionization echoes.''}
 \label{fig:oiiihbratio}
\end{figure*}

\citet[][]{binette+1987} examined how the properties of a photoionized cloud change over time after a sharp reduction in the intensity of the ionizing photon field. They sought to determine how long the nebula could continue to glow and maintain its excitation signatures after the central AGN becomes inactive. They found that [OIII] is rapidly quenched by an efficient charge transfer combination, which proceeds as soon as the gas begins to recombine and the fraction of neutral hydrogen increases. However, hydrogen recombines relatively slowly, and its timescale can be thousands of years in EELRs.
If the AGN is turned off suddenly, the [OIII] luminosity of the EELR would fade quickly, leading to a low [OIII]/H$\beta$ ratio. However, the fading progresses through the ionized cloud in a timescale that depends on the light crossing time and, thus, on the size of the EELR. Fig. \ref{fig:oiiihbratio} shows the [OIII]/H$\beta$ for all the sources, presenting an overall trend of higher [OIII]/H$\beta$ ratio along the EELR, in contrast with lower ratios near the nucleus. This is consistent with the interpretation of the EELRs being ``ionization echoes.'' Closer to the nucleus, the [OIII] emission has already declined, but this fading has not yet reached the outermost regions. Therefore, the EELR farthest from the center still retains information about the ionizing power before the shutdown.

\section{Conclusions}
\label{sec:conclusions}

We have presented in this work integral field spectroscopic VLT/MUSE observations for a small sample of nearby galaxies presenting prominent EELRs at large physical extensions. We have used the method outlined in \cite{finlez+2022} to analyze these sources in detail. Our main results can be summarized as follows:

   \begin{enumerate}
      \item We detect EELRs extending to 14, 27, and 14 kpc for SDSS1510+07, UGC 7342, and SDSS1524+08, respectively. The presence of tidal tails in the stellar component and the morphologies and kinematics of the ionized clouds indicate the presence of recent galaxy interactions in the sources' histories.  

 \item  The main ionizing mechanism, as shown through a BPT diagnostic, is consistent with AGN photoionization. In SDSS1510+07 and SDSS 1524+08, the Seyfert-like photoionization seems to follow a clear biconical distribution, with a combination of LINER and Composite emission in the central region and perpendicular to the possible bicone. For UGC 7342 effectively all the ionized gas is consistent with Seyfert ionization.

\item The kinematics of the ionized gas and stellar components show complex and different characteristics for each source. SDSS1510+07 seems to show the least perturbed kinematics of the sample, with a clearly rotating stellar component. The ionized gas shows signs of rotation on a different major axis than the stellar component, while the gaseous disk appears to be warped. Ionized gas has been hauled along the stellar tidal tails, where it was likely illuminated by the nuclear source. UGC 7342 shows weaker signs of rotation and a high degree of perturbation. Signatures of a biconical outflow are observed. The gas outside the galactic disk seems to be following the tidal tails and, therefore, is located outside the galactic plane. The low velocity dispersion values along these areas of the EELR, further from the nucleus, indicate that the gas may not be outflowing and is kinematically turbulent due to the past galaxy interaction.
For SDSS1524+08, there are weak signs of rotation only from a nuclear disk, both in the stellar and ionized gas components. However, at larger scales, the galaxy lacks rotational support. The velocities in the LOS and the velocity dispersion indicate a possible outflow along the bicone formed by the NLR. The main feature of the EELR, a large ionized cloud at the SE, shows small velocity dispersion and is located at the same spatial location as the tidal arms. Given that this cloud is in the direction of the bicone, this gas was possibly hauled by the tidal arms or even pushed by the biconical outflow and later illuminated by the AGN.

\item We use the CLOUDY photoionization code to generate a grid of models covering a range of ionization parameters to derive the luminosity required to ionize the gas at every aperture to the observed state. For all the sources in our sample, we see an increase in luminosity with increasing radius. This suggests that the AGN has faded between $\sim 0.2-3$ dex on time scales of $4-8\times 10^{4}$ yr. These variability amplitudes and related timescales are in good agreement with previous luminosity history studies and AGN variability models \citep[e.g.][]{lintott+2009,keel+2012a,gagne+2014,sartori+2018}.

\item The dramatic changes in luminosities described above occur in timescales that are in agreement with a scenario in which the AGN duty cycle is broken down in shorter phases of $10^{4-5}$ yr, resulting in a total $10^{7-9}$ yr "on" lifetime \citep[][]{schawinski+2015}. This "flickering" of the nuclear source could be responsible for the observed timescale variability. Considering that there remains significant ionized gas (and likely neutral and molecular as well) in the vicinity, it is reasonable to expect that a portion will eventually reach the nucleus within time frames of $< 10^{7-9}$ yr.

 \end{enumerate} 

Future observations of a larger and potentially less biased sample are necessary to further consolidate the results reported here regarding large timescale AGN variability. 
Considering the biases and selection effects discussed in \ref{Sec:caveats}, different regions of the AGN parameter space and their host galaxies must be explored to understand the reach of our conclusions.
We aim to explore this in a subsequent study, using a sample selected from the BAT AGN Spectroscopic Survey \citep[BASS;][]{koss+2017}, which is the first large all-sky survey of hard X-ray selected AGN with over 1,000 sources, providing the most detailed and complete census of SMBHs in the local Universe. One of our sources, The Teacup, is part of this sample. This large sample of bright AGN, selected uniformly and robustly, provides an unbiased sample of currently luminous AGN in the local Universe with well-measured SMBH masses and Eddington ratios.

\begin{acknowledgements}
We acknowledge support from FONDECYT through Postdoctoral grant 3220751 (CF), Regular 1190818 (ET, FEB), 1200495 (ET, FEB), 1250821 (ET, FEB), and 3230310 (YD); ANID grants CATA-Basal AFB-170002 (ET, FEB, CF), FB210003 (ET, FEB, CF) and ACE210002 (CF, ET, and FEB); Millennium Nucleus NCN19\_058 (TITANs; ET, CF, NN); Millennium Science Initiative Program – ICN12\_009 (FEB);  and European Union’s HE ERC Starting Grant No. 101040227 - WINGS (GV).
\end{acknowledgements}

\bibliography{refs}{}
\bibliographystyle{aa}

\begin{appendix}

\section{CLOUDY models parameters}
\label{sec:appendixA}

To assess the impact of the column density (N$_{H}$) on our CLOUDY models, we generated a suite of models with N$_{H}$ values spanning \( 10^{19} - 10^{23} \, \text{cm}^{-2} \). This range was chosen to cover the expected values for both radiation-bounded and matter-bounded clouds, allowing us to evaluate the influence of this parameter on our results.

\begin{figure*}
	\includegraphics[width=\textwidth]{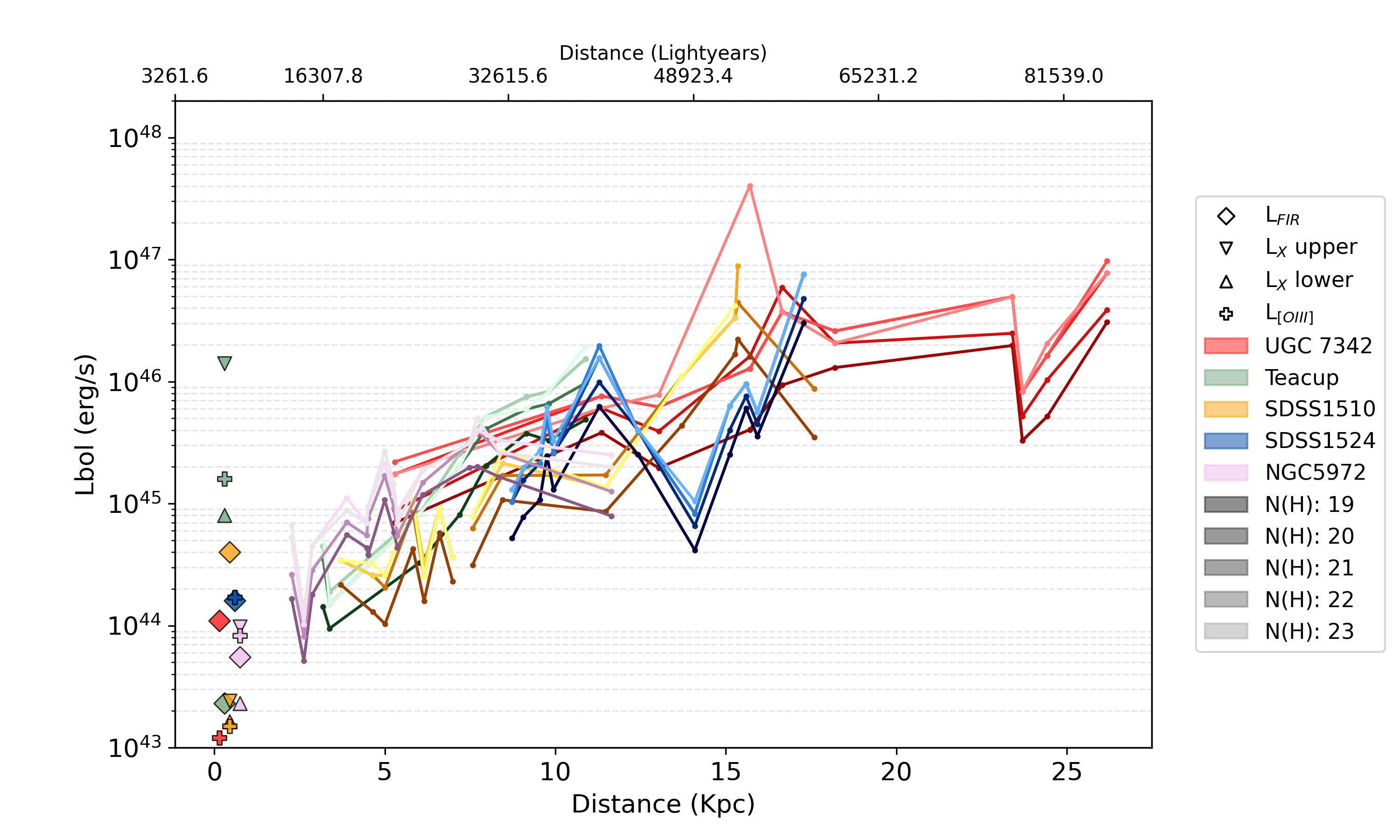}
    \caption{Derived bolometric luminosity as a function of the projected distance from the galaxy nucleus, as described in Fig. \ref{fig:lbol_all}. The model outlined in Sect. \ref{sec:lbolmodels} was run with five different values of N(H), ranging from $10^{19}$cm$^{-2}$ (darker color shades) to $10^{23}$cm$^{-2}$ (lighter color shades). The trend of increasing luminosity with projected distance persists across all values of N(H).}
    \label{fig:lbol_allmodels}
\end{figure*}

To illustrate the effect of N$_{H}$ on the derived bolometric luminosity (L$_{bol}$), Figure \ref{fig:lbol_allmodels} presents a modified version of Figure \ref{fig:lbol_all}, displaying five lines per object corresponding to \( \log(\text{N$_{H}$}) \) values of 19, 20, 21, 22, and 23.

We find that while variations in N$_{H}$ influence the ionization parameter (\( U \)) and, consequently, the estimated L$_{bol}$, the overall trend in L$_{bol}$ remains consistent across all N$_{H}$ values. Specifically, the estimated L$_{bol}$ increases with radius, and the difference in L$_{bol}$ between the nearest and furthest radii remains largely unchanged. However, differences in N$_{H}$ do affect the fading factor when comparing the earliest L$_{bol}$ estimate with the current AGN L$_{bol}$, derived from observations at different wavelengths. As a result, we consider the difference between the L$_{bol}$ estimated for N$_{H}$ = \( 10^{19} \, \text{cm}^{-2} \) and our earliest L$_{bol}$ estimate to represent a lower limit on the discrepancy between the current and earlier L$_{bol}$ values.

\begin{figure*}
	\includegraphics[width=\textwidth]{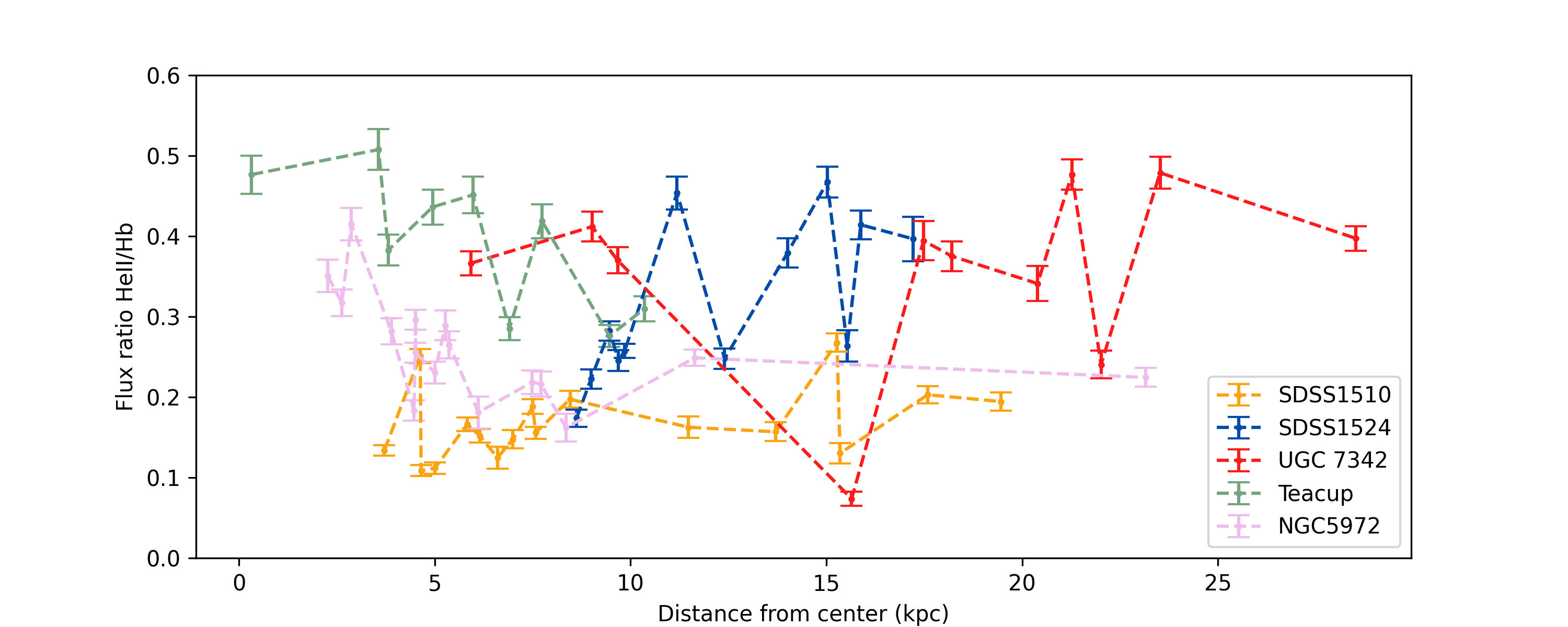}
    \caption{ He II/H$\beta$ flux ratio over projected distances from the center. Colors are the same as Fig. \ref{fig:lbol_all}.  }
    \label{fig:allHeIIHbratios}
\end{figure*}

\begin{figure*}
	\includegraphics[width=\textwidth]{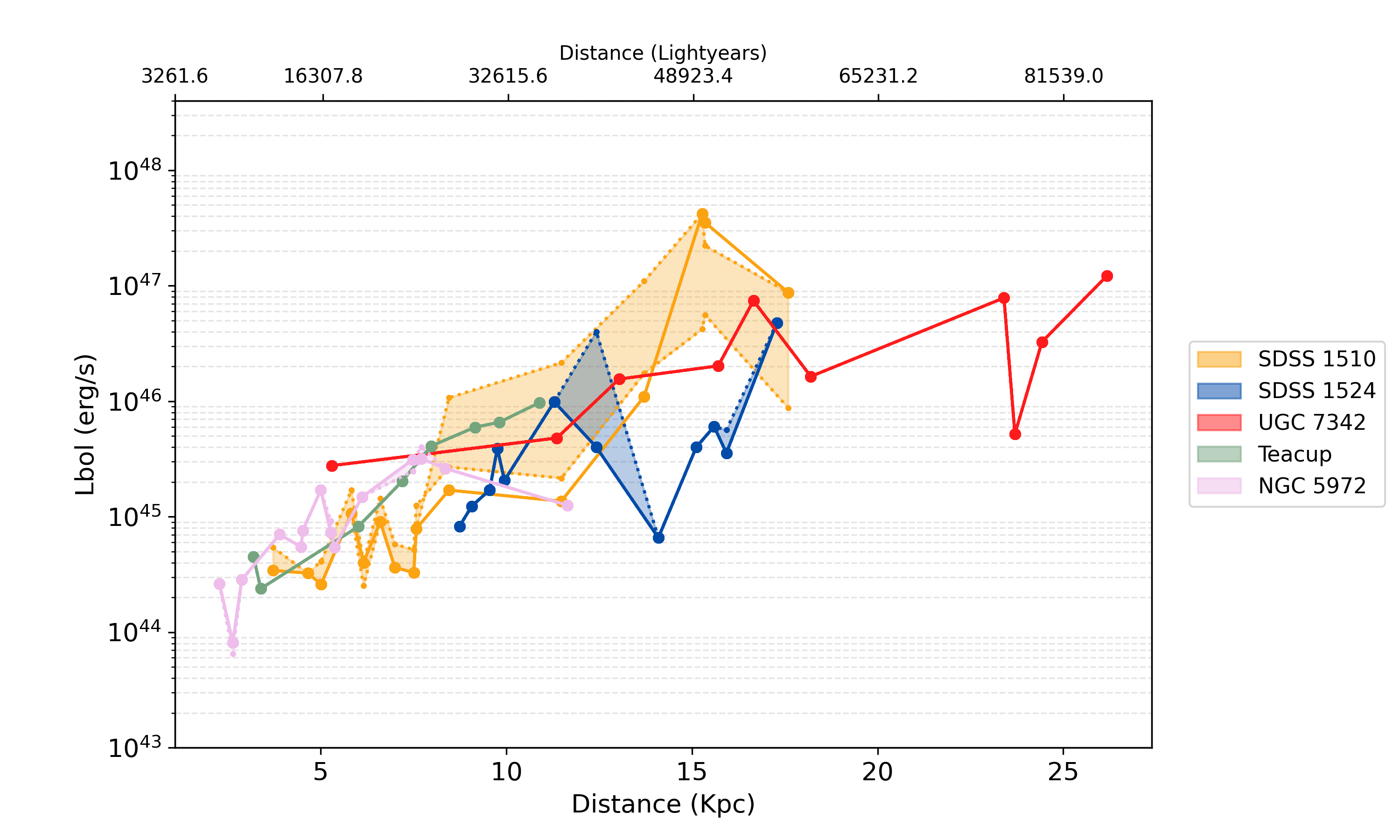}
    \caption{Derived bolometric luminosity as a function of the projected distance from the galaxy nucleus, as described in Fig. \ref{fig:lbol_all}. The model outlined in Sect. \ref{sec:lbolmodels} was run with different values of $alpha$ in the range 13.6--500 eV. This change in the shape of the continuum correspond to the dispersion of values in the He II/H$\beta$ ratio, showing as a solid line the median value, and in dotted lines (shaded in between) the $\pm 1 \sigma$ difference. For most of the galaxies in the sample, the change in the shape of the continuum translates into a small difference in the luminosity curve. SDSS 1510, which despite showing a larger spread in L$_{bol}$, maintains the trend of increasing luminosity with projected distance.}
    \label{fig:lbol_allmodels_alphas}
\end{figure*}

To estimate the impact of the shape of the continuum chosen, we have run a suite of models with different values of $\alpha$ in the range 13.6--500 eV, based on the values of the He II/H$\beta$ ratio observed along the VPs. In Fig. \ref{fig:allHeIIHbratios} we show the values of the He II/H$\beta$ along the VP of every source, at different projected distances from the galaxy nucleus. In Table \ref{table:heiihbratios} we tabulate the median values and the 1$\sigma$ dispersion of these values. We calculate the $\alpha$ values corresponding to the median and 1$\sigma$ values of the He II/H$\beta$ ratio, following \cite{penston+1978}. 
In Fig. \ref{fig:lbol_allmodels_alphas} we plot the luminosity curves corresponding to the median and $\pm 1\sigma$ values of the He II/H$\beta$ ratio, for every galaxy. We find that the variations observed in the He II/H$\beta$ ratio translate into mostly minor variations to the luminosity curve. The exception is SDSS 1510+07, for which the variations are larger. However, the trend of increasing luminosity with radius remains.

\begin{table}[]
\caption{He II/H$\beta$ ratios along the VP}             
\label{table:heiihbratios}      
\centering          
\begin{tabular}{lccc}   
\hline\hline   
Source   & Median & 1$\sigma$ \\
(1)   & (2) & (3) \\
\hline
SDSS1510 & 0.16                    & 0.043                                   \\
SDSS1524 & 0.31                    & 0.095                                   \\
UGC 7342 & 0.38                    & 0.131                                   \\
Teacup   & 0.10                    & 0.054                                   \\
NGC 5972 & 0.25                    & 0.064                                   \\

\hline                  
\end{tabular}
\tablefoot{Columns: (1) Target name, (2) Median value of He II/H$\beta$ along the VP, (3) 1$\sigma$ dispersion of the He II/H$\beta$ ratio along the VP.}
\end{table}

\end{appendix}
\end{document}